\documentclass[twocolumn]{aa}
\usepackage{amsmath}
\usepackage{amssymb}
\usepackage{epsfig}
\usepackage{txfonts}
\usepackage{natbib}
\usepackage{graphicx}
\usepackage{float}
\usepackage{here}
\usepackage[figuresright]{rotating}
\usepackage{subfig} 
\def\deg{$^{\circ}$}

\newcommand{\arcs} 	{$^{\prime\prime}$}
\newcommand{\arcm} 	{$^{\prime}$}

\newcommand{\mr}   	{\mathrm}
\newcommand{\tco}   	{$^{13}$CO}
\newcommand{\ceo}   	{C$^{18}$O}
\newcommand{\cts}   	{C$^{34}$S}
\newcommand{\nht}   	{NH$_{3}$}
\newcommand{\hk}		{$(H-K_{\mr{S}})$}
\newcommand{\jh}		{$(J-H)$}

\begin{document}
\title{Physical Characteristics of a Dark Cloud in an Early Stage of Star Formation 
toward NGC 7538\thanks{Based on observations obtained with the 100-m telescope of 
MPIfR (Max-Planck-Institut f\"{u}r Radioastronomie) at Effelsberg and
the IRAM 30 m telescope. IRAM is supported by INSU/CNRS (France), MPG 
(Germany) and IGN (Spain).}
}
\subtitle{An Outer Galaxy Infrared Dark Cloud?}

\author{Wilfred W.F. Frieswijk\inst{1,2} \and Marco Spaans\inst{1}
  \and Russell F. Shipman\inst{2} \and David Teyssier\inst{2,3} \and Pierre
  Hily-Blant\inst{4}}

   \offprints{W.W.F. Frieswijk}

   \institute{Kapteyn Astronomical Institute, University of Groningen, PO Box 800, 9700\ \ AV Groningen, The Netherlands
            \and
     SRON, National Institute for Space Research, PO Box 800, 9700\ \ AV Groningen, The Netherlands
             \and
     ESAC, European Space Astronomy Centre, Urb. Villafrancadel Castillo, P.O. Box 50727, Madrid 28080, Spain
         \and
     IRAM, Institut de Radio Astronomie Millim\'{e}trique, 300 rue de la Piscine, St-Martin d'\`Heres, France\\
     \email{frieswyk@astro.rug.nl}
     }

   \date{Received - -, -; accepted - -, -}

   \abstract{In the inner parts of the Galaxy the Infrared Dark Clouds (IRDCs) are 
     presently believed to be the progenitors of massive stars and star clusters. 
     Many of them are predominantly devoid of active star formation and for now
     they represent the earliest observed stages of massive star formation. Their 
     Outer Galaxy counterparts, if present, are not easily identified because of a 
     low or absent mid-IR background.}{We characterize the ambient conditions by determining
     physical parameters in the Outer Galaxy IRDC candidate G111.80+0.58,
     a relatively quiescent molecular core complex in the vicinity of active 
     star forming regions such as NGC 7538 and S159.}{We
     conduct molecular line observations on a number of dense cores
     in G111.80+0.58. We analyze the data in terms of excitation
     temperature, column and volume density, mass and stability.}{The
     temperatures we find (15\,--\,20\,K) are higher than expected from 
     only cosmic ray heating, but are comparable to those found in 
     massive cores, such as IRDCs. Star forming activity
     could be present in some cores, as indicated by the presence of warm gas
     (\nht, \tco\ self-absorption) and Young Stellar Object candidates. 
     The observed super-thermal line-widths are typical for star forming regions.
     The velocity dispersion is
     consistent with a turbulent energy cascade over the observed size
     scales of the complex. We do not find a correlation between the
     gas temperature and the line-width. The LTE masses we derive are much larger
     than the thermal Jeans mass. Therefore, fragmentation is expected and may
     have occurred already, in which case the observed lines represent the combined
     emission of multiple unresolved components.}
     {We conclude that G111.80+0.58 is a molecular core complex with bulk properties
     very similar to IRDCs in an early, but not pristine, star forming state. The individual 
     cores are close to virial equilibrium and some contain sufficient material to form 
     massive stars and star clusters. 
     The ambient conditions suggest that turbulence is involved in supporting the 
     cores against gravitational collapse, at least down to the observed sizes. 
     Additional high resolution data are necessary to resolve and analyze the 
     smaller scale properties.}
 
     \keywords{Molecular data -- Stars: formation -- ISM: clouds, molecules, structure}

\authorrunning{W.W.F. Frieswijk et al.}
\titlerunning{Physical characteristics of an Outer Galaxy dark cloud}
   \maketitle
%__________________________________________________________________
\section{Introduction}
\label{intro}
Infrared Dark Clouds (IRDCs) were discovered a decade ago as dark
silhouettes against a bright mid-infrared background by the Midcourse
Space Experiment \citep[MSX, ][]{1998ApJ...494L.199E} and the Infrared
Space Observatory \citep[ISO, ][]{1996A&A...315L.165P}.
Many IRDCs contain compact (sub-) millimeter cores 
\citep{2000ApJ...543L.157C,2004ApJ...610..313G,2005A&A...439..613O,2005ApJ...630L.181R}
and the current picture is that these massive 
dense cores represent the early, cold stages of clustered star formation 
\citep[][ and references therein]{2006ApJ...641..389R,2005IAUS..227...23M}
and that they are potentially the birth sites of massive stars. 
In order to assess the role of IRDCs in the process of star formation,
their ambient physical conditions need to be determined. Only then
insight may be gained into the differences between low- and high-mass
star formation, the nature of the initial mass function and the impact
of environment on star (cluster) formation. That is, the putative
early stage that IRDCs represent implies that their physical state
provides a direct record of the initial conditions pertinent to stellar birth.\\
A complete picture of the star forming properties in massive dark clouds, specifically 
the effects of external conditions on the formation process, requires a study of
similar objects in different environments. The Outer Galaxy obviously represents the
place in our Galaxy where the conditions, such as metallicity, density, interstellar 
radiation field and overall star formating activity are different compared to the inner 
regions of the Galaxy \citep[e.g.,][]{1995A&A...303..851B,2006ApJS..162..346R}. 
Variations of these properties from cloud to cloud may affect 
the local star formation rate, the mass distribution (IMF) or the star forming efficiency.
In fact, only a direct comparison between Inner and Outer Galaxy star forming regions
can asses in what way star formation differs with Galactic radius, if at all.
However, due to a lack of bright background emission elsewhere, IRDCs were identified
only in the inner parts of the Galaxy, mainly toward the Molecular Ring and the inner 
spiral arms \citep{2006ApJ...653.1325S}. Can we identify massive dark clouds in an early
evolutionary stage in the Outer Galaxy using a different approach?\\
\citet[][ in prep., FS07 hereafter]{frieswijk2007} propose a list of candidate IRDC-like objects in the 
Outer Galactic Plane in their investigation of the distribution of highly reddened point 
sources observed in the Two Micron All Sky Survey 
\citep[2MASS,][]{2006AJ....131.1163S}. They suggest that a number of these 
clustered red sources are reddened due to foreground extinction in the form of 
dark clouds. As a verification of the presence of molecular material, they
identified CO structures, observed in the Five College Radio Astronomy Observatory 
CO Survey of the Outer Galaxy \citep[FCRAO,][]{1998ApJS..115..241H}, that match the global morphology 
of the regions. Even though CO traces mainly the outer layers of molecular clouds, the amount of 
extinction they find indicates the presence of large column densities. 
A subset of the structures in this near-IR study are not associated with MSX or IRAS point sources
and this would also suggest an early star forming nature. Additional data, e.g., of molecular lines
and dust continuum are required to confirm if some of these near-IR structures are 
indeed the Outer Galaxy counterparts to the Inner Galaxy IRDCs.\\
In this paper we present follow-up molecular line observations of the Outer Galaxy dark cloud 
candidate G111.80+0.58 in the direction of Cepheus in the Perseus spiral arm. We derive physical 
properties for a  number of dense cores located along the filamentary cloud structure, such as 
temperature, density and mass and we characterize the nature of the cloud by comparing
the results to existing studies.
The cloud is part of the Cas OB2 complex at a radial velocity around $-$55\,km\,s$^{-1}$ 
\citep[e.g.,][]{1982ApJS...49..183B}. There are a few well-known star forming regions located nearby,
such as NGC 7538 and S159 . NGC 7538 is actively forming stars and extensive 
studies have been conducted and reveal various stages of evolution including compact 
dense cores, HII regions, several massive proto-stellar objects and many lower mass 
Young Stellar Objects (YSO's)
\citep[e.g.,][]{1979MNRAS.188..463W,1990ApJ...355..562K,2004ApJ...600..269S}.
S159 is a bright reflection nebulae illuminated by an optically visible O-type star
\citep{1984A&A...139L...5C}. Part of the cloud is associated with IRAS 23133+6050 and seen 
as an emission nebula.
In the immediate surrounding of S159 several compact radio continuum sources and (compact) HII 
regions have been identified \citep[e.g.,][]{2001ApJ...560..806L}.\\
Section \ref{context} describes how the object was identified and gives a brief overview of the
the appearance of the region and its immediate environment in MSX, 2MASS and FCRAO data. 
The observational setup and the target positions are given in Section \ref{obssetup} and 
Section \ref{obsprop} describes the observational properties.
The derived physical properties of the cores are given in Section \ref{results}. In Section 
\ref{discussion} we discuss the results and present our conclusion on the
nature of dark cloud G111.80+0.58. Section \ref{conclusions} ends with
some concluding remarks.
%__________________________________________________________________
\section{G111.80+0.58: Source selection}
\label{context}
G111.80+0.58 was selected from a list of candidate dark clouds.
This list comprises all sources that were identified in the
2MASS Point Source Catalog (PSC) as extended red features in the
Outer Galactic Plane (FS07).
In this work, a statistical measure, using the Mann-Whitney-U-test, is performed on a 
60\arcs $\times$\,60\arcs\, grid covering the entire 
Outer Galactic Plane. Adjacent cells where the \hk\ color distribution of stars is different from the 
local surroundings on a high ($>$\,99\%) confidence level are selected as initial targets. The dark cloud candidates 
are then chosen on the basis of their abnormal red color and an absence of counterparts in SIMBAD. In 
this Section we briefly describe the appearance of the G111.80+0.58 region in MSX, 2MASS and 
FCRAO data. Details of interest for the individual target positions which were selected for single pointing 
observations (see Sec. \ref{point}) are discussed in Section \ref{individual}.\\
We selected this region specifically for a follow-up study because it was identified as an excessively 
red, filamentary region in a pilot-survey covering blindly about ten square degrees in the second Galactic quadrant.
The color scaling in Figures 1, 2 and 3 shows the MSX 8\,$\mu$m 
emission toward the location of the complex and its immediate surroundings. The two prominent emission 
regions labeled A and B correspond to NGC 7538 and S159, respectively. The weaker emission near
label C is associated with several masers and radio sources. The outlined region depicts 
the area mapped in \ceo\ (Fig. 4). The target positions for single pointing observations (Table 1)
are indicated by the black open stars. Compared to the active regions in the field, only faint 8\,$\mu$m 
emission ($<$\,$5\times10^{-6}$\,Wm$^{-2}$sr$^{-1}$) is associated with this area. Within the box, several 
IRAS sources are present, indicated by the filled red stars.\\
The blue contours in Figure 1 display the red tail of the \hk\ colors of stars observed 
in 2MASS on an oversampled 30\arcs $\times$\,30\arcs\, grid. The solid and dashed contour represent 
the values 0.52 and 0.68\,mag, respectively. Assuming that the \hk\ colors are related to the extinction they 
can be analyzed using the Near Infrared Color Excess technique \citep[e.g.,][]{1994ApJ...429..694L}. The
near-IR extinction in a cell is then given by
\begin{equation}
<E(H-K_{\mr{S}})_{\mr{cell}}> = \frac{\sum^N_{i=1} E(H-K_{\mr{S}})_{i}}{N}\ \ \ ,
\end{equation}
where the summation is over $N$ stars present in a grid-cell. The color excess 
$E(H-K_{\mr{S}})_{i}$ per star is the difference between the observed color and the
intrinsic color, where the latter is derived from an off-position chosen to represent the
color distribution free of extinction.
Note that foreground stars will play a significant role and reduce the average reddening
per cell at large cloud distances. A correction for this, e.g., by comparing with the off-position, 
is required to determine the extinction accurately. This is the main reason why the identification
is based on a statistical color distribution instead of the color excess, because a priori a distance to 
the objects is unknown and off-positions were not chosen while processing the entire
Outer Galactic Plane. Keeping this in mind and adopting an intrinsic \hk\ value of 0.14\,mag,
the contour values correspond to $\sim$\,6 and $\sim$\,8.5\,mag extinction, for convenience
converted to $A_{\mr{V}}$ using a standard extinction law 
\citep[i.e., $A_{\mr{V}}$=15.9$\times$\,$<$\,$E(H-K_{\mr{S}})_{\mr{cell}}$\,$>$\,;][]{1985ApJ...288..618R}. The peak 
extinction that is measured this way corresponds to $\sim$\,15\,mag in $A_{\mr{V}}$.
Note that these values are similar to those reported for Inner Galaxy IRDCs using the same 
color excess method \citep[e.g., G48 toward W51 with peak $A_{\mr{V}} \sim$10\,--\,20\,mag;][]{2003ASPC..287..252S}.\\
The contours in Figure 2 give a measure of the 2MASS stellar distribution. The mean
distribution of the 2 by 2 square degree surrounding field is $\sim$\,8 stars per 60\arcs $\times$\,60\arcs\,
cell with a 1$\sigma$ spread of 3 stars. The red contours depict the regions deficit in stars 
(4 and 3 stars per cell) whereas the blue contours show a surplus (11 and 13 stars per cell).
The star counts are also related to the extinction along the line of sight, but here, a direct
translation is difficult because they represent data from an incomplete 2MASS catalog,
i.e., the Point Source Catalog including the faint extension. Therefore, they merely give an 
indication of the stellar distribution and the number of stars that is used for the identification and 
color excess method.\\
\begin{figure}[htbp]
\begin{minipage}[c]{1\linewidth}
\label{msxhk}
\centering \includegraphics[width=8.5cm]{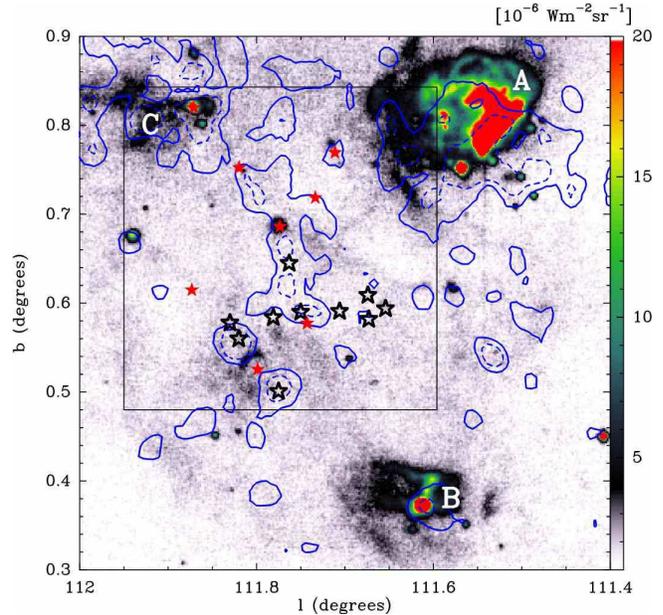} 
%\end{minipage}
\caption{8\,$\mu$m emission of the region toward G111.80+0.58. The two prominent 
sources labeled A and B are NGC 7538 and S159, respectively. Region C is associated with 
several masers and radio sources. The region of interest for the current work is outlined by the 
box and is mostly devoid of bright 8\,$\mu$m emission. The contours show the \hk\ color distribution.
The solid and dashed contours correspond to 0.52 and 0.68 mag, respectively. The black open stars 
represent the target positions (Table 1) where we conducted single pointing observations. The red 
filled stars correspond to IRAS point sources.}
\end{minipage}
\end{figure} 
\begin{figure}[htbp]
\begin{minipage}[c]{1\linewidth}
\label{msxsc}
\centering \includegraphics[width=8.5cm]{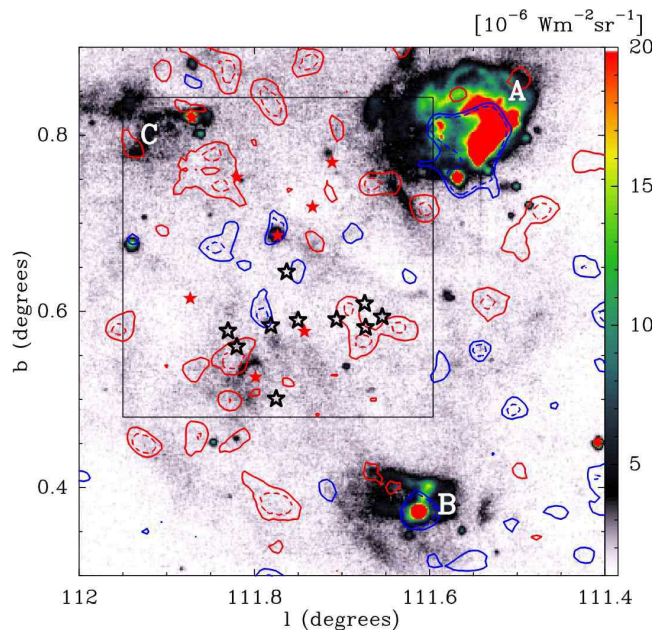} 
\caption{Same as Figure 1, except the contours represent the 2MASS source distribution on a
60\arcs $\times$\,60\arcs\, grid. The red dashed and solid contours correspond to 3 and 4 stars per 
cell, respectively. The blue dashed and solid contours correspond to 13 and 11 stars per cell, respectively. 
The average source count per cell is 8.}
\end{minipage}
\end{figure}
Based on the 2MASS data, four different appearances can be distinguished in the area of interest
and may be explained intuitively by the following:\\
1) The combination of color excess and a deficit in star counts. The measured extinction is due to
background and, if present, embedded stars and the column of foreground material is large
resulting in fewer background stars compared to the off-position.\\
2) The combination of color excess and a surplus in star counts. The measured extinction is due
to background and embedded sources. The embedded objects result in an enhancement in
the star count distribution.\\
3) Only star count contours. A large column of material prevents the observation of sufficient 
background reddened stars resulting in a decline of the stellar distribution. The average
color distribution is normal either due to a complete absence of red background stars or
due to the presence of too many foreground stars.\\
4) Only color excess contours. Red background and embedded sources account for the average
red color and the embedded sources may compensate for the deficit of background stars
due to the extinction.\\
Note that S159 and NGC 7538 both show up as reddened regions and have a surplus of stars, 
as can be expected for these active star forming regions where multiple embedded sources are
present.\\
G111.80+0.58 is not an entirely unknown object. It was identified in the FCRAO survey and
cataloged as part of a molecular cloud \citep{2003ApJS..144...47B}. 
The blue contours in Figure 3 display the CO data integrated approximately between $-$45 and 
$-$60\,km\,s$^{-1}$.
\begin{figure}[htbp]
\begin{minipage}[c]{1\linewidth}
\label{msxcgps}
\centering \includegraphics[width=8.5cm]{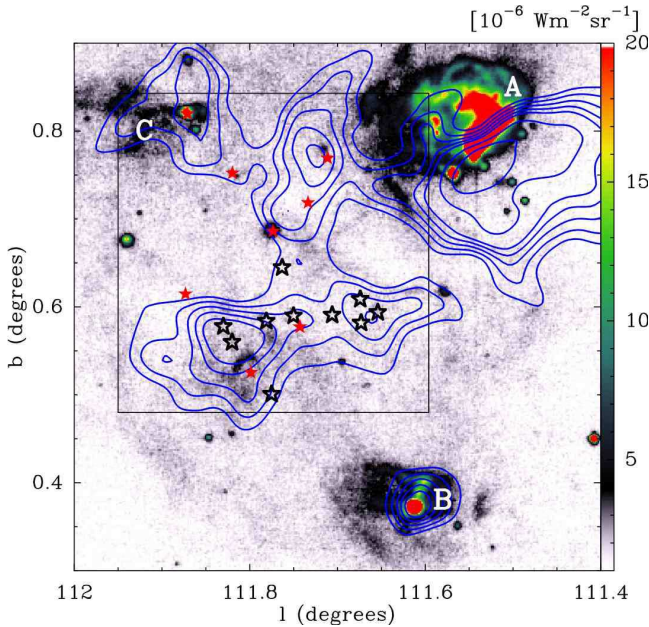} 
\end{minipage}
\caption{Same as Figure 1, except the contours represent the integrated CO emission between
$-$45 and $-$60 km\,s$^{-1}$ as observed in the FCRAO survey. The contours range from 60 to 100 
K\,km\,s$^{-1}$ in steps of 10 K\,km\,s$^{-1}$.}
\end{figure}
The cloud is in velocity space as well as in projection on the sky part of a large 
star forming molecular cloud complex in the Perseus spiral arm which also includes the
aforementioned regions NGC 7538 and S159. The kinematic distance 
to the G111.80+0.58 complex, assuming a flat rotation curve and the IAU standard constants 
$V_{\odot}$\,=\,220\,km\,s$^{-1}$ and $R_{\odot}$\,=\,8.5\,kpc, is $\sim$\,5.0\,kpc.
S159 has a similar central velocity ($-$56 km\,s$^{-1}$) as the object presented in this 
paper. However, a distance of 3.1\,kpc was reported by \cite{1993A&A...275...67B} for this cloud,
based on spectrophotometric observations of associated stars. Also, the distance to NGC 7538 is 
generally considered to be less \citep[2.8\,kpc; e.g.,][]{2000ApJ...537..283V} than the kinematic 
distance. It appears that this part of the Perseus spiral arm is actually closer to us 
than suggested by the radial velocity. For the analysis in this paper we adopt a distance of
3.1$\pm$0.2\,kpc, but this may be scaled accordingly.\\
More extensive details on the identification of candidate IRDCs in the Outer Galactic Plane are 
given in FS07, but the spectroscopic data presented in this paper indeed 
confirm the presence of a filamentary molecular cloud complex at the identified position.

\section{Observations}
\label{obssetup}
The observations presented in this paper were carried out with the
Effelsberg 100-m telescope in Germany and the IRAM 30-m telescope in
Spain early 2005. An overview of the observed positions and the main
parameters of the observed lines are given in Tables~\ref{tab1}
and \ref{tab2}.\\
\begin{table}
\begin{minipage}[t]{\columnwidth}
\caption{Source list}
\ \\
\label{tab1} 
\centering  
\renewcommand{\footnoterule}{}  
\begin{tabular}{c c c c} 
\hline\hline\\[-1.8ex]
Name & $l$       & $b$      & $S$\footnote{Size of the corresponding core (see Sec. \ref{size}), derived from the
\ceo\ map displayed in Fig. 4 and assuming a distance $D$ of 3.1\,kpc, where $D$ is adopted from S159 \citep{1993A&A...275...67B}.} \\  
            & (\deg) & (\deg) & ($\frac{D}{3.1}$ pc) \\[1.0ex]
\hline\\[-1.8ex]
G111.80+0.58 & 111.80 & 0.575 &  \\
P1 & 111.65 & 0.59 & 1.04  \\
P2 & 111.67 & 0.58 &  0.80  \\
P3 & 111.67 & 0.61 &  1.14  \\
P4 & 111.71 & 0.59 &  0.77  \\
P5 & 111.75 & 0.59 &  1.01  \\
P6 & 111.76 & 0.65 &  1.01  \\
P7 & 111.78 & 0.50 &  0.90  \\
P8 & 111.78 & 0.58 &  1.19  \\
P9 & 111.82 & 0.56 &  0.60  \\
P10 & 111.83 & 0.58 &  1.13  \\
S159 (reference) & 111.61 & 0.38 & 0.60 \\
\hline\\[-1.5ex]
\end{tabular}
\end{minipage}
\end{table}

\begin{table}
\begin{minipage}[t]{\columnwidth}
\caption{Line parameters} 
\ \\
\label{tab2} 
\centering
\renewcommand{\footnoterule}{}  
\begin{tabular}{c c c c c c c}
\hline\hline\\[-1.8ex]
\multicolumn{2}{c}{Line} & Frequency & HPBW & $T_{\mr{sys}}$ & $F_{\mr{eff}}$\footnote{Forward efficiency, only for IRAM observations} & $B_{\mr{eff}}$\footnote{Beam efficiency}   \\ 
                  &                   &  (GHz)            & (\arcs)         & (K) & & \\[1.0ex] 
\hline\\[-1.8ex]
\tco           &  1-0 & 110.201 & 22.5 & 100-200 & 0.95 & 0.75 \\
                  &  2-1 & 220.399 & 11.2 & 400-900 & 0.91 & 0.77\\
\ceo          &  1-0 & 109.782 & 22.5 & 110-200 & 0.95 & 0.75\\
                  &  2-1 & 219.560 & 11.2 & 200-600\footnote{200\,--\,350\,K for the map, 400\,--\,600\,K for single pointing observations} & 0.91 & 0.77\\
\cts           &  2-1 & 96.413 & 25.5 & 115-135 & 0.95 & 0.55\\
\nht           &  1,1 & 23.694 & 40 & 120-400 & -& 0.58\\
                  &  2,2 & 23.723 & 40 & 120-400 & -& 0.58\\
                  &  3,3 & 23.870 & 40 & 120-400 & -& 0.58\\
%\tco           &  1-0 & 110.201353 & 22.5 & 100-200 & 0.95 & 0.75 \\
%                  &  2-1 & 220.398686 & 11.2 & 400-900 & 0.91 & 0.77\\
%\ceo          &  1-0 & 109.782160 & 22.5 & 110-200 & 0.95 & 0.75\\
%                  &  2-1 & 219.560319 & 11.2 & 200-600\footnote{200\,-350\,K for the map, 400\,-600\,K for single pointings} & 0.91 & 0.77\\
%\cts           &  2-1 & 96.4129495 & 25.5 & 115-135 & 0.95 & 0.55\\
%\nht           &  1,1 & 23.694495 & 40 & 120-400 & &\\
%                  &  2,2 & 23.722633 & 40 & 120-400 & &\\
%                  &  3,3 & 23.870129 & 40 & 120-400 & &\\
\hline\\[-1.8ex]
\end{tabular}
\end{minipage}
\end{table}
%$F_{eff}$=\,0.95 for \tco\ 1-0, \ceo\ 1-0 and \cts\ 2-1; $F_{eff}$=\,0.91
%for \tco\ 2-1 and \ceo\ 2-1; $B_{eff}$=\,0.75 for \tco\ 1-0 and \ceo\ 1-0;
%$B_{eff}$=\,0.77 for \tco\ 2-1 and \ceo\ 2-1, and $B_{eff}$=\,0.55 for \cts\ 2-1.
%-----------------------------------------------------------------------
\subsection{Effelsberg 100-m observations}
\label{effel}
Single pointing observations of the \nht\ (1,1), (2,2) and (3,3)
inversion transitions were performed using the Effelsberg 100-m telescope 
of the Max-Planck-Institut f\"{u}r Radioastronomie to assess the gas kinetic 
temperatures at the positions listed in Table \ref{tab1}. The first set of data
were taken under reasonable winter conditions in January and February
2005 using the 18\,--\,26\,GHz frontend in position switch mode. The
receiver was tuned to a central frequency of 22\,GHz and the
auto-correlator with 2 sub-units of 20\,MHz bandwidth covered the 3
transitions simultaneously. After smoothing the data the spectral
resolution was $\approx$\,0.2\,km\,s$^{-1}$. At the 100-m telescope, no
chopper wheel calibration is available, so that the atmospherical
opacity and the gain response of the receiver need to be calibrated
with photometric references. The details of the procedure used to
calibrate our data are given in Appendix A. The resulting system
noise temperatures are of order 120\,--\,400\,K. A beam efficiency of $\eta$\,=\,0.58
is used to convert the antenna temperature $T_{\mr{A}}^*$ to the main beam 
temperature $T_{\mr{mb}}$. Additional data were taken in the following months 
and calibration was performed using a position observed in the first run. 
Data reduction was done using the CLASS package  \citep{Forveille1989}.
For the analysis of the kinetic temperature the absolute scaling is not 
a concern because it depends only on the ratio of the hyperfine (1,1) 
lines and the (1,1) to (2,2) brightness temperature ratios. The spectra are 
displayed in Figures \ref{figA1}, \ref{figA2} and \ref{figA3} in the Appendix.
%-----------------------------------------------------------------------
\subsection{IRAM 30-m observations}
\label{iram}
Observations at the IRAM-30m telescope were conducted under excellent
winter conditions (zenith opacity at 220\,GHz $\lesssim$\,0.05)
in February 2005. A fully-sampled map of 18\arcm $\times$\,18\arcm\, in
\ceo\ 2-1, covering the main complex, was obtained using the
HERA multi-beam instrument \citep{2004A&A...423.1171S}. The
corresponding SSB receiver and system temperatures were
$\approx$100\,--\,250\,K and $\approx$\,200\,--\,350\,K, 
respectively. The on-the-fly observing mode was used to allow
continuous data acquisition as the antenna was moving, with
a scanning velocity of 1\arcs\,s$^{-1}$. Frequency-switching mode
was used to subtract the sky background contribution. The
HERA matrix projected on the sky was rotated by 9.6$^\circ$\
to provide a 4\arcs\ spatial sampling in both directions.
Maps were done in lambda and
beta directions to minimize striping effects due to temporal
drifts. Each spectrum has 896 channels of 80\,kHz or 0.1\,km\,s$^{-1}$
width thanks to the VESPA autocorrelator facility
backend. Data were reduced using the \texttt{GILDAS}
software\footnote{\texttt{URL:
http://www.iram.fr/IRAMFR/GILDAS/}}. The bandpass of the
system was removed from each spectrum prior to folding, by
subtracting a low-order ($\le$\,3) polynomial. The spectra
were then resampled on a 6\arcs\ grid by Gaussian
convolution with a final rms of 0.17\,K in each 0.1\,km\,s$^{-1}$
channel.\\
In addition, pointing observations were performed toward the integrated 
intensity peaks in \ceo\,1-0 and 2-1 (for column density), \tco\,1-0 and 2-1 
(for temperature) and \cts\,2-1 (as a high density tracer). The receivers were 
connected to an autocorrelator with a resolution of 80\,kHz resulting
in velocity channels of 0.2\,--\,0.25 and 0.1\,km\,s$^{-1}$ at 3 and 1.3\,mm,
respectively. The rms noise levels are of the order of 0.3 and 0.5\,K at 3 and 
1.3\,mm, respectively. The absolute
calibration to obtain the main beam temperature $T_{\mr{mb}}$ involved a
standard conversion of the antenna temperature $T_{\mr{A}}^*$ using the
forward and beam efficiencies listed in Table \ref{tab2}. 
The observed intensities, and hence all the derived parameters, are overestimated
due to a contribution of the emission present in the error beam. We adopt a very 
conservative estimate for this contribution of 20\%.
%-----------------------------------------------------------------------
\section{Observational properties}
\label{obsprop}
\subsection{Spatial distribution: $\mr{C}^{18}\mr{O}$ 2-1  map}
\label{spatprop}
The high-resolution ($\sim$\,12\arcs) intensity map of the \ceo\ 2-1
transition (integrated between $-$48 and $-$56\,km\,s$^{-1}$) is displayed
in Figure 4.
The \ceo\ molecule, being optically thin under typical ISM conditions,
traces the gas column along the line of sight and reveals the filamentary
structure in great detail. Several high column density regions are identified along the 
filaments and a number of these were selected for single pointing observations (P1\,--\,P10).
The extend of the \ceo\ emission in the map is used to determine the size of the
cores (Sec. \ref{size}).\\
The reddening contours derived from 2MASS in Figure 1 show a similar structure, 
but they do not trace the \ceo\ everywhere.
This can be explained by both a lower resolution and the limitations of the color 
excess method, which relies completely on the presence of background and embedded
red objects along the line of sight. If the extinction is too high, no red background 
sources are detected and the method fails. However, in such a case, a decrease in 
star counts (Figure 2) can be expected and indeed this is seen for the regions near 
P1, P2, P3 and P4.\\
Filamentary structure is frequently observed in IRDCs 
\citep[e.g.,][]{1998ApJ...508..721C,2003ApJ...588L..37J} and is predicted by 
theories of cloud evolution and clump formation \citep[e.g.,][]{2004ASPC..322..299K}.\\
In this paper we focus on the physical properties of a number of \ceo\ intensity 
peaks, indicated by the open stars and corresponding to the positions listed in
Table \ref{tab1}. The filled red stars represent the location of IRAS 
point sources in the region. Except for P5 and possibly P9 and P10 
(see Section \ref{individual}), the IRAS sources do not coincide or appear to interact 
with any of the target intensity peaks and though they may be associated with the 
complex, their connection is not considered further in the work presented here.
\begin{figure*}[htbp]
\label{c18o}
\centering 
\begin{minipage}[c]{1\linewidth}
\centering \includegraphics[width=17cm]{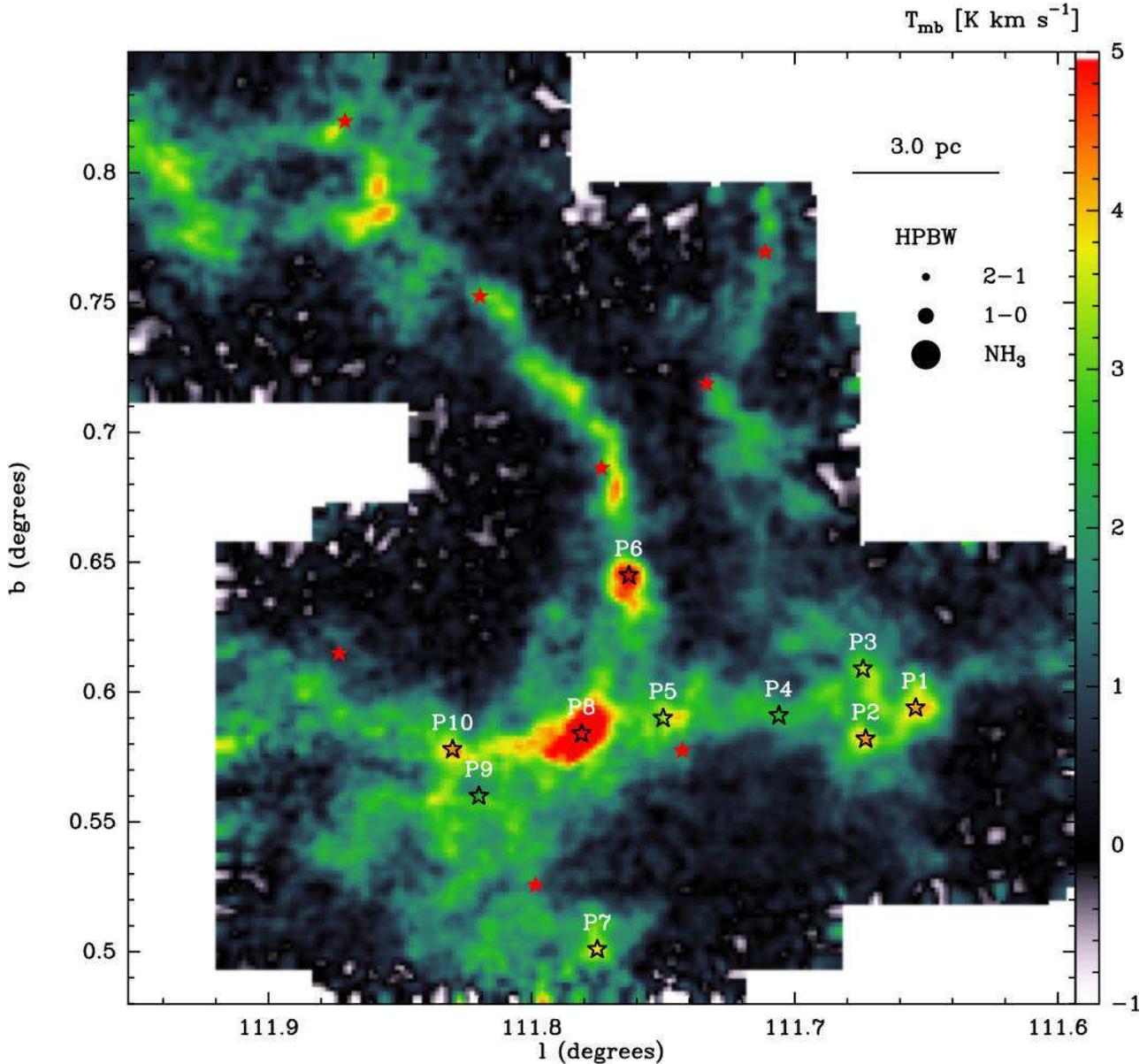} 
\end{minipage}
\caption{This Figure shows the \ceo\ 2-1 map of the filamentary dark cloud complex
  G111.80+0.58 observed with the HERA instrument on IRAM, integrated between $-$56 and $-$46\,km\,s$^{-1}$. 
  The cores listed in Table 1 are given and correspond to the positions indicated by the open black stars, 
  where additional single pointing observations were conducted. The filled red stars represent the IRAS 
  point sources. The linear scale is indicated assuming a distance of 3.1\,kpc. The approximate half-power-beam-widths
  (HPBW) represent the \tco\ and \ceo\ 2-1, the \cts\ 2-1, \tco\ and \ceo\ 1-0 and the \nht\ beams, respectively.}
\end{figure*} 
%-----------------------------------------------------------------------
\subsection{Velocity distribution: $\mr{C}^{18}\mr{O}$ 2-1 channel maps}
\label{velprop}
Though it is beyond the scope of this paper to fully analyze the
velocity maps of the region it is worth showing the spectral velocity
channels to reveal the complexity and different components present
in the region. The velocity 
maps are given in Figure 5 and below is a list of characteristic 
features that can be identified:  \\
$a$) the horizontal structure just below the center (e.g., at $-$51.5\,km\,s$^{-1}$) 
containing several of the observed positions (P4, P5, P8, P10); \\
$b$) the curved, vertical structure (e.g., at $-$53.5\,km\,s$^{-1}$) containing 
position P6 and moving up at nearer velocities; \\
$c$) the main intensity peak (corresponding to P8), possibly enhanced
in intensity because of the superposition of the filamentary structures
mentioned above, though the core itself is seen already at $-$51\,km\,s$^{-1}$;\\
$d$) the positions of P1, P2 and P3 on the right side are dominant at
the nearest velocities ($-$48 to $-$51\,km\,s$^{-1}$) and appear to be the end of
the horizontal structure, bending towards us;\\
$e$) the structures in the upper left corner, where no target positions 
are located, may be connected with the vertical curved structure.\\
\begin{figure*}[htbp]
\label{figvel}
\centering 
\begin{minipage}[c]{1\linewidth}
\centering \includegraphics[width=17cm]{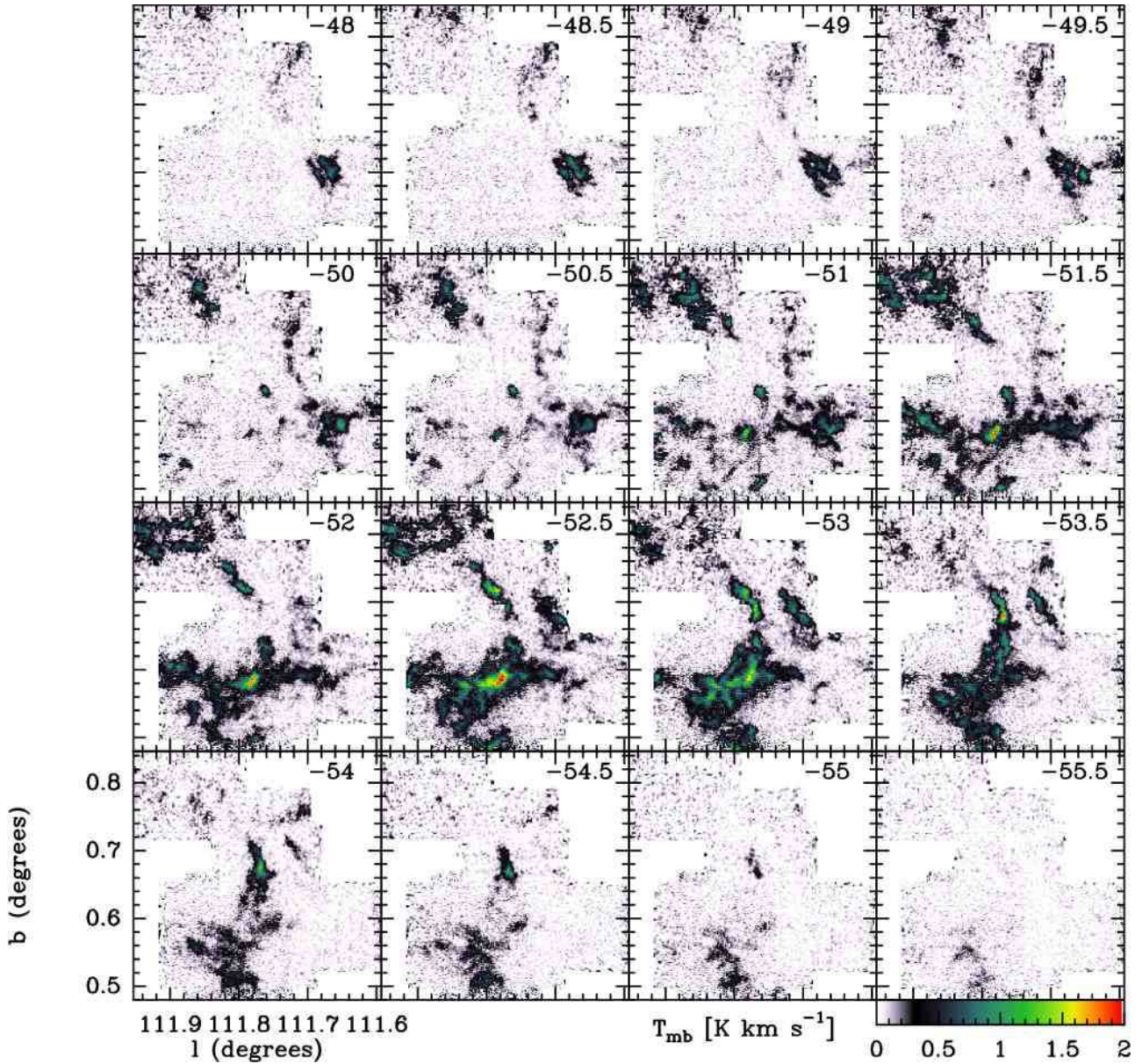} 
\end{minipage}
\caption{This Figure displays the velocity channels centered on the
  velocities given in the upper right corners in the range between $-$48
  and $-$55.5\,km\,s$^{-1}$ and integrated over 0.5\,km\,s$^{-1}$. The region shows a
  complex velocity distribution and several main features can be
  identified (see text).}
\end{figure*} 
%-----------------------------------------------------------------------
\subsection{Single pointings}
\label{point}
The observed spectra of \tco, \ceo\ and \cts\ toward the main \ceo\ 2-1
intensity peaks in the map are given in Appendix \ref{appen2}. Furthermore, 
the Gaussian components that were fitted using the \texttt{GILDAS} 
software are displayed as dashed lines. 
Some positions have additional components in the CO lines that do not correspond 
to a \cts\ detection. 
The focus of this paper however, is on dense cores, i.e., cores that do have a \cts\ detection. 
Therefore, the apparently lower density components, likely associated with lower density cores, 
cloud envelopes or specific phenomena such as outflows are not included in the analysis.\\
We used the following approach to define the dense components; 
1) Gauss-fitting the \cts\ with $N$ components, 2) Gauss-fitting the \ceo\ and \tco\ 
species with at least $N$ lines, 3) imposing as constraints the positions that correspond 
to the \cts\ fits. The different components are numbered according to increasing central
velocity (Column 2, Table 3). Some of the \tco\ lines show high optical depth effects and are impossible 
to match in velocity with the optically thin lines. In that case fits are produced with the 
central velocity as a free parameter. Hence the \ceo\ is used to determine the column 
densities and masses.\\
%-----------------------------------------------------------------------
\section{Results: Physical properties of the clouds}
\label{results}
The physical characteristics of the cores that are determined from the molecular lines are
listed in Tables 3 and 4. The integrated properties (column density and derived
parameters) as well as the \tco\ temperature are determined by using the Gaussian fits
of the individual components. %\ref{tabA1} and \ref{tabA2}. 
A brief description of the tables and their content is given
here. More details on the derivation of the values are presented in
the following sub-sections:\\
\ \\
Table 3-\,$a$) Physical properties assuming the kinetic temperature 
given by \tco.\\
Column (1): Identification number of the cores sorted according to 
Galactic coordinates.\\
Column (2): Identification number $n$ of the component along the line of 
sight; corresponds to the $n^{th}$ component in the spectra 
counting from the left, where $n$ is the same for the different molecules if
they have a similar central velocity. Note that only those components that 
correspond to \cts\ emission are part of the analysis and only these are listed.\\
Column (3): Kinetic temperature $T_{\mr{kin}}$ (K) derived from \tco.\\
Column (4-5): Column density (in units of 10$^{14}$\,cm$^{-2}$) of the CO molecules 
assuming LTE and optically thin emission. In brackets the optical depth $\tau_{\mr{c}}$ at the line 
center is given. Note that the \tco\ emission is in general
optically thick in dense cores and for some positions only a lower limit could be determined
for both $\tau_{\mr{c}}$ and $N_{\mr{mol}}$.\\
Column (6): Column density (in units of 10$^{21}$\,cm$^{-2}$) of H$_2$ derived from 
the \ceo\ emission using a standard abundance of $1.67\times10^{-7}$ 
\citep{1982ApJ...262..590F}.\\
Column (7): Average H$_2$ volume density (in units of 10$^2$\,cm$^{-3}$) derived 
from the H$_2$ column and using the size.\\
Column (8): LTE Mass (M$_{\odot}$) given by Equation \ref{eqmlte} assuming a 
constant volume density and using the size from Table \ref{tab1}.\\
\ \\
Table 3-\,$b$) Physical properties assuming the kinetic temperature 
given by \nht.\\
Column (1-2): Same as part $a$.\\
Column (3): Kinetic temperature $T_{\mr{kin}}$ (K) derived from \nht.\\
Column (4-8): Same as part $a$.\\
\ \\
Table 4)
Line-widths and mass characteristics derived from the IRAM data.\\
Column (1-2): Same as Table 3.\\ %\ref{tabA1}\\
Column (3-5): Line-width (FWHM) of the observed molecules. The \tco\ and \ceo\ values 
are an error-weighted mean of the 1-0 and 2-1 transitions.\\
Column (6): The error weighted mean FWHM of the optical thin \ceo\ and \cts\ lines that is 
used to derive the virial mass.\\
Column (7): Virial mass (M$_{\odot}$) as given by Equation \ref{eqmvir}
assuming a constant volume density $n$(H$_2$) and using the line-width $\Delta V$
given in Column 6.\\
Column (8): LTE mass (M$_{\odot}$). Same as Column (8), Table 3-a. %\ref{tabA1}. 
Listed here for comparison with the other masses.\\
Column (9): Thermal Jeans mass as given by Equation \ref{eqmjeans}.\\
Column (10): Thermal over gravitational energy ratio $\eta$ as given by Equation \ref{eqenratio}.

\begin{table*}[htbp]
%\begin{sidewaystable*}
\label{tabA1}
\begin{minipage}[t]{\textwidth}
%\begin{tiny}
\caption{Physical properties}             % title of Table
\centering                          % used for centering table
\begin{tabular}{cccccrccccc}        % centered columns (10 columns)
\hline
$$&ID  & CP & $T_{\mr{kin}}$$^{13}$CO\footnote{The kinetic temperature is assumed equal to the excitation temperature.} & \multicolumn{4}{c}{$N_{\mr{mol}}$\ \ \ [$\tau_{\mr{c}}$]}& $N_{\mr{H_2}}$ & $n_{\mr{H_2}}$ & $M_{\mr{LTE}}$ \\ 

$$&$$ &  &  & \multicolumn{2}{c}{$^{13}$CO (1-0)} & \multicolumn{2}{c}{C$^{18}$O (1-0)} &($\frac{1.67\times 10^6}{A}\times$ & ($\frac{3.1~\mr{kpc}}{D}\times$& ($\frac{D^2}{9.6~\mr{kpc^2}}$ \\ 

$(a)$&$$ &  & (K) &  \multicolumn{2}{l}{($10^{14}$ cm$^{-2}$)} &  \multicolumn{2}{l}{($10^{14}$ cm$^{-2}$)} & $10^{21}$ cm$^{-2}$) & $10^{2}$ cm$^{-3}$) & M$_{\odot}$) \\
$$&(1)&(2)&(3)&\multicolumn{2}{c}{(4)}&\multicolumn{2}{c}{(5)}&(6)&(7)&(8)\\[0.5ex]
\hline\\[-1.8ex]
$$&                S159                   &                   2                   &     29.1~($\pm$5.1)                   &     1340~($\pm$751) &  [1.27]                  &
        97~($\pm$31) &  [0.09]                    &        58~($\pm$19)                   &      313~($\pm$100)                   &       204~($\pm$65)                  \\
$$&                  P1                   &                   2                   &     10.6~($\pm$1.6)                   &     $>$\,496 &  [$>$\,4.79]       &
         11~($\pm$4) &  [0.11]                    &          6~($\pm$3)                   &         20~($\pm$8)                   &        67~($\pm$28)                  \\
$$&                 P1                   &                   3                   &      9.8~($\pm$1.4)                   &       125~($\pm$29) &  [1.44]                      &
        45~($\pm$10) &  [0.41]                    &         27~($\pm$6)                   &        85~($\pm$18)                   &       287~($\pm$62)                  \\
$$&                  P2                   &                   2                   &      6.8~($\pm$0.9)                   &      $>$\,195 &  [$>$\,2.88]    &
         23~($\pm$5) &  [0.39]                    &         14~($\pm$3)                   &        56~($\pm$13)                   &        87~($\pm$20)                  \\
$$&                  P2                   &                   3                   &     10.2~($\pm$1.5)                   &      589~($\pm$228) &  [4.38]                      &
         24~($\pm$6) &  [0.22]                    &         15~($\pm$4)                   &        59~($\pm$16)                   &        91~($\pm$24)                  \\
$$&                 P3                   &                   1                   &      8.7~($\pm$1.3)                   &       107~($\pm$26) &  [2.02]                    &
          9~($\pm$4) &  [0.14]                    &          5~($\pm$2)                   &         15~($\pm$6)                   &        66~($\pm$29)                  \\
$$&                  P3                   &                   2                   &     13.1~($\pm$2.1)                   &      656~($\pm$547) &  [4.51]                    &
         38~($\pm$9) &  [0.20]                    &         23~($\pm$5)                   &        65~($\pm$15)                   &       291~($\pm$67)                  \\
$$&                  P4                   &                   1                   &      7.5~($\pm$1.0)                   &        51~($\pm$13) &  [1.30]                    &
         16~($\pm$4) &  [0.27]                    &         10~($\pm$3)                   &        41~($\pm$11)                   &        57~($\pm$15)                  \\
$$&                  P4                   &                   2                   &     10.0~($\pm$1.4)                   &       $>$\,488 &  [$>$\,3.68]    &
         21~($\pm$5) &  [0.18]                    &         13~($\pm$3)                   &        53~($\pm$13)                   &        73~($\pm$18)                  \\
$$&                  P5                   &                   2                   &      9.7~($\pm$1.4)                   &       227~($\pm$36) &  [1.84]                    &
         26~($\pm$5) &  [0.19]                    &         15~($\pm$3)                   &        50~($\pm$10)                   &       154~($\pm$31)                  \\
$$&                  P6                   &                   1                   &     12.4~($\pm$1.9)                   &       168~($\pm$53) &  [1.54]                    &
         17~($\pm$7) &  [0.15]                    &         10~($\pm$4)                   &        32~($\pm$13)                   &        99~($\pm$41)                  \\
$$&                  P6                   &                   2                   &     12.1~($\pm$1.9)                   &      $>$\,655 &  [$>$\,4.60]   &
        48~($\pm$10) &  [0.28]                    &         29~($\pm$6)                   &        93~($\pm$20)                   &       289~($\pm$62)                  \\
$$&                  P7                   &                   2                   &     13.2~($\pm$2.1)                   &       199~($\pm$57) &  [1.24]                   &
         24~($\pm$7) &  [0.14]                    &         14~($\pm$4)                   &        52~($\pm$16)                   &       114~($\pm$34)                  \\
$$&                  P7                   &                   3                   &     10.2~($\pm$1.5)                   &      $>$\,318 &  [$>$\,5.13]   &
         18~($\pm$8) &  [0.28]                    &         11~($\pm$5)                   &        39~($\pm$17)                   &        86~($\pm$38)                  \\
$$&                  P8                   &                   2                   &     13.9~($\pm$2.4)                   &     $>$\,625 &  [$>$\,3.69]    &
       124~($\pm$26) &  [0.54]                    &        74~($\pm$16)                   &       202~($\pm$43)                   &     1025~($\pm$217)                  \\
$$&                  P9                   &                   2                   &      9.4~($\pm$1.3)                   &        67~($\pm$18) &  [1.00]                   &
         10~($\pm$3) &  [0.11]                    &          6~($\pm$2)                   &        33~($\pm$11)                   &         22~($\pm$7)                  \\
$$&                  P9                   &                   3                   &     13.0~($\pm$1.9)                   &       250~($\pm$88) &  [2.29]                   &
         11~($\pm$5) &  [0.10]                    &          6~($\pm$3)                   &        34~($\pm$17)                   &        22~($\pm$11)                  \\
$$&                 P10                   &                   3                   &     11.2~($\pm$2.0)                   &      $>$\,389 &  [$>$\,2.29]    &
        46~($\pm$11) &  [0.36]                    &         28~($\pm$7)                   &        80~($\pm$20)                   &       347~($\pm$85)                  \\
\hline
\ \\[-2.2ex] 
$$&$$ &  & $T_{\mr{kin}}$(NH$_3$) & \multicolumn{2}{c}{ } &   &   & & & \\ 
$(b)$\footnote{Part $b$ is the same as part $a$, except here the kinetic temperature is derived from \nht. Note that the results
are very similar.}&$$ &  & (K) &  &  &  & & & &   \\[0.5ex] 
\hline\\[-1.8ex]
$$&                S159                   &                   2                   &     29.3~($\pm$3.4)                   &     1341~($\pm$742) &  [1.25]                    &
        97~($\pm$25) &  [0.09]                    &        58~($\pm$15)                   &       315~($\pm$82)                   &       205~($\pm$54)                  \\
$$&                  P1                   &                   2                   &     14.4~($\pm$2.6)                   &      434~($\pm$108) &  [1.50]                    &
         12~($\pm$6) &  [0.07]                    &          7~($\pm$4)                   &        23~($\pm$11)                   &        76~($\pm$38)                  \\
$$&                  P2                   &                   2                   &     19.0~($\pm$8.4)                   &       149~($\pm$39) &  [0.36]                    &
        30~($\pm$20) &  [0.08]                    &        18~($\pm$12)                   &        73~($\pm$48)                   &       113~($\pm$74)                  \\
$$&                  P5                   &                   2                   &     18.2~($\pm$2.7)                   &       217~($\pm$46) &  [0.45]                    &
         34~($\pm$9) &  [0.08]                    &         20~($\pm$5)                   &        65~($\pm$17)                   &       201~($\pm$53)                  \\
$$&                  P8                   &                   2                   &     16.7~($\pm$1.4)                   &      470~($\pm$159) &  [1.63]                    &
       131~($\pm$29) &  [0.40]                    &        79~($\pm$17)                   &       215~($\pm$47)                   &     1090~($\pm$240)                  \\
$$&                  P9                   &                   2                   &     20.0~($\pm$2.3)                   &        78~($\pm$27) &  [0.26]                    &
         15~($\pm$7) &  [0.04]                    &          9~($\pm$4)                   &        47~($\pm$24)                   &        31~($\pm$16)                  \\
$$&                  P9                   &                   3                   &     18.0~($\pm$2.6)                   &       219~($\pm$92) &  [0.90]                    &
         13~($\pm$8) &  [0.06]                    &          8~($\pm$5)                   &        41~($\pm$26)                   &        26~($\pm$17)                  \\
$$&                 P10                   &                   3                   &     13.5~($\pm$2.1)                   &      435~($\pm$141) &  [2.68]                    &
        49~($\pm$13) &  [0.07]                    &         29~($\pm$8)                   &        84~($\pm$22)                   &       366~($\pm$94)                  \\
\hline\hline
\end{tabular}
\vfill
\end{minipage}
%\end{sidewaystable*}
\end{table*}

\begin{table*}[htbp]
%\begin{sidewaystable*}
\label{tabA2}
\begin{minipage}[t]{\textwidth}
%\begin{tiny}
\caption{Line-width and mass properties}             % title of Table
\centering                          % used for centering table
\begin{tabular}{ccccccccccc}        % centered columns (17 columns)
\hline
ID  & CP &  \multicolumn{4}{c}{$\Delta V$} & $M_{\mr{vir}}$ & $M_{\mr{LTE}}$ & $M_{\mr{Jeans}}$ & $\eta$ \\ 
$$ & & $^{13}$CO & C$^{18}$O & C$^{34}$S & virial & & & \\ 
$$ &  &  (km~s$^{-1}$) & (km~s$^{-1}$) & (km~s$^{-1}$) & (km~s$^{-1}$) & ($\frac{D}{3.1~\mr{kpc}}$~M$_{\odot}$) & ($\frac{D^2}{9.6~\mr{kpc^2}}$~M$_{\odot}$) &  (M$_{\odot}$) &  $\left[\frac{E_{\mr{T}}}{E_{\mr{grav}}}\right]$ \\[1ex]
(1)&(2)&(3)&(4)&(5)&(6)&(7)&(8)&(9)&(10)\\[0.5ex]
\hline\\[-1.8ex]
                S159                   &                   2                   &    2.84~($\pm$0.01)                   &    2.40~($\pm$0.02)                   &    2.25~($\pm$0.04)                   &    2.36~($\pm$0.02)                   &       352~($\pm$59)                   &       204~($\pm$65)                   &         15~($\pm$5)                   &    0.09~($\pm$0.03)
                  \\
                  P1                   &                   2                   &    3.00~($\pm$0.06)                   &    1.07~($\pm$0.15)                   &    1.57~($\pm$0.25)                   &    1.22~($\pm$0.13)                   &       162~($\pm$38)                   &        67~($\pm$28)                   &         13~($\pm$4)                   &    0.17~($\pm$0.08)
                  \\
                  P1                   &                   3                   &    1.77~($\pm$0.08)                   &    1.74~($\pm$0.05)                   &    2.27~($\pm$0.50)                   &   1.77~($\pm$0.06)   &       342~($\pm$40)                   &       287~($\pm$62)                   &          5~($\pm$1)                   &    0.04~($\pm$0.01)
                  \\
                  P2                   &                   2                   &    2.19~($\pm$0.02)                   &    1.95~($\pm$0.03)                   &    1.33~($\pm$0.33)                   &    1.91~($\pm$0.04)                 &       307~($\pm$40)                   &        87~($\pm$20)                   &          4~($\pm$1)                   &    0.07~($\pm$0.02)
                  \\
                  P2                   &                   3                   &    3.98~($\pm$0.02)                   &    1.62~($\pm$0.02)                   &    1.67~($\pm$0.85)                   &    1.62~($\pm$0.03)       &       220~($\pm$29)                   &        91~($\pm$24)                   &          7~($\pm$2)                   &    0.09~($\pm$0.03)
                  \\
                  P3                   &                   1                   &    1.34~($\pm$0.02)                   &    1.17~($\pm$0.04)                   &    1.47~($\pm$0.29)                   &    1.20~($\pm$0.05)      &       172~($\pm$20)                   &        66~($\pm$29)                   &         11~($\pm$3)                   &    0.16~($\pm$0.07)
                  \\
                  P3                   &                   2                   &    2.91~($\pm$0.02)                   &    2.07~($\pm$0.03)                   &    1.06~($\pm$0.34)                   &   2.02~($\pm$0.03)    &       489~($\pm$46)                   &       291~($\pm$67)                   &         10~($\pm$3)                   &    0.05~($\pm$0.01)
                  \\
                  P4                   &                   1                   &    1.28~($\pm$0.02)                   &    1.63~($\pm$0.05)                   &    0.82~($\pm$0.19)                   &    1.51~($\pm$0.05)      &       185~($\pm$27)                   &        57~($\pm$15)                   &          5~($\pm$1)                   &    0.10~($\pm$0.03)
                  \\
                  P4                   &                   2                   &    2.52~($\pm$0.02)                   &    1.74~($\pm$0.05)                   &    1.26~($\pm$0.12)                   &     1.63~($\pm$0.05)      &       215~($\pm$30)                   &        73~($\pm$18)                   &          7~($\pm$2)                   &    0.11~($\pm$0.03)
                  \\
                  P5                   &                   2                   &    2.83~($\pm$0.02)                   &    2.25~($\pm$0.04)                   &    2.62~($\pm$0.24)                   &    2.29~($\pm$0.05)                   &       556~($\pm$59)                   &       154~($\pm$31)                   &          7~($\pm$2)                   &    0.07~($\pm$0.02)
                  \\
                  P6                   &                   1                   &    1.49~($\pm$0.01)                   &    1.14~($\pm$0.03)                   &    0.96~($\pm$0.16)                   &      1.12~($\pm$0.03)       &       132~($\pm$15)                   &        99~($\pm$41)                   &         13~($\pm$4)                   &    0.13~($\pm$0.06)
                  \\
                  P6                   &                   2                   &    2.88~($\pm$0.01)                   &    2.09~($\pm$0.02)                   &    1.20~($\pm$0.01)                   &    2.08~($\pm$0.03)      &       459~($\pm$47)                   &       289~($\pm$62)                   &          7~($\pm$2)                   &    0.04~($\pm$0.01)
                  \\
                  P7                   &                   2                   &    1.91~($\pm$0.02)                   &    1.49~($\pm$0.11)                   &    1.62~($\pm$0.30)                   &     1.52~($\pm$0.11)      &       218~($\pm$40)                   &       114~($\pm$34)                   &         11~($\pm$3)                   &    0.11~($\pm$0.04)
                  \\
                  P7                   &                   3                   &    1.41~($\pm$0.01)                   &    1.05~($\pm$0.02)                   &    1.68~($\pm$0.77)                   &      1.06~($\pm$0.02)      &       106~($\pm$13)                   &        86~($\pm$38)                   &          9~($\pm$3)                   &    0.11~($\pm$0.05)
                  \\
                  P8                   &                   2                   &    2.32~($\pm$0.02)                   &    2.22~($\pm$0.01)                   &    2.08~($\pm$0.05)                   &    2.20~($\pm$0.01)              &       604~($\pm$51)                   &     1025~($\pm$217)                   &          6~($\pm$2)                   &    0.02~($\pm$0.00)
                  \\
                  P9                   &                   2                   &    1.55~($\pm$0.02)                   &    1.48~($\pm$0.08)                   &    1.57~($\pm$0.22)                   &    1.50~($\pm$0.08)                   &       142~($\pm$28)                   &         22~($\pm$7)                   &          8~($\pm$2)                   &    0.27~($\pm$0.10)
                  \\
                  P9                   &                   3                   &    1.64~($\pm$0.02)                   &    1.23~($\pm$0.06)                   &    1.82~($\pm$0.28)                   &    1.31~($\pm$0.07)                &       109~($\pm$21)                   &        22~($\pm$11)                   &         13~($\pm$5)                   &    0.36~($\pm$0.19)
                  \\
                 P10                   &                   3                   &    2.22~($\pm$0.71)                   &    1.60~($\pm$0.02)                   &    1.99~($\pm$0.08)                   &    1.66~($\pm$0.02)                &       326~($\pm$30)                   &       347~($\pm$85)                   &          7~($\pm$2)                   &    0.04~($\pm$0.01)
                  \\
\hline\hline
\end{tabular}
%\end{tiny}
\vfill
\end{minipage}
%\end{sidewaystable*}
\end{table*}

%-----------------------------------------------------------------------
\subsection{Kinetic temperatures}
\label{kineticT}
The kinetic temperature of the cores can be estimated from both the
\nht\ and the \tco\ lines. In general, considering the conditions in dense
cores, collisions dominate the excitation process and the levels are
thermalized. In this case, the kinetic temperature equals the
excitation temperature, i.e., $T_{\mr{\mr{kin}}}$\,=\,$T_{\mr{ex}}$. A brief description of
the determination of $T_{\mr{ex}}$ for
both molecules is given below. \\
A comparison between the temperatures derived from \nht\ and \tco\ can be 
misleading because they likely represent different regions within the cores. \nht\ is a good
temperature tracer and because the molecules do not freeze out onto
the dust grains for densities below $\sim10^6$\,cm$^{-3}$ \citep{1997ApJ...486..316B}
they are observed even in very dense regions. In
the inner, dense parts of cores the heating is due to cosmic rays and,
if present, internal sources. The \tco\ molecule is likely depleted in dense and
cold cores and it is optically thick due to its much larger abundance. 
Thus it traces mostly the outer parts of molecular clouds where heating is due to the 
interstellar radiation field penetrating the outer layers.
\subsubsection{Ammonia}
\begin{table*}[htbp]
\label{tab3}
\begin{minipage}[t]{\textwidth}
\caption{NH$_3$ derived properties}
\centering                      
\begin{tabular}{cccccccccccccc}   
\hline\hline\\[-1.8ex]
ID  & CP & $V_{\mr{LSR}}$ &$\Delta V$ & $T_{\mr{mb}} (1,1)$ & E$_{T_{\mr{mb}} (1,1)}$ & $\tau$(1,1) & E$_{\tau(1,1)}$ & $T_{\mr{mb}} (2,2)$ & E$_{T_{\mr{mb}} (2,2)}$ & $T_{\mr{rot}}$ & E$_{T_{\mr{rot}}}$ & $T_{\mr{kin}}$ & E$_{T_{\mr{kin}}}$ \\ 
& & (km~s$^{-1}$) & (km~s$^{-1}$) & (K) & (K) &  & & (K) & (K) & (K) & (K) & (K) & (K) \\ 
(1)&(2)&(3)&(4)&(5)&(6)&(7)&(8)&(9)&(10)&(11)&(12)&(13)&(14)\\[0.5ex]
\hline\\[-1.8ex]
S159 &1&      -58.01   & 1.13~($\pm$0.17)       &        0.59           &        0.06           & 1.3&0.7 &        0.19           &
        0.05           &        17.1           &         2.0           &        18.2           &         2.0          \\
$"$   &2&      -55.87    & 2.16~($\pm$0.25)       &        0.45           &        0.06           & 0.1&0.1 &        0.32           &
        0.05           &        25.7           &         3.4           &        29.3           &         3.4          \\
P1   &2&      -50.12      & 2.15~($\pm$0.19)     &        0.74           &        0.10           & 0.3&0.3 &        0.13           &
        0.07           &        13.7           &         2.6           &        14.4           &         2.6          \\
P2   &2&      -50.95     & 1.50~($\pm$0.00)      &        0.41           &        0.20           & -&- &        0.14           &
           0.14           &        17.6           &         8.4           &        19.0           &         8.4          \\
P5   &2&      -52.76      & 2.25~($\pm$0.05)     &        1.58           &        0.09           & 1.1&0.1 &        0.50           &
        0.19           &        17.0           &         2.7           &        18.2           &         2.7          \\
P6   &-&           -          & - &        0.12           &           0.12           & -&- &        0.08           &
           0.08           &       26.4           &       23.9           &       30.2           &       23.9          \\
P8   &2&      -52.83     & 2.07~($\pm$0.08)      &        2.11           &        0.15           & 1.0&0.2 &        0.55           &
        0.12           &        15.8           &         1.4           &        16.7           &         1.4          \\
P9   &2&      -54.18    & 2.82~($\pm$0.82)       &        0.54           &        0.06           & -&- &        0.21           &
        0.05           &        18.5           &         2.3           &        20.0           &         2.3          \\
$"$   &3&      -52.03     & 1.40~($\pm$0.82)      &        0.47           &        0.06           & 1.0&0.1 &       0.14           &
        0.05           &        16.8           &         2.6           &        18.0           &         2.6          \\
P10  &3&      -52.23    & 1.65~($\pm$0.11)       &        1.35           &        0.14           & 0.8&0.3 &        0.20           &
        0.10           &        13.0           &         2.1           &        13.5           &         2.1          \\
\hline\\[-1.8ex]
\hline
\end{tabular}
\vfill
\end{minipage}
\end{table*}
The rotational temperatures of the \nht\ molecule, characterizing the
level populations, were determined by fitting the main and hyperfine
components of the (1,1) transition and the main component of the
(2,2). The reduction of the (1,1) transition was done using "METHOD
NH3(1,1)" in CLASS to fit the hyperfine structure, whereas for the
(2,2) line a standard Gauss fitting procedure was sufficient. A
detailed description of the standard analysis is given by \citet{1987A&A...173..324B}. 
The kinetic temperature is derived using the analytical expression given by 
\citet{2004A&A...416..191T};
\begin{equation}
\label{Tkin_nh3}
T_{\mr{kin}} = \frac{T_{\mr{rot}}}{1 - \frac{T_{\mr{rot}}}{42}\mr{ln}[1 + 1.1\mr{exp}(-16/T_{\mr{rot}})]}~(\mr{K})~,
\end{equation}
where $T_{\mr{rot}}$ is the rotational temperature. Table 5
lists the properties derived from the \nht\ observations. Columns 1 to 4 give
the target position, line component, central velocity and line-width (FWHM) of the (1,1) transition, 
respectively. Columns 5\,--\,6 and 9\,--\,10
give the main beam temperatures and the noise of the spectra for the (1,1) and (2,2)
transition, respectively. The optical depth $\tau$ and the corresponding error derived from 
the hyperfine structure fitting for the (1,1) transition are listed in Columns 7 and 8, respectively. 
Typical values for the optical depth are around unity with the exception of P1 and S159 CP2. Note 
however that the error is of the order of $\tau$ itself and the hyperfine fitting 
is considered dubious.
The emission of the hyperfine structure for P2, P6 and P9 CP2, if present at all, is too weak to be 
detected at the observed signal to noise levels. Hence the missing entries.
The rotational temperatures are given in Column 11 with the corresponding error in Column 12. 
Column 13 and 14 give the kinetic temperatures derived using Equation \ref{Tkin_nh3}
and the corresponding error, respectively. Typically, the kinetic temperatures range from 13\,--\,30\,K.
Note that P2 is detected only in the (1,1) transition and P6 not at all, hence the
missing entry of a central velocity for the latter. The $T_{\mr{mb}}$ values for
P2 and P6 are derived from the noise of the observations and the temperature 
parameters merely give an upper limit.
\subsubsection{Optically thick \tco}
\label{tcoT}
From basic radiative transfer \citep[e.g.,][]{1978ApJS...37..407D}, the 
observed radiation temperature for a certain molecule toward a core is 
given by the expression
\begin{equation}
T_{\mr{R}}^* = [J_{\nu}(T_{\mr{ex}}) - J_{\nu}(T_{\mr{bg}})][1 - \mr{exp}(-\tau_{\nu})]~,
\end{equation}
in which
\begin{equation}
J_{\nu}(T) = \frac{h\nu}{k}\frac{1}{\mr{exp}(\frac{h\nu}{kT})-1}~.
\end{equation}
The constants $h$, $k$, $T_{\mr{ex}}$, $T_{\mr{bg}}$ and $\tau_{\nu}$ are the
Planck constant, the Boltzmann constant, the excitation temperature,
the background temperature ($\sim$\,2.7\,K) and the optical depth at
frequency $\nu$, respectively.  In the dense cores considered in this
paper, the \tco\ lines are usually optically thick and the results derived in 
Section \ref{cdens} confirm indeed that $\tau_{\nu}$ exceeds unity at the line centers.
The radiation temperature of the molecule is approximately given by
\begin{equation}
T_{\mr{R}}^* \mr{(^{13}CO)}= [J_{\nu}(T_{\mr{ex}}) - J_{\nu}(T_{\mr{bg}})]~,
\end{equation}
where $T_{\mr{R}}^* \mr{(^{13}CO)}$ is the observed peak intensity of the
\tco\ line (here we use the 1-0 transition).  From this equation the
excitation temperature can be derived:
\begin{equation}
T_{\mr{ex}} = \frac{5.29}{\mr{ln}[1 + \frac{5.29}{T_{\mr{R}}^*\mr{(^{13}CO)} + 0.868}]}~(\mr{K})~. 
\end{equation}
The values are listed in Column (3) of Table 3-a.\\
In the following analysis the \tco\ 1-0 temperature is adopted as the
kinetic temperature of the clouds. When the data allow, the
analysis is also performed using the kinetic temperature derived
from the \nht\ lines.
%__________________________________________________________________
\subsection{Core sizes}
\label{size}
We define a size for each core as the average of the minor
and major axis of the 50\% peak intensity level observed
in the \ceo\ 2-1 map. The sizes are given in parsecs in Column 4 of
Table \ref{tab1}, assuming a distance to the cores of 3.1\,kpc. 
Several positions have multiple components along the line of
sight and obviously the size is then determined from a superposition 
of several cores. However, a distinction between the components is difficult
because of the small separation in velocity so that the same
size is adopted for all components. Position P9 is
not identified as a coherent core in the map and a size of 40\arcs\ is
adopted. P4 is part of a filamentary structure and the
size is assumed equal to the projected width of the filament.
%__________________________________________________________________
\subsection{Column densities}
\label{cdens}
The following approach \citep[see e.g., ][]{1997ApJ...476..781B} assumes these 
conditions: a) the cores are isothermal, and the kinetic temperature, $T_{\mr{kin}}$, 
is given by the excitation temperature derived from \tco\ or \nht; b) the cores have a
constant density; c) the cores are in local thermodynamic equilibrium
(LTE). Then, using the observed \ceo\  radiation temperature $T_{\mr{R}}^*$
(K), the optical depth can be calculated as
a function of the LSR velocity $V$ (km\,s$^{-1}$) parameterized by the
frequency $\nu$:
\begin{equation}
\tau_{V}(\mr{C^{18}O}) = -\mr{ln}\left\{1 - \frac{kT_{\mr{R}}^*}{h\nu}  \left[ \frac{1}{\mr{exp}(\frac{h\nu}{kT_{\mr{ex}}}) -1} - \frac{kJ_{\nu}(T_{\mr{bg}})}{h\nu} \right]^{-1} \right\}~.
\end{equation}
The corresponding column density can be estimated using
\begin{equation}
N(\mr{C^{18}O}) = 2.42\times10^{14} \sum \frac{\tau_{V}(\mr{C^{18}O})~\Delta V~T_{\mr{ex}}}{1 - \mr{exp}(-\frac{h\nu}{kT_{\mr{ex}}})} ~(\mr{cm^{-2}})~,
\end{equation}
where $h$, $\nu$, $k$, $J_{\nu}(T_{\mr{bg}})$ are the same as before and
$\Delta V$ is the velocity step size in km\,s$^{-1}$. 
The column densities derived for \ceo\ 1-0 are given in Column (5) of Table 3. %\ref{tabA1}.1
The values listed in brackets corresponds to the optical depth, $\tau_{\mr{c}}$, at the line center and
are typically much less than unity for \ceo. \\
An analogous derivation is done for the \tco\ 1-0 transition and these values are listed
in Column (4). However, as noted before, in the dense cores the \tco\ is usually optically thick
(i.e., $\tau_{\mr{c}} > 1$) and the lines cannot be used to determine a reliable column density. 
Nonetheless, the results are listed and for the cases where the optical depth becomes too large, 
a lower limit is given.
The total molecular column densities, $N$(H$_2$), are derived from
the \ceo\ 1-0 results using a canonical abundance of $1.67\times10^{-7}$ 
\citep{1982ApJ...262..590F}.
These values are given in Column 6 of Table 3.%\ref{tabA1}.\\
%__________________________________________________________________
\subsection{Volume densities}
\label{vdens}
Column 7 (Table 3) %\ref{tabA1}) 
lists the volume density derived
from the H$_2$ column density, assuming the depth of the core equals
the size. The density is by definition an average over the
actual density profile along the line of sight. In the inner parts of
the cores the density is considerably higher, as suggested by the 
presence of \cts.
%__________________________________________________________________
\subsection{Mass estimates}
\label{mass}
\subsubsection{LTE mass}
\label{lte}
Given the size and adopting a constant volume density, the
molecular mass derived from the lines, referred to as the LTE mass, is
determined using
\begin{equation}
\label{eqmlte}
%	 totmassteff=4./3. * !pi * ((diam*3.086d18)/2.)^3 * (nvolteff*2.33*1.6724d-24) / 1.99d33
%M_{\mr{LTE}} = \frac{1}{6}\pi~\mu~n(\mr{H_2})~S^3~\mr{(M_{\odot})}~,
M_{\mr{LTE}} = \frac{4}{3}\pi~\mu~m_{\mr{H}}~n(\mr{H_2})~(\frac{1}{2}S)^3 \mr{C}~\mr{(M_{\odot})}~,
\end{equation}
where $n$(H$_2$) is the volume density, $S$ is the size, $m_{\mr{H}}$ is the mass of a hydrogen 
atom, $\mu$\,=\,2.33 is the mean molecular weight consistent with a 25\%
mass fraction of helium and C\,=\,1pc$^3$/\, 1M$_{\odot}$\,$\sim$\,1.48$\times$10$^{22}$\,cm$^3$/\,g, the 
conversion to solar masses. The values are listed in Column 8 (Table 3). % \ref{tabA1}). 
Because the volume density is derived from the
column density ($n(\mr{H_2}) \sim N(\mr{H_2})/S$), the LTE mass of the
cores scales with the square of the size and thus the distance (i.e.,
$M_{\mr{LTE}}\sim S^2$ and $\sim D^2$).
%__________________________________________________________________
\subsubsection{Virial mass}
\label{vir}
Comparing the LTE mass with the virial mass enables an evaluation of
the dynamical state of the cores. Assuming a spherical core with
uniform density and considering only thermal and dynamical broadening
(neglecting e.g., magnetic fields, internal heating) the virial mass
can be derived using the expression \citep[e.g., ][]{1988ApJ...333..821M}
\begin{equation}
\label{eqmvir}
%M_{vir} = 0.509~D~\theta~(\Delta V)^2 ~\mr{(M_{\odot})}~,
M_{\mr{vir}} = 210~R~(\Delta V)^2 ~\mr{(M_{\odot})}~,
\end{equation}
where $R$ (=$\frac{S}{2}$) is the radius of the core in pc  %($R=\frac{1}{2}S$) and
%$\theta$ is the size of the core in arcmin and 
and $\Delta V$ is the FWHM of the line in km\,s$^{-1}$. In our case a weighted 
average of the observed optically thin lines, given in Column 6 of Table 4,
is used. The resulting virial masses
are given in Column 7 of Table 4. %\ref{tabA2}. 
Equation \ref{eqmvir} assumes a constant density. The constant may be 
replaced by 190 or 126 for a density profile given by $\rho \sim r^{-1}$ 
or $\rho \sim r^{-2}$, respectively.\\
%__________________________________________________________________
\subsubsection{Jeans mass}
\label{jeans}
Thermal Jeans fragmentation, often referred to as the mechanism to set
the mass scale for star formation \citep{1985MNRAS.214..379L}, is
approximately given by
\begin{equation}
\label{eqmjeans}
M_{\mr{Jeans}} \approx \frac{90}{\mu^2} \times T^{\frac{3}{2}} n^{-\frac{1}{2}} ~\mr{(M_{\odot})}~,
\end{equation}
where $T$ is the gas temperature and $n$ is the total volume density. The
values are listed in Column 9 of Table 4. %\ref{tabA2}. 
The typical
value is of the order of 10\,M$_{\odot}$, much smaller than the derived
LTE masses of the cores. Therefore, structure on much smaller scales
may be expected within the cores.
%__________________________________________________________________
\subsubsection{Stability}
\label{enratio}
The energy ratio $\eta$ listed in Column 10 (Table 4) is defined as the thermal energy 
over the gravitational energy:
\begin{equation}
\label{eqenratio}
\eta = \frac{E_{\mr{T}}}{E_{\mr{grav}}} = \frac{\frac{3}{2}\frac{M_{\mr{LTE}}}{\mu m_{\mr{H}}}kT}{\frac{3}{5}\frac{G M^2_{\mr{LTE}}}{0.5~S}}~,
\end{equation}
%$E_{grav} = \frac{3}{5}\frac{G M^2_{\mr{LTE}}}{0.5~S}$, 
where $M_{\mr{LTE}}$ is given
in Column 8 (Table 4), $T$ is the kinetic temperature listed in Column 3 (Table 3-a), $S$ is the size of the 
core, described in Sec. \ref{size}, $\mu$, $m_{\mr{H}}$, $k$ and 
$G$ are the mean molecular weight (=\,2.33), the mass of a hydrogen atom, the 
Boltzmann constant and the gravitational constant, respectively. Typically, $\eta$ is much less than
unity and it suggests that the thermal energy by itself is insufficient to support the cores against 
gravitational collapse.
%__________________________________________________________________
\section{Discussion and conclusion}
\label{discussion}
What kind of object is G111.80+0.58? \\
Before we attempt to answer this question it is worthwhile
to discuss some key properties derived from the observations and compare the results 
with other studies.
\subsection{Gas temperature}
\label{gasT}
The gas temperatures derived from \nht\ (13\,--\,20\,K) and, less obvious also from \tco\ 
(7\,--\,14\,K), are higher than expected inside molecular cores if only cosmic ray ionization is 
considered as a heating source \citep[a mean Galactic value of 
$\zeta_{\mr{CR}}\sim3\times10^{-17}$ s$^{-1}$
yields a temperature of $T$\,$\sim$\,8\,--\,10\,K,][]{2000A&A...358L..79V}.
The \nht\ temperatures suggest that deep inside the cores additional heating sources are
present. Because the \tco\  is usually optically thick at the observed positions, it traces
the temperature of the material at an optical depth of about unity. The
actual location of this $\tau$\,=\,1 surface in the cores cannot
be determined, but it is likely that the two molecules trace the temperatures
at different depths in the cores.\\
The derived temperatures are in agreement with values found for massive and dense
cores \citep[e.g., $<$\,20\,K for IRDCs,][]{1998ApJ...508..721C}. The temperature found
for S159 is significantly higher (29\,K), which is consistent with its more advanced star forming 
state.
%__________________________________________________________________
\subsection{Core masses and total mass}
\label{totmass}
The individual core masses ($M_{\mr{LTE}}$) vary from about 20 to
1000\,M$_{\odot}$ (approximately 100\,--\,600\,M$_{\odot}$ for $M_{\mr{vir}}$) with
an average of around 190\,M$_{\odot}$ (about 280\,M$_{\odot}$ for 
$M_{\mr{vir}}$). Most cores are massive when compared to low-mass dark clouds 
\citep[e.g.,][ and references therein]{1999ARA&A..37..311E}, but correspond well with masses 
found for high-mass protostellar objects \citep[HMPOs; e.g., ][]{2004yCat..34170115W}.
The LTE mass we find for S159 is in agreement with previous high resolution studies
\citep{2001ApJ...560..806L} where the core is resolved into multiple substructures.\\
A rough estimate for the total mass of the complex, including only the observed cores in 
this paper, is 3000\,M$_{\odot}$ (about 4500\,M$_{\odot}$ for $M_{\mr{vir}}$).
This is comparable to masses found for Inner Galaxy IRDCs 
\citep[e.g., ][]{2000ApJ...543L.157C}.
%__________________________________________________________________
\subsection{Physical characterization of the cores}
\label{physchar}
The observations suggest that the thermal energy is insufficient to support the cores against
gravitational collapse. Additional support is seen in the observed line-widths toward the cores.
These are much broader than expected from purely thermal motion 
(see Column 3\,--\,5, Table 4 and Column 4, Table 5). %\ref{tabA2}). 
The average line-width is about $1.8\pm0.4$
km\,s$^{-1}$. The sound speed $\Delta V_{\mr{s}}$, considering that the cores 
have a typical temperature $T$ of 15\,--\,20\,K, is given by
\begin{equation}
\Delta V_s = \left(\frac{8~\mr{ln}(2)kT}{\mu m_{\mr{H}}}\right)^{\frac{1}{2}} \approx 0.6~ \mr{km~s^{-1}}~,
\end{equation}
and thus the velocities are supersonic.\\
The broad lines could be due
to turbulent motions in the clouds and are indicative of regions of
high-mass star formation \citep[e.g.,][]{1993ApJ...402..635M}, in contrast to 
mostly thermal broadened lines often seen toward low-mass pre-stellar cores 
\citep[e.g.,][]{1998ApJ...504..223G}.
%Remarkably, the broad lines are present even on the smaller 
%scales and at the higher densities which are traced by the \cts\ lines.\\
The line-width across the filament, extending over 10\,pc (at a
distance of 3.1\,kpc), is about 5\,km\,s$^{-1}$. The relation between
velocity dispersion of substructures ($\Delta V_l$) and the velocity
dispersion over the larger complex ($\Delta V _{L}$) in a turbulent
medium can be expressed using the empirically derived Larson's law 
\citep{1981MNRAS.194..809L};
\begin{equation}
\label{eqlarson}
\Delta V_l  \sim \Delta V _{L} \left(\frac{l}{L}\right)^{0.38}~,
\end{equation}
where $L$ is the size of the filament and $l$ the size of the
sub-structure. For a typical core size in our sample of $\approx$\,1\,pc, the 
expected line-width is then $\approx$\,2\,km\,s$^{-1}$ for individual cores. 
The values we find are consistent with this relationship. The standard relations
of mass versus line-width are, however, for much smaller cores and the
larger cores presented here may be a superposition of such small cores.
Extending the relation to smaller cores of 0.1\,pc, the expected line-width using 
Equation \ref{eqlarson} is about 0.8 to 0.9\,km\,s$^{-1}$. This is still in agreement
with values found for typical massive starless cores \citep[e.g.,][]{1995ApJ...446..665C}.\\
A decay of the turbulence is expected on short timescales \citep[see
][ for a review on star formation and turbulence]{2006astro.ph..3357B}
and some mechanism is needed to sustain the turbulent
support. Internal sources, e.g., deeply embedded YSO's may be present
and stir up the material from the inside through outflows. Indicative
of the presence of a heating source can be the high temperatures traced 
by \nht. However, we find no correlation between the line-width and the 
gas temperature.\\
The line-width could also be the result of a number of small dense clumps
at slightly different velocities within the spatial resolution element
of the observations. This idea is supported by the presence of
multiple components along the line of sight toward many of the
observed positions. In addition, the derived Jeans mass (Column 9, Table 4) %Table \ref{tabA2}) 
for the cores is typically 
$\approx$\,10\,M$_{\odot}$ and therefore, fragmentation is expected to occur on 
smaller mass scales compared to the observed core masses, if thermal processes 
dominate stability.\\
Figure 6 shows a plot of the virial mass versus the LTE mass.
\begin{figure}[htbp]
\label{figmass}
\centering 
\begin{minipage}[c]{1\linewidth}
\centering \includegraphics[width=9cm]{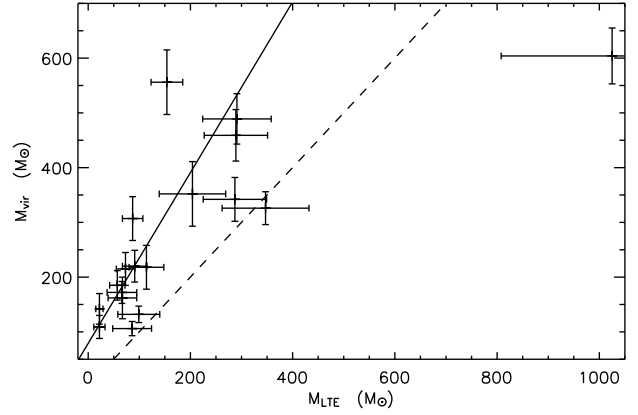} 
\end{minipage}
\caption{The virial mass is plotted against the LTE mass. The dashed line
  corresponds to equal masses, where the virial parameter
  $\alpha=1$. The solid line is the fit to the data points.}
\end{figure}
The virial parameter $\alpha$ can be defined as the ratio of the two and a 
value of unity corresponds to a virial equilibrium state (dashed line).
The observed values for $\alpha$ range from about 1 to 3 and thus suggest that, on a 
global scale, the cores are close to virial equilibrium. Note that depletion
of the \ceo\ molecule may play an important role for densities in excess of $10^4$\,cm$^{-3}$
if $T_{\mr{dust}}$\,$\sim$\,$T_{\mr{gas}}$ \citep{1995ApJ...441..222B} and
consequently, the LTE mass could be underestimated.
Core P8 is an exceptional case where $M_{\mr{LTE}}>M_{\mr{vir}}$. As seen from the
channel maps (Fig. 5, Sec. \ref{velprop}), in projection P8 appears to be a superposition
of two crossing filaments. The material traced by the \ceo\ may just be
the line of sight sum of two components. Alternatively,
because the components have a similar LSR velocity, a collision may be taking
place between the two filaments. This would lead to the formation of a
dense (perhaps gravitationally unstable) central core. If so, then we 
may be witnessing this scenario prior to the system being fully virialized.
\subsection{Star forming activity}
The \nht\ lines are very sensitive to the temperature and therefore the presence of the
higher transitions, particularly of the (3,3) line, is a good indicator
for the occurrence of warmer gas, possibly indicating internal heating sources.
The (3,3) line is clearly seen toward S159, as expected from work by \citet{2001ApJ...560..806L}, 
and also position P5, where it may result from the presence of the associated IRAS point source. 
A very weak detection is seen toward positions P8 and P9.\\
Signatures of self-absorption in the \tco\ lines can be an additional indication of the presence of
the warmer gas inside the cores. This may explain also the failure in fitting the \tco\ lines
at the central velocities of the optically thin lines. Most noticeable are the positions toward S159, P5 and P8.\\
If embedded sources are present, they may have been identified in the 2MASS data. Figure 7
shows the \hk\ versus \jh\ colors of the sources present in the 2$\times$\,2 square
degree field. The grey scaling represents the overall distribution of the field and the majority of 
the sources are located near the dwarf and giant sequence, indicated by the 
black solid curves. The spread in the colors is significant but can be addressed entirely by the
use of the very faint sources in the 2MASS catalog (the faint extension) where the photometric 
errors are larger. The dashed lines give the reddening vectors \citep{1985ApJ...288..618R} for 
the dwarf and giant sequence and the arrow corresponds to an $A_{\mr{V}}$ of 10 magnitudes. 
In addition, the figure shows for each of the cores the associated 2MASS sources in different 
symbols. Note that most star colors toward the cores resemble main sequence stars with or 
without reddening, and these are likely reddened background or 'normal' foreground stars. 
However, the very red \hk\ area right of the reddening vectors corresponds to colors of 
proto-stellar objects \citep[e.g.,][]{2006ApJS..167..256R}. Most promising candidates for YSO's 
correspond to sources associated with S159, P5 and P8.
\begin{figure}[htbp]
\label{2mcolcol}
\centering 
\begin{minipage}[c]{1\linewidth}
\centering \includegraphics[width=8.5cm]{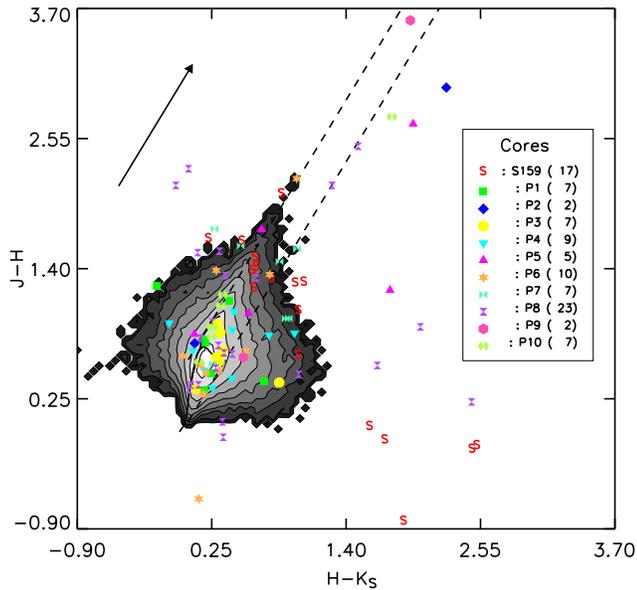} 
\end{minipage}
\caption{2MASS color-color plot. The grey-scales represent the location of colors of all stars
in the 2 by 2 square degree field. Lighter colors correspond to a higher occurrence. The
majority of the stars are situated with some spread around the dwarf and giant sequences overlaid 
as black curves. The arrow corresponds to a visual extinction of 10 magnitudes, in the direction of the 
reddening vectors \citep[dashed lines,][]{1985ApJ...288..618R}. The sources associated with the cores 
are plotted using the symbols listed in the legend. The number of associated sources is given in brackets.}
\end{figure}

\subsection{Notes on individual cores}
\label{individual}
P1, P2, P3 and P7: These are regions of the molecular cloud where multiple components with 
enhanced densities (based on \cts) appear to be present along the line of sight. There are no 
clear indications for star forming activity from \tco\ self absorption, 2MASS colors or \nht\ heating 
(\nht\ is only available for P1 and P2).\\
P4: This is part of the horizontal filamentary structure. No clear core is seen in the \ceo\ 2-1 map but
high density gas is present, indicated by the \cts\ line. Gravity may have induced a density enhancement 
along the filament. Two components are fitted by the software to the spectra, but possibly opacity effects 
cause a non-Gaussian behavior of a single component. Manual fitting of one component would result in 
an LTE mass of about 138\,M$_{\odot}$, a velocity width of 1.8\,km\,s$^{-1}$ and a virial mass of 
262\,M$_{\odot}$.\\
P5: This position may be associated with, and heated by, the nearby source IRAS23137+6105, which can explain
the presence of the higher \nht\ transitions. One object in the 2MASS data has typical colors of a YSO. 
Also here, the spectra suggest that a single manual fit may be favored to two components. 
This would result in an LTE mass of about 220\,M$_{\odot}$, a velocity 
width of 2.57\,km\,s$^{-1}$ and a virial mass of 702\,M$_{\odot}$.\\
P6: Two dense (\cts) components are present along the line of sight. The stellar density appears 
slightly enhanced but this may be a chance encounter along the line of sight. There is no indication for 
the presence of YSO's and none of the \nht\ transitions is detected which suggests a quiescent, cold stage.\\
P8: This is the most profound core in the region and the best candidate for being an object similar to 
Inner Galaxy IRDCs (see Section \ref{irdc}). Automatic fitting of the CO lines resulted in two components,
but in \cts, a single profile is seen. Manual fitting of a single component to the CO lines results in an LTE 
mass of about 1250\,M$_{\odot}$, a velocity width of 2.28\,km\,s$^{-1}$ and a virial mass of 647\,M$_{\odot}$.
This makes the case for being an Outer Galaxy IRDC even stronger. Star formation may 
have started here, indicated by the \nht\ lines and the presence of YSO candidates in the 2MASS color plot.
Opacity effects such as self absorption due to colder foreground gas can explain the non-Gaussian
appearance of the CO lines. The stellar density derived from 2MASS is significantly higher (11-13 stars per cell), 
compared to the rest of the field. \\
P9,10: There is arc-like MSX 8$\mu$m emission nearby these cores (see Figures 1-3). This may be caused 
by heating due to radiation from the neighboring source IRAS23143+6103. Such heating may also explain the 
detection of \nht\ (2,2) and faint \nht\ (3,3) emission. Compression of the material due to a shock may then lead to the 
observed high densities. No indication of embedded sources is seen from the 2MASS data.

\subsection{Conclusion: Is G111.80+0.58 similar to Inner Galaxy IRDCs ?}
\label{irdc}
The global characteristics of the G111.80+0.58 complex that are presented in this paper,
such as its size ($\sim$\,10\,pc), column density (peak\,$>$\,10$^{22}$\,cm$^{-2}$) and LTE mass
($\sim$\,3000\,M$_{\odot}$) indicate a clear resemblance to the bulk properties of Inner 
Galaxy IRDCs \citep{2006ApJ...653.1325S}. However, except for P8, the individual core properties
(e.g., size: $\sim$1\,pc, LTE mass: 20\,--\,350\,M$_{\odot}$, density: 10$^3$\,--\,10$^4$\,cm$^{-3}$)
fall short of the values found for compact (sub)mm cores in IRDCs 
\citep[size: 0.02\,--\,0.8\,pc, mass: 10\,--\,10$^3$\,M$_{\odot}$, density: 10$^3$\,--\,10$^7$\,cm$^{-3}$][]{2006ApJ...641..389R}.\\
Would this cloud be seen in extinction were it observed toward a bright mid-IR background?\\
A typical column density found for the cores is 10\,--\,20\,$\times$\,10$^{21}$\,cm$^{-2}$, which corresponds
to an $A_{\mr{V}}$ of 5\,--\,10\,mag \citep[][]{1978ApJ...224..132B}. The corresponding extinction at 8\,$\mu$m can be expressed as
$A_{8\mu\mr{m}}\sim$\,0.04$A_{\mr{V}} < 0.4$\,mag, using conversions adopted from \citet{2005ApJ...619..931I} and 
\citet{1985ApJ...288..618R}.
Typical extinctions around 8\,$\mu$m for Inner Galaxy IRDCs are 1\,--\,2\,mag \citep{2000ApJ...543L.157C}. Thus, based
on this simple reasoning the cloud would not meet the IRDC criterion of mid-IR extinction.
However, following the same analysis, the extinction toward position P8 (74\,$\times$\,10$^{21}$\,cm$^{-2}$)
is about 1.4\,mag at 8 $\mu$m. Considering that the above values represent the densest component of the
two CO fits toward P8, the actual 8\,$\mu$m extinction may even be higher when using a single fit (1.8\,mag).
This central part of the cloud makes the region a very promising Outer Galaxy IRDC candidate. The presence of 
possible embedded heating sources at this position may indicate that the core is not in a very early,
quiescent state, but already forming stars. Recent studies 
\citep[][ v.d. Wiel, privat comm.]{2005ApJ...630L.181R,2005A&A...439..613O},
using submm and Spitzer data indicate that some Inner Galaxy IRDCs are presently forming 
stars in their cores and along the filaments as well.\\
Obviously, the dark cloud candidate presented in this paper is in the vicinity of star forming
activity (NGC 7538, S159) and in that sense the boundary conditions may not be that deviant from 
Inner Galaxy IRDCs. It should be noted however, that there are environmental differences in the Outer 
Galaxy \citep[e.g., radiation field, density, abundance;][ and references therein]{1995A&A...303..851B,2006ApJS..162..346R} 
when compared to the inner spiral arms and the Molecular Ring, where most of the IRDCs 
are found. \\
The super-thermal line-widths and the presence of massive, cold and dense cores in G111.80+0.58 are 
both indicative of a high-mass star forming complex. The combined mass of the cores is perhaps too low to 
form a massive star cluster (like Orion). However, this region is part of a much larger molecular cloud 
complex. We conclude that the G111.80+0.58 complex likely belongs to a category of objects similar to 
intermediate-mass IRDCs \citep[e.g., IRDC G48 toward W51; ][]{2005A&A...439..613O} in an early, but 
not pristine stage of star formation. 

%__________________________________________________________________
\section{Concluding remarks}
\label{conclusions}
Radiative transfer codes can provide a better understanding of the
properties of the gas, e.g., radial density and temperature profiles
can be investigated, but require more detailed observations. We did conduct
tests with a multi-zone escape probability code 
\citep[$\beta$\,3D,][]{2006A&A...453..615P,2005A&A...440..559P} 
and these resulted in similar temperature, density and mass estimates as presented in this
paper. We therefore conclude that for the integrated properties these codes have no
additional contribution to the results. The main advantage when using these models is
getting a better understanding of the line profiles. In particular, effects such as self absorption or 
velocity structures, like infall motion may explain some of the double peaked and non-Gaussian 
features ({see e.g., Evans 1999}).
In the work presented here, non-Gaussian features are treated as separate components. 
Note in this  respect that different optically thin lines (e.g., \ceo\ and \cts\ for 
core P1) show their peak emission at opposite sides of the central velocity $V_{\mr{LSR}}$.\\
High spatial and spectral resolution in addition with radiative transfer models are required
in the future to get better constraints on the physical properties on smaller scales.
A spatial resolution at sub-arcsecond scales may resolve the presence of small high 
density clumps. Resolving this substructure will allow an investigation of the Dense Clump Mass 
Function (DCMF). It is of great importance to conduct these studies to put constraints on the
origin of the stellar mass spectrum. Not only in nearby, generally low
mass star forming clouds \citep[e.g., Pipe Nebula;  ][]{2006A&A...454..781L}, 
but particularly in the more massive and distant IRDCs, including the Outer Galaxy region 
presented in this paper.\\
A different way of characterizing IRDCs is by using (sub)mm continuum observations.
All Inner Galaxy IRDCs show strong submm emission \citep[e.g.,][]{2000ApJ...543L.157C}
and some contain bright, centrally peaked cores. In addition, these data allow an independent 
measurement of properties such as temperature, mass and luminosity. It will be worthwhile
to observe the G111.80+0.58 cloud in dust continuum and compare the data with
the molecular line results presented here and with existing submm studies in the Inner Galaxy.

%__________________________________________________________________
\begin{acknowledgements}
We thank the anonymous referee for his/her careful reading of the manuscript and his/her 
constructive remarks.
\end{acknowledgements}
%__________________________________________________________________
\bibliographystyle{aa}
\bibliography{master}
%__________________________________________________________________
\newpage
\appendix
\section{Calibration of data obtained at the 100-m telescope}
\label{appen1}
\begin{figure}[htbp]
\label{calibeffel}
\centering 
%\begin{minipage}[r]{0.5\linewidth}
%\hspace{-3cm}
\centering \includegraphics[width=9cm]{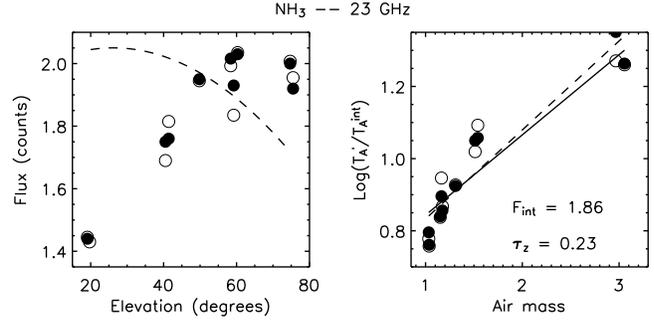} 
%\end{minipage}% 
%\begin{minipage}[c]{0.5\linewidth} 
%\centering \includegraphics[width=9cm]{Images/DC111_c18o.ps} 
%\end{minipage} 
\caption{Example of Effelsberg calibration measurements taken on NGC 7027 for one day at the two frequencies considered here. {\bf Left}: Flux measurements for azimuthal (black circles) and horizontal (white circles) scans. The elevation gain curve is plotted, normalized to the highest flux of the graph. {\bf Right}: fit of eq.~\ref{eq cal 100m 2} on data collected during a single day for azimuthal (full line) and horizontal (dashed lines) data. The resulting $\tau_z$ and F$_{\rm int}$ are the mean of these two fits.}
\end{figure}

At the frequency considered here (23\,GHz), the Effelsberg
telescope offers a calibration based on noise diodes, so that the
received signal is not converted into K but into internal receiver
counts. In order to both convert the receiver counts into antenna
temperature, as well as to correct for the atmospherical absorption,
we have observed a photometric calibrator (here NGC 7027) at various
elevations in the course of each observing night. This approach is
similar to the so-called ``antenna tipping'' used to measure the
sky opacity during e.g., bolometer observations. The weather conditions
were stable enough to assume that the opacity derived through this
technique was reasonably representative of the daily set of data. If
F$_{\rm int}$ is the conversion factor between internal counts and Jy above
the atmosphere and corrected for backward losses, the measured
receiver counts are (in a given spectrometer channel):
\begin{eqnarray}
\label{eq cal 100m}
T^{\rm int}_{\rm A} = G(\rm el) \times {\rm e}^{(-\tau_z\,A)}~,
\times \frac{T^{*}_{\rm A}\,\eta_\ell}{F_{\rm int}}
\end{eqnarray}
where $A$ is the airmass at elevation el, $\tau_z$ the zenith opacity
and $G(\rm el)$ the normalized elevation gain. Observing a photometric
calibrator of known flux $S_{\nu,{\rm ref}}$ one has:
\begin{eqnarray}
\label{eq cal 100m 2}
T^{*}_{\rm A,ref} & = &\frac{A_{\rm geom}\,\eta_A}{2k\,\eta_\ell}\,S_{\nu,{\rm ref}}
        = \frac{G_{\rm (K/Jy)}}{\eta_\ell}\,S_{\nu,{\rm ref}}
\end{eqnarray}
and
\begin{eqnarray}
{\rm Log}\left(\frac{G_{\rm (K/Jy)}\,S_{\nu,{\rm ref}}}{T^{\rm int}_{\rm A}}\right)& =&
{\rm Log}\left(\frac{F_{\rm int}}{G(\rm el)}\right) + \frac{\tau_z}{\rm sin(el)}~.
\end{eqnarray}
Using the calibrated flux published by \citet{1994A&A...284..331O} and observing
the reference at various elevations, the calibration procedure thus
consists of fitting the previous expression with a pair of ($F_{\rm
int}$,$\tau$), assuming that the opacity has not significantly changed
in the course of the observations. The antenna temperature is then
given by:
\begin{eqnarray}
\label{eq cal 100m 3}
T^{*}_{\rm A} = \frac{T^{\rm int}_{\rm A} \times G_{\rm (K/Jy)}\times F_{\rm int}}{G(\rm el)\times
        {\rm e}^{(-\tau_z\,A)} \times \eta_\ell}~.
\end{eqnarray}
The NGC 7027 flux at 23\,GHz was taken as 5.11\,Jy.
We derived a conversion gain of 1.86$\pm$0.1 and the
opacity was around 0.2, consistent with expected
atmospherical transmission in the centrimetric domain. Figure~\ref{calibeffel}.1
illustrates the steps involved in the calibration process. It
shows the combined effect of antenna gain and opacity for data taken
at various elevations. Finally, assuming that our sources emit in a
solid angle comparable to the main beam, the derived fluxes are
translated into antenna main beam temperatures using the antenna
efficiencies provided by the 100-m staff (A. Kraus, priv. comm.).
%%%%%%%%%%%%%%%%%%%%%%%%%%%%%%%%%%%%%%%%%%%%%%%%%%%%%%%%%%%%%%%%%%%%%%
\newpage
\section{Figures}
\label{appen2}
\begin{figure*}[!H]
\centering
\subfloat{
\includegraphics[bb=75 375 660 900,width=0.25\linewidth,clip]{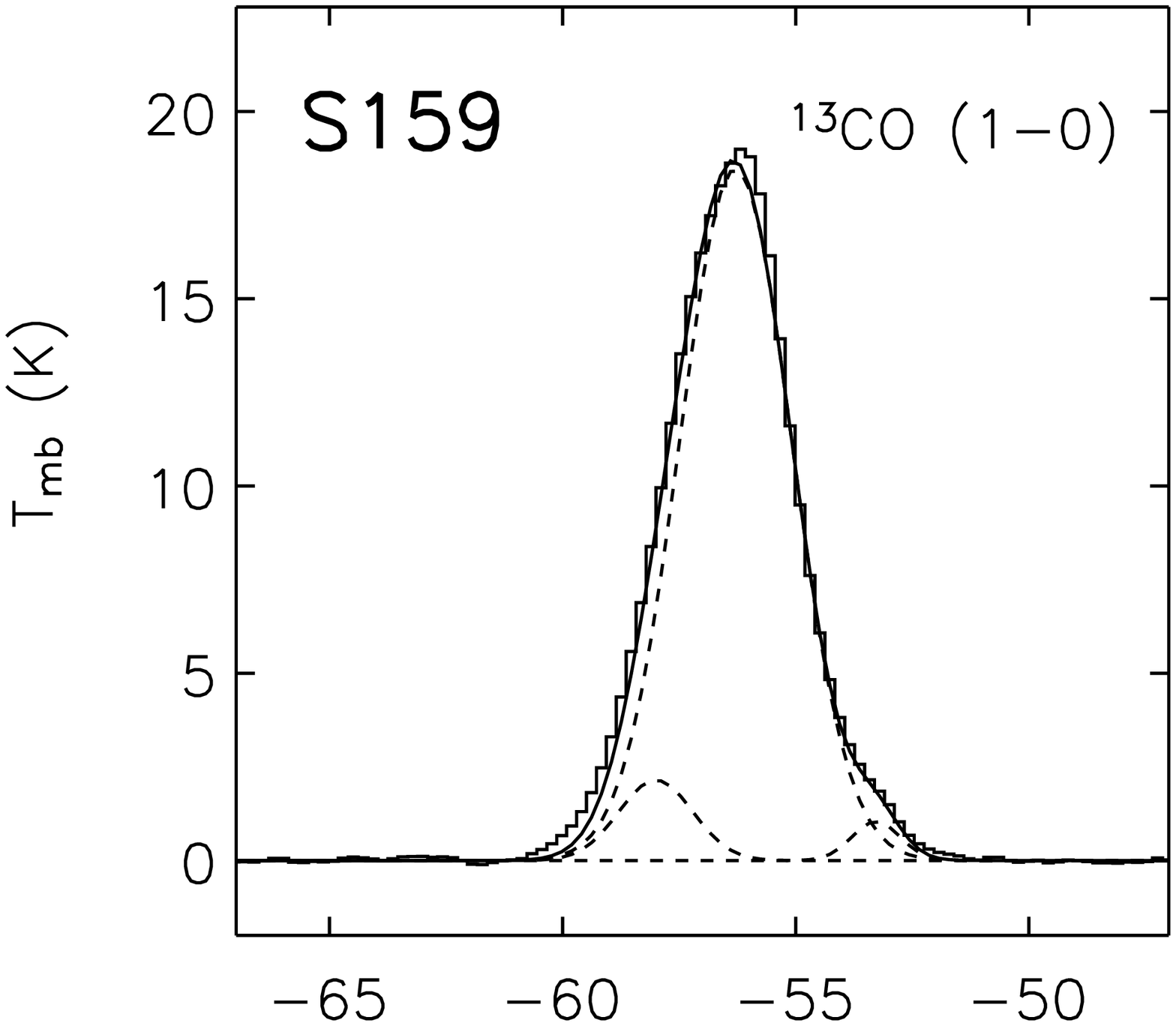}}
\hspace{-0.04\linewidth}
\subfloat{
\includegraphics[bb=75 375 660 900,width=0.25\linewidth,clip]{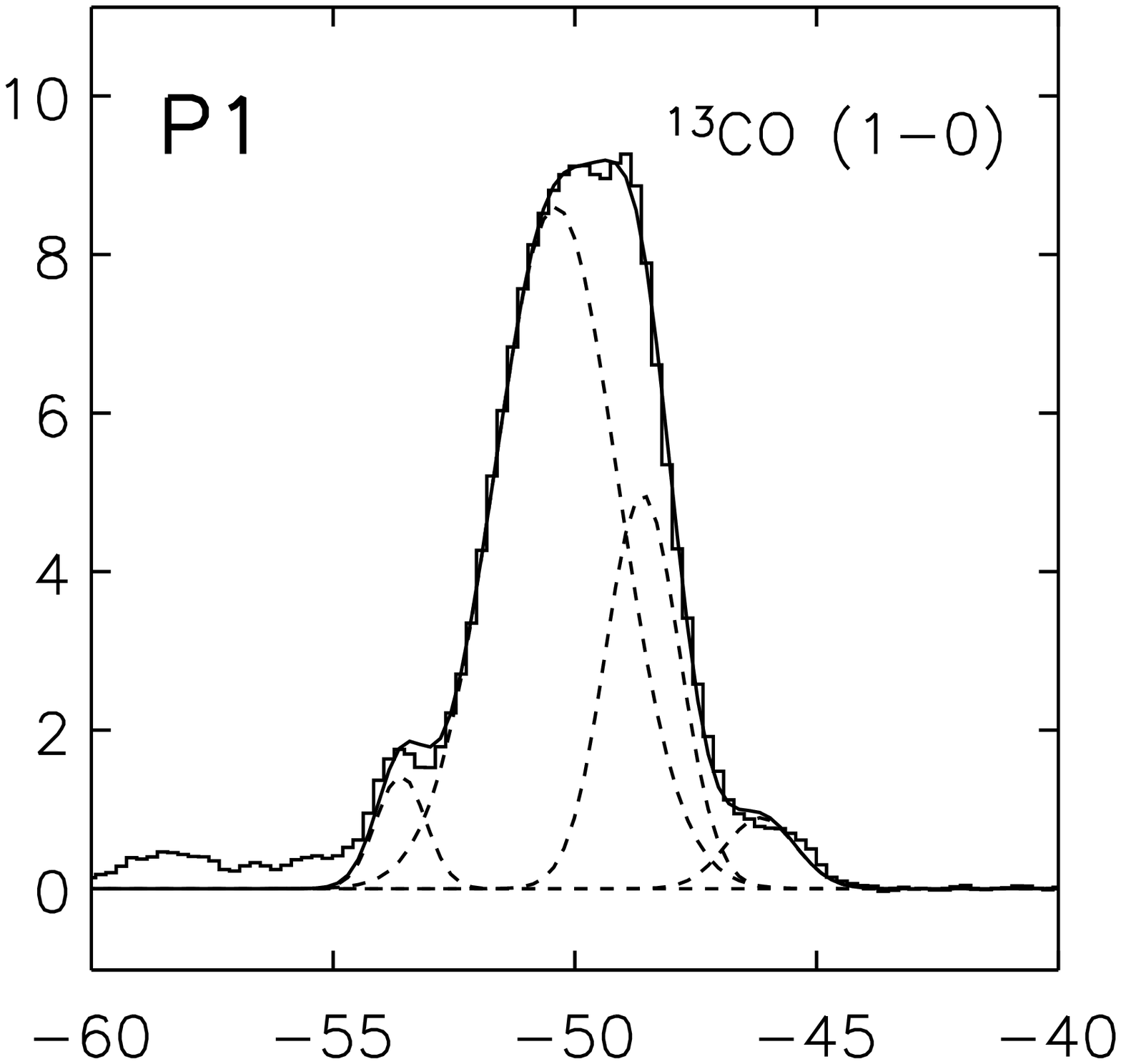}}
\hspace{-0.04\linewidth}
\subfloat{
\includegraphics[bb=75 375 660 900,width=0.25\linewidth,clip]{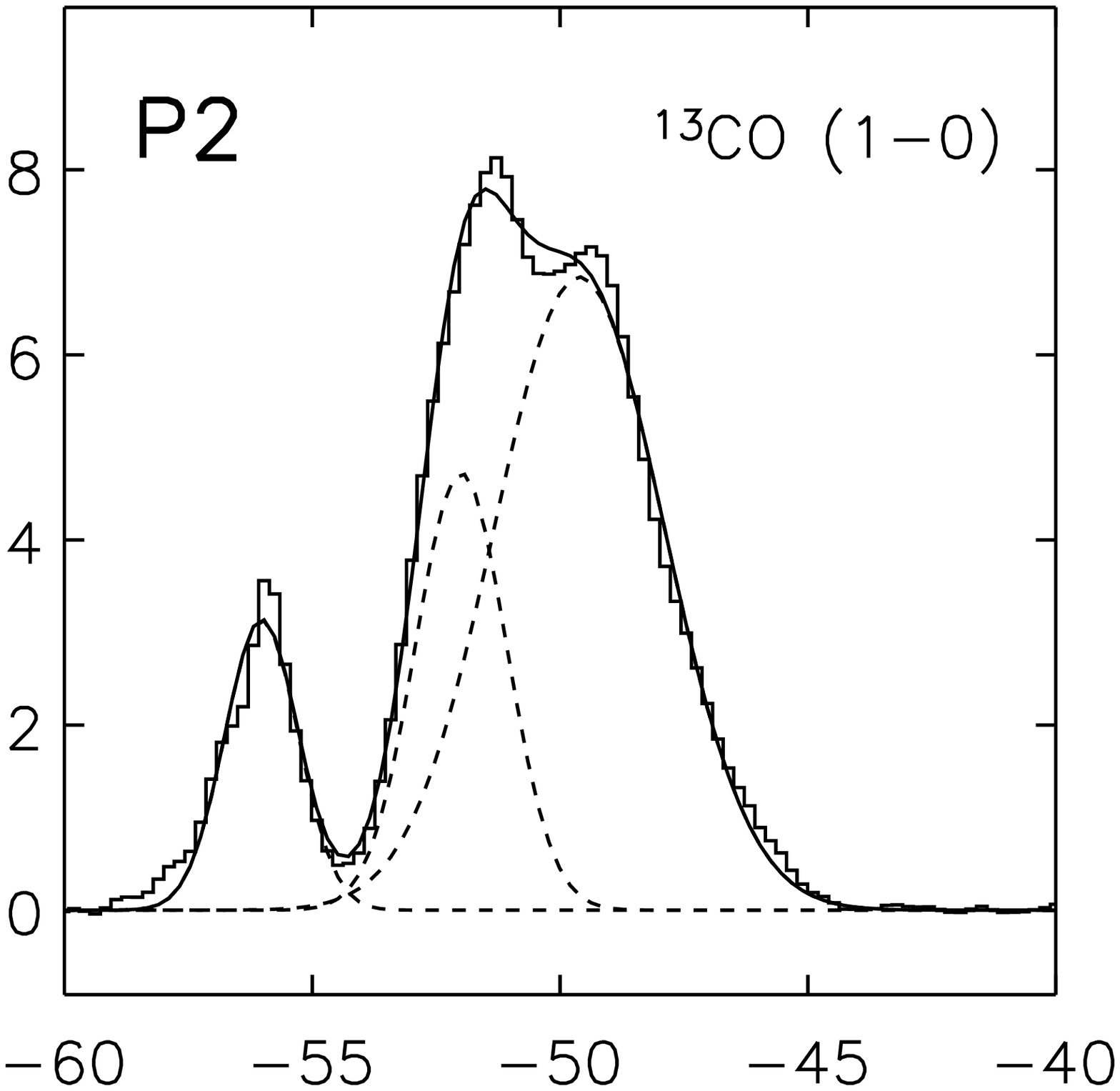}}
\hspace{-0.04\linewidth}
\subfloat{
\includegraphics[bb=75 375 660 900,width=0.25\linewidth,clip]{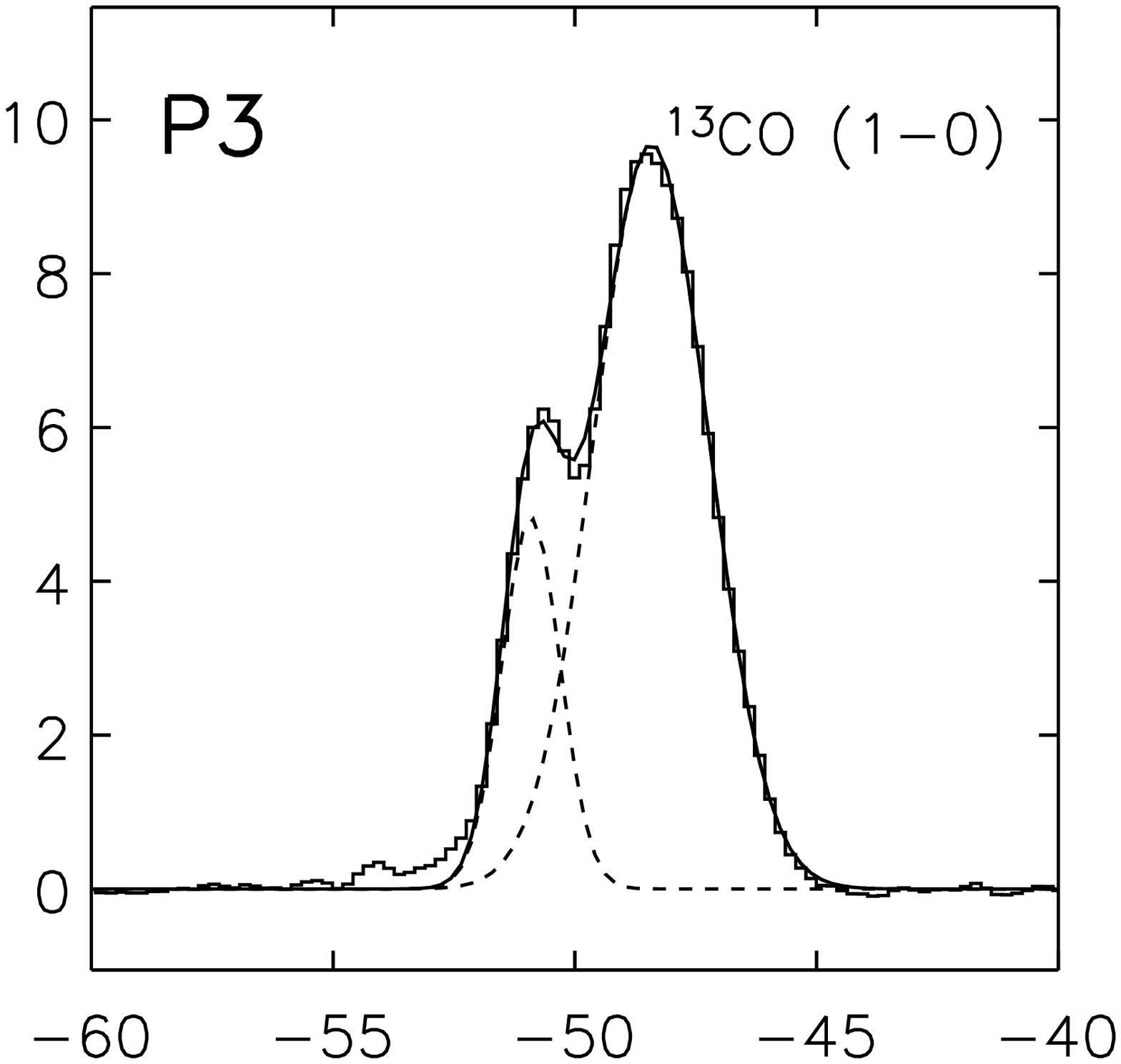}}
\\[-10pt]
\subfloat{
\includegraphics[bb=75 375 660 900,width=0.25\linewidth,clip]{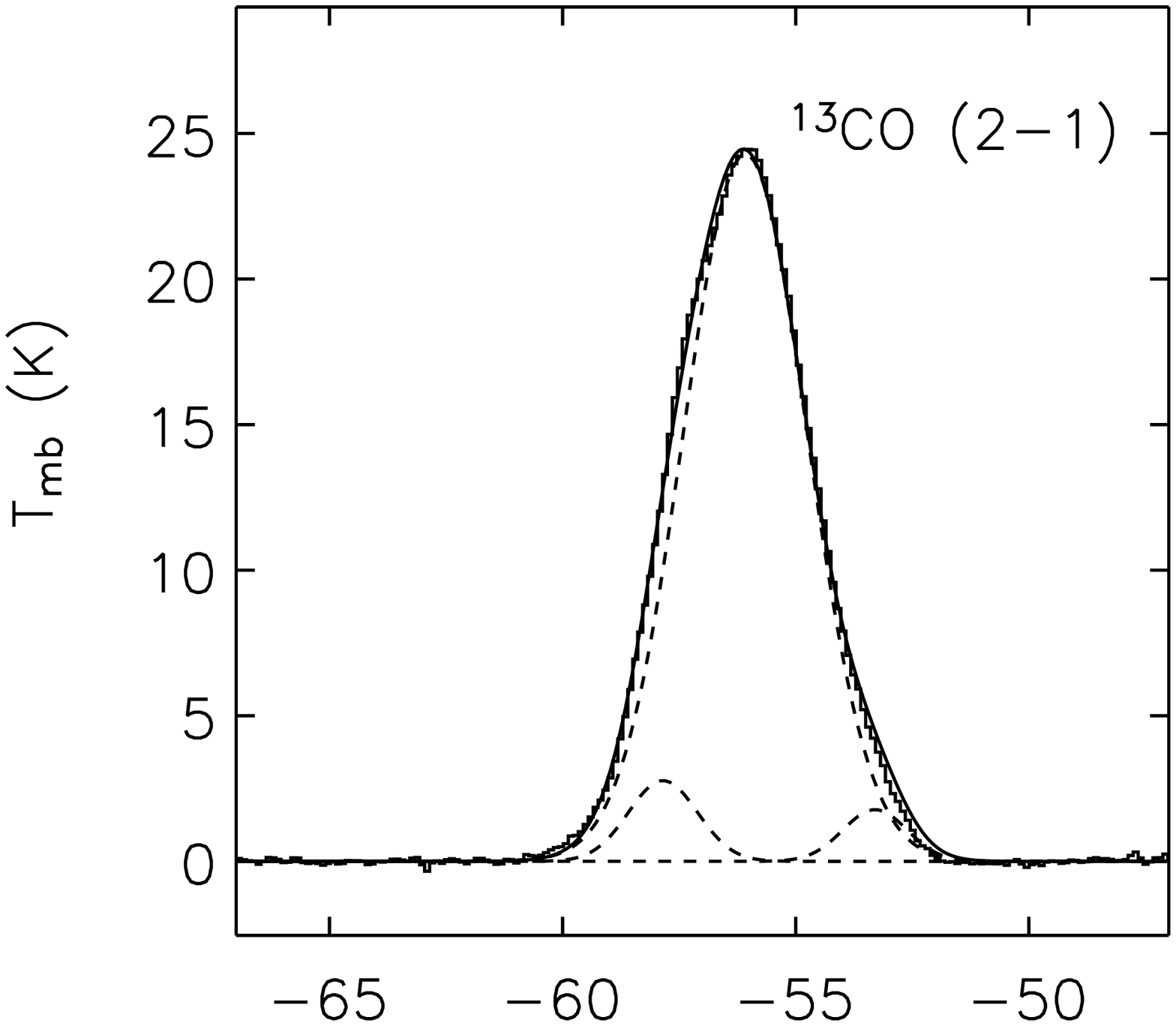}}
\hspace{-0.04\linewidth}
\subfloat{
\includegraphics[bb=75 375 660 900,width=0.25\linewidth,clip]{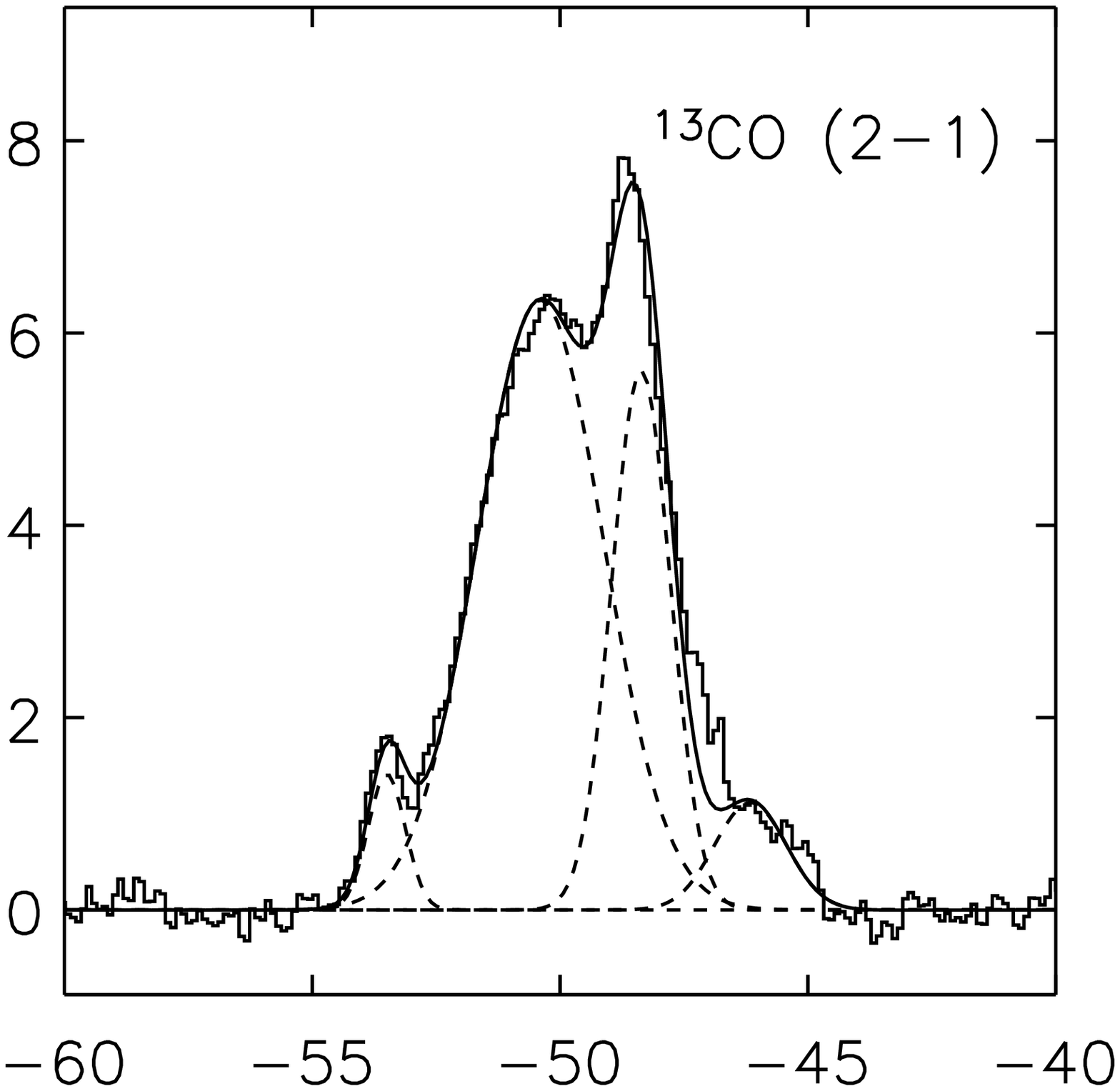}}
\hspace{-0.04\linewidth}
\subfloat{
\includegraphics[bb=75 375 660 900,width=0.25\linewidth,clip]{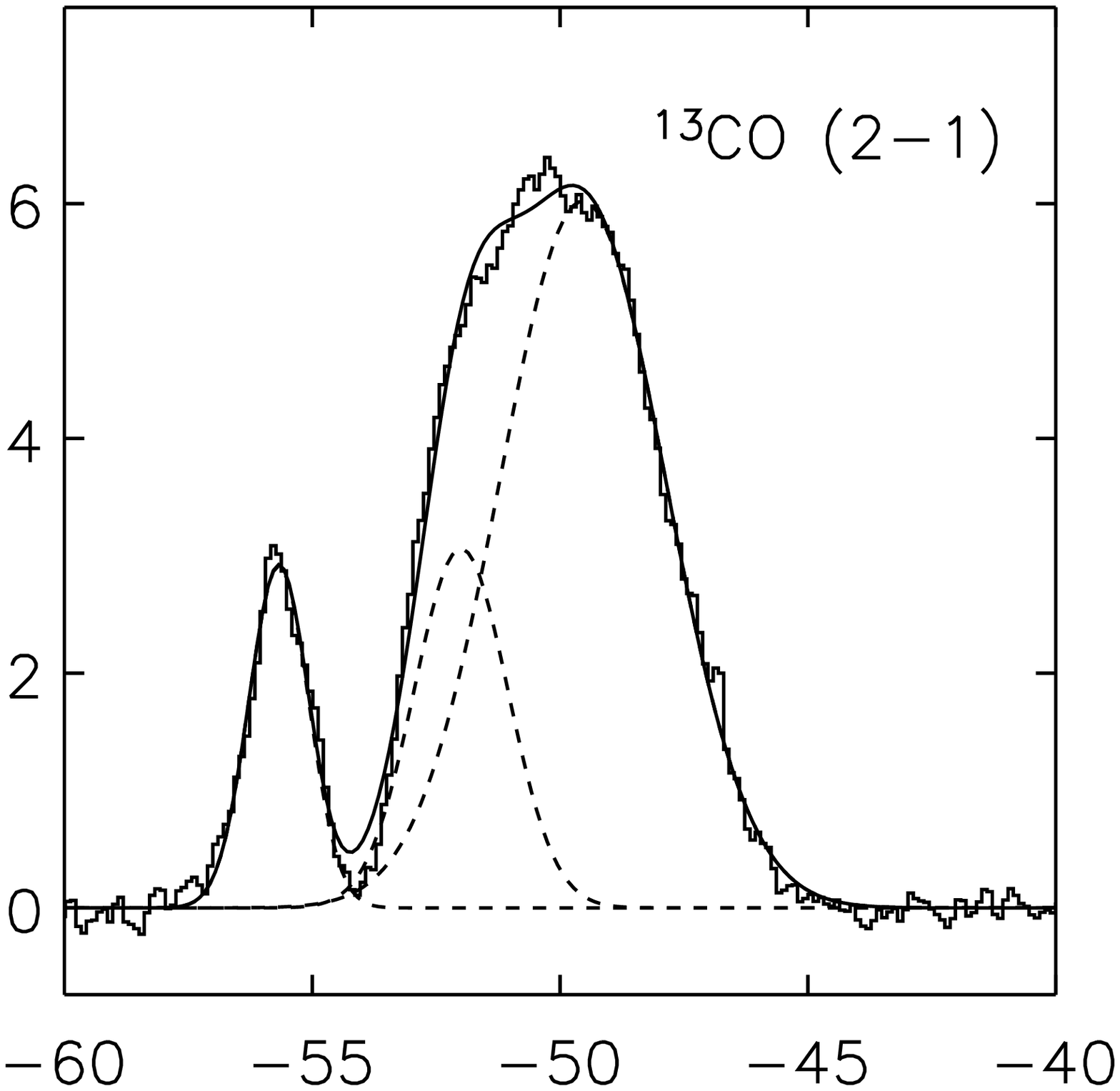}}
\hspace{-0.04\linewidth}
\subfloat{
\includegraphics[bb=75 375 660 900,width=0.25\linewidth,clip]{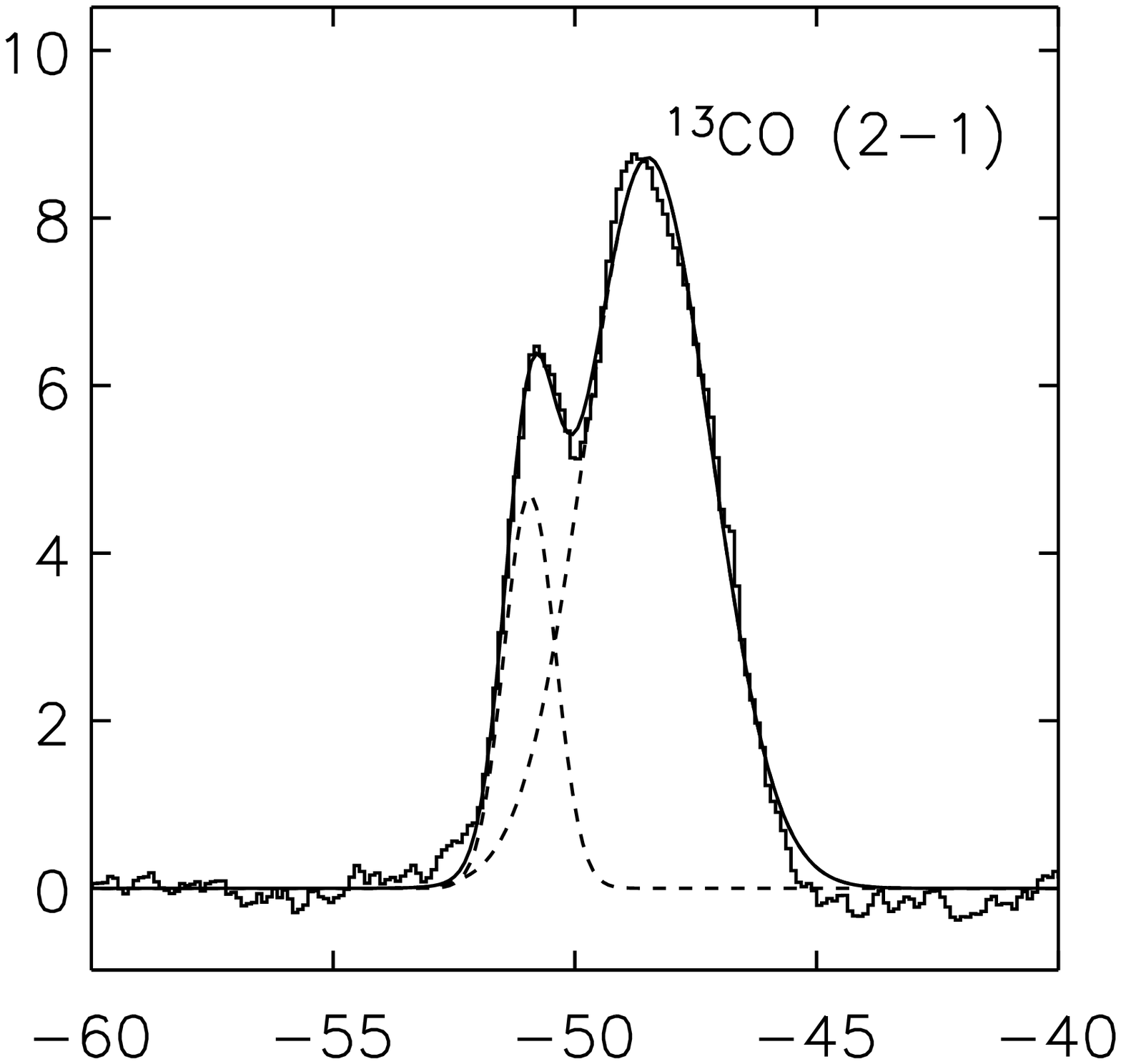}}
\\[-10pt]
\subfloat{
\includegraphics[bb=75 375 660 900,width=0.25\linewidth,clip]{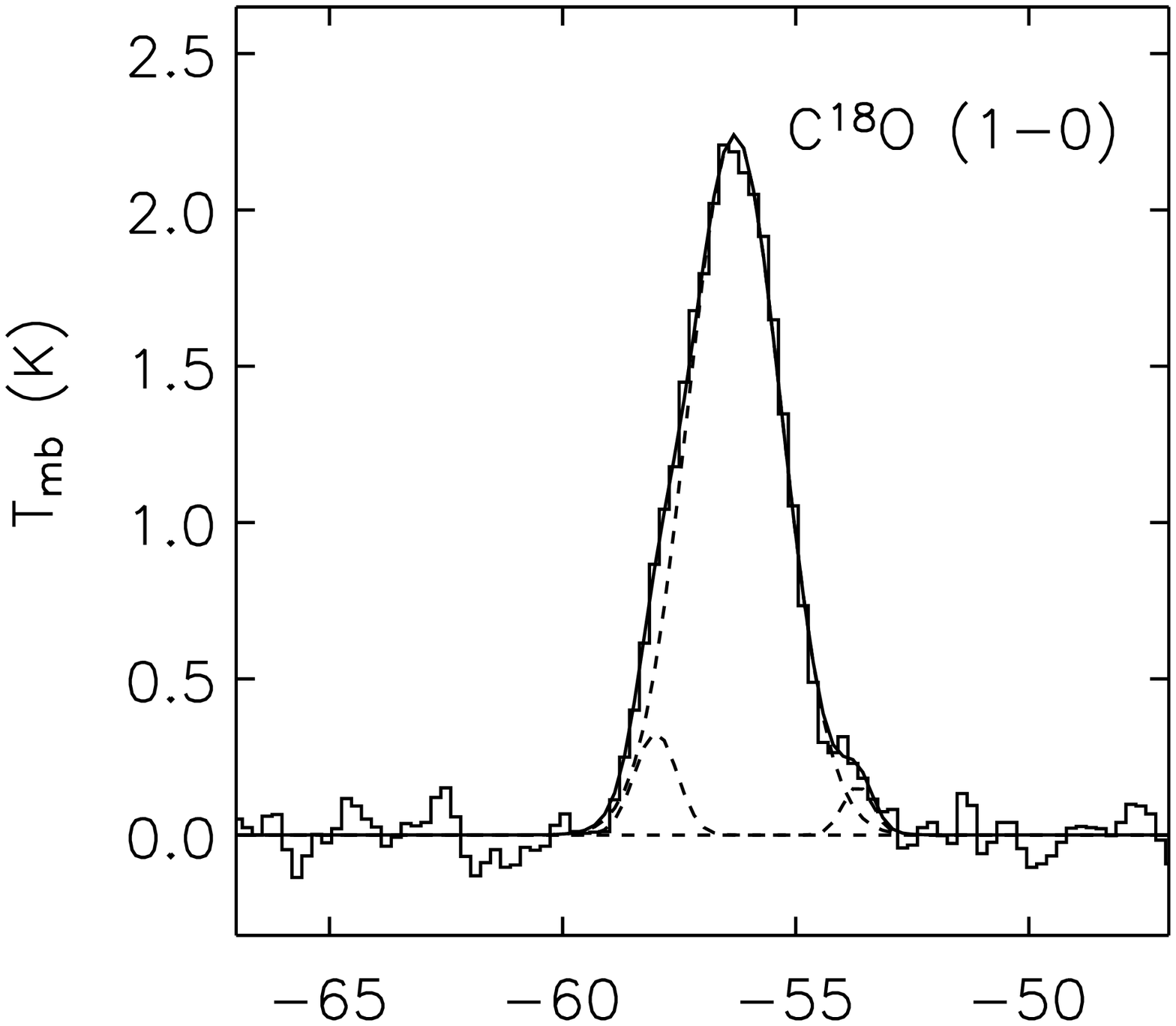}}
\hspace{-0.04\linewidth}
\subfloat{
\includegraphics[bb=75 375 660 900,width=0.25\linewidth,clip]{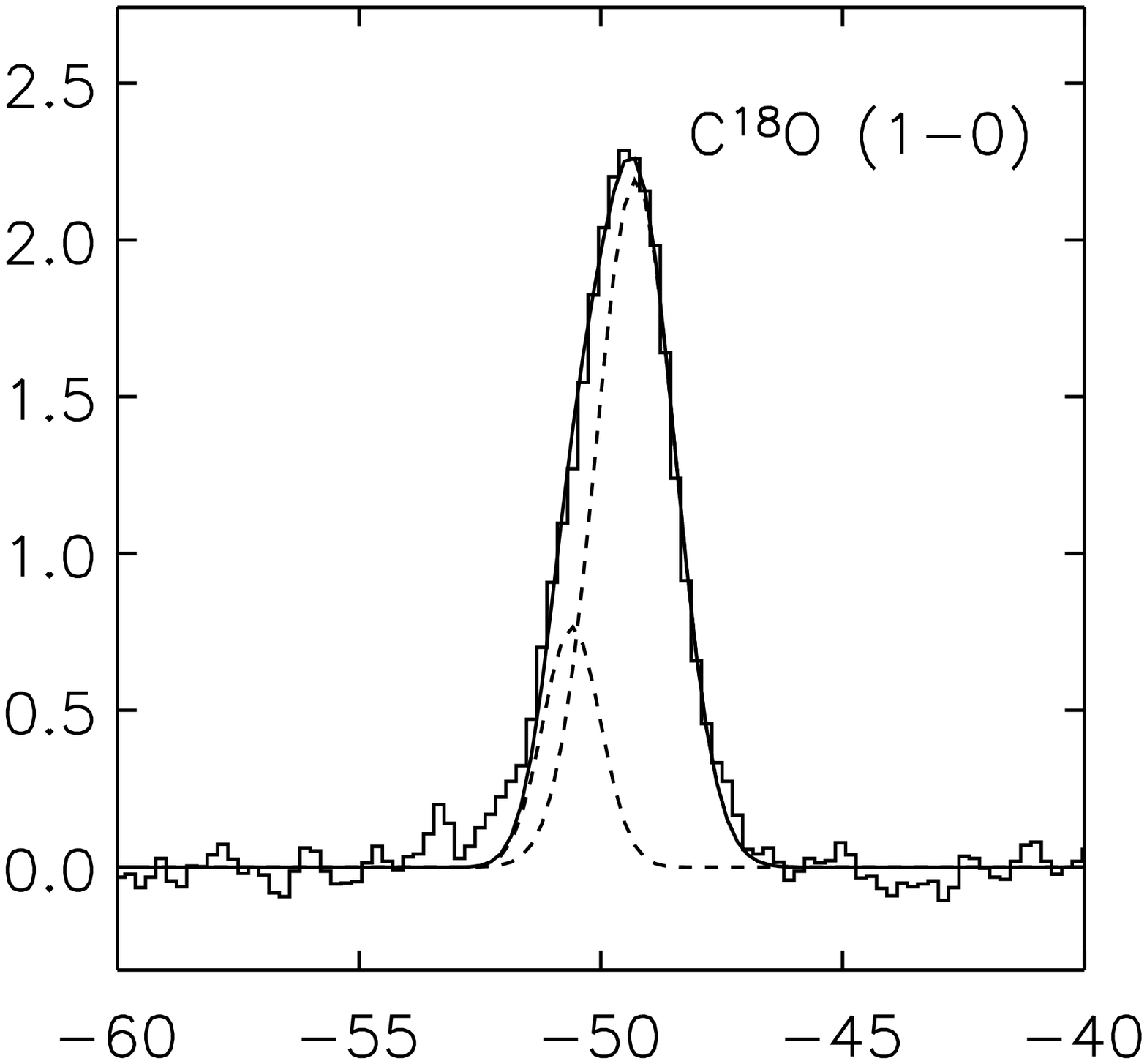}}
\hspace{-0.04\linewidth}
\subfloat{
\includegraphics[bb=75 375 660 900,width=0.25\linewidth,clip]{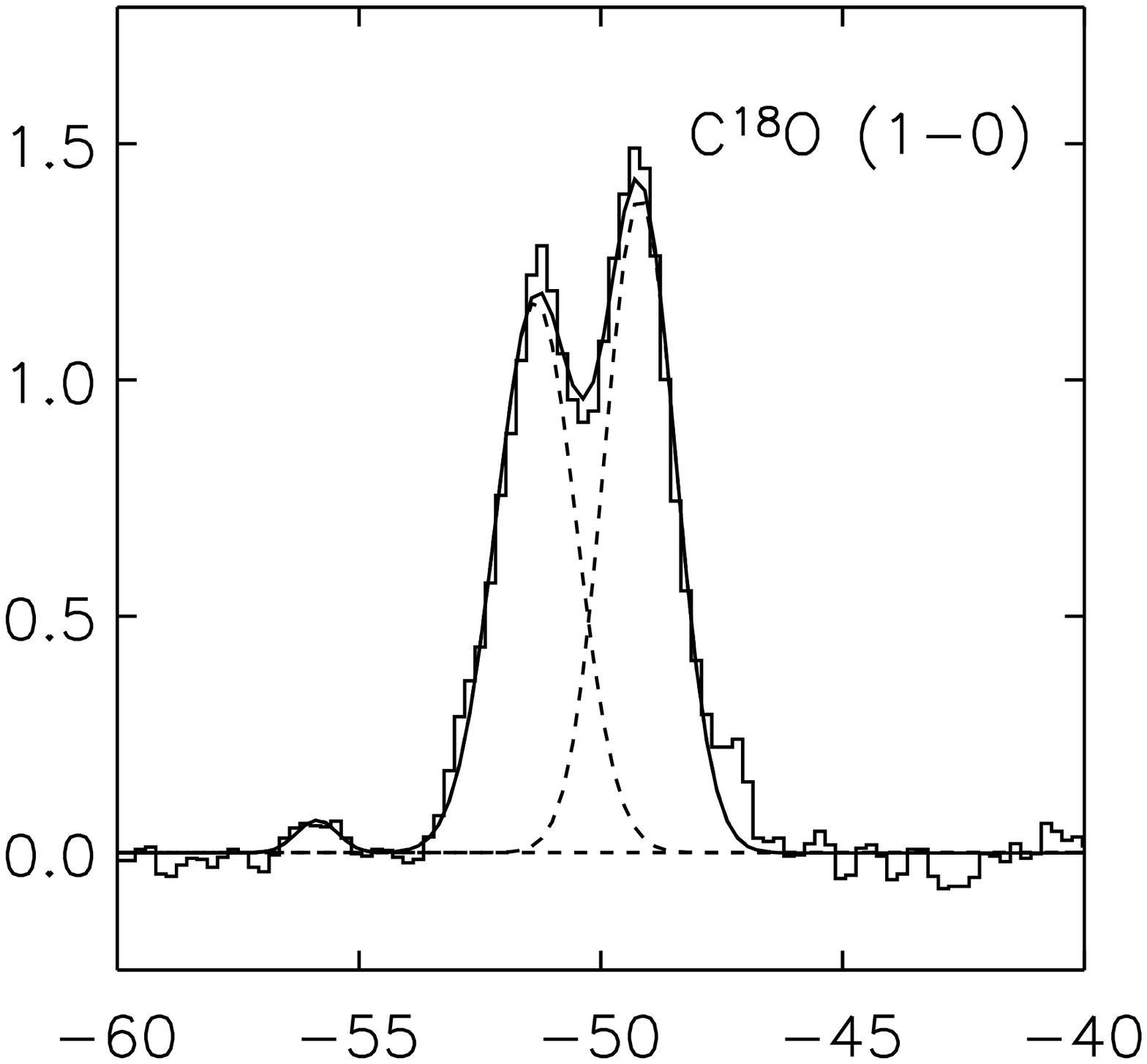}}
\hspace{-0.04\linewidth}
\subfloat{
\includegraphics[bb=75 375 660 900,width=0.25\linewidth,clip]{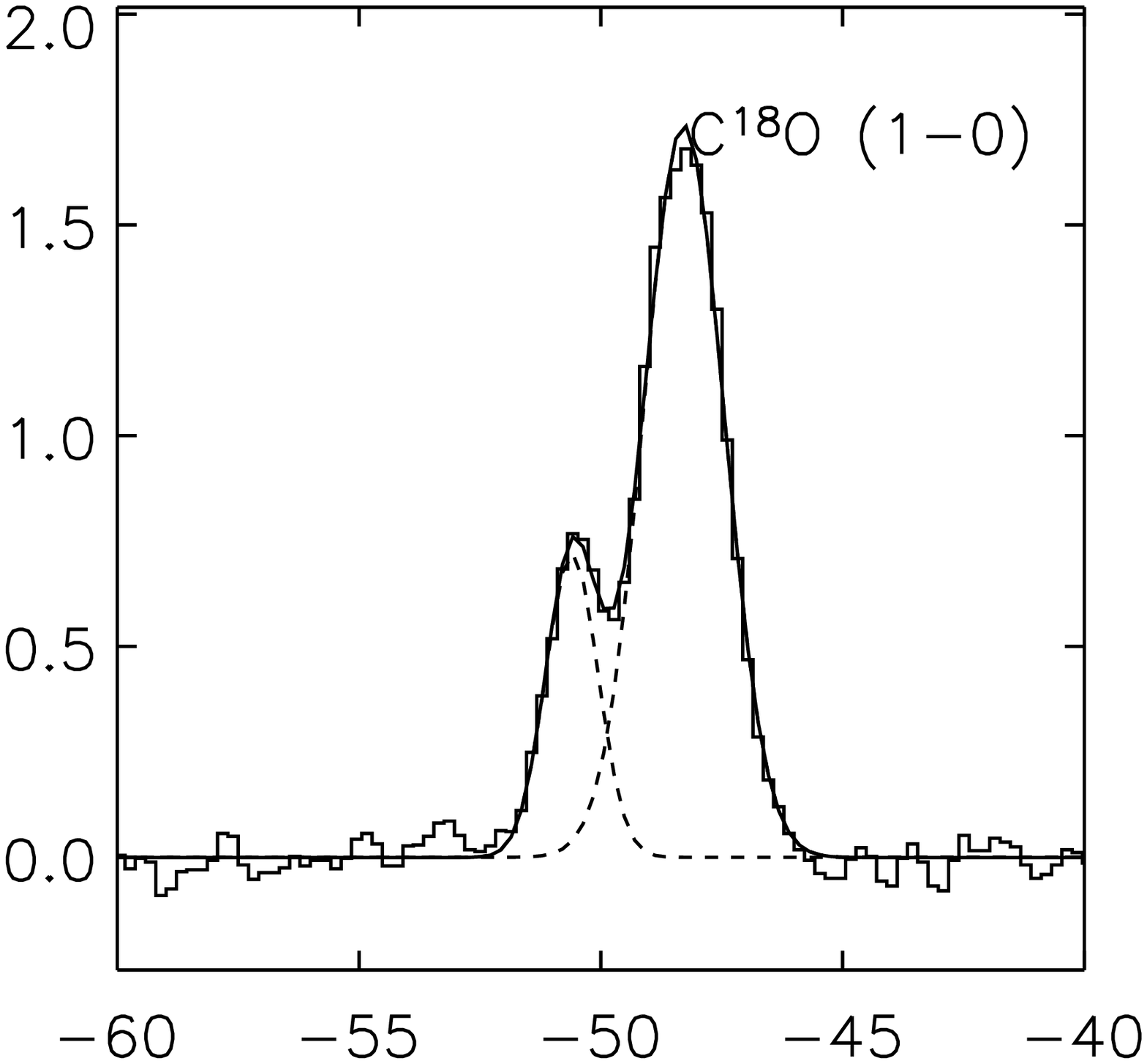}}
\\[-10pt]
\subfloat{
\includegraphics[bb=75 375 660 900,width=0.25\linewidth,clip]{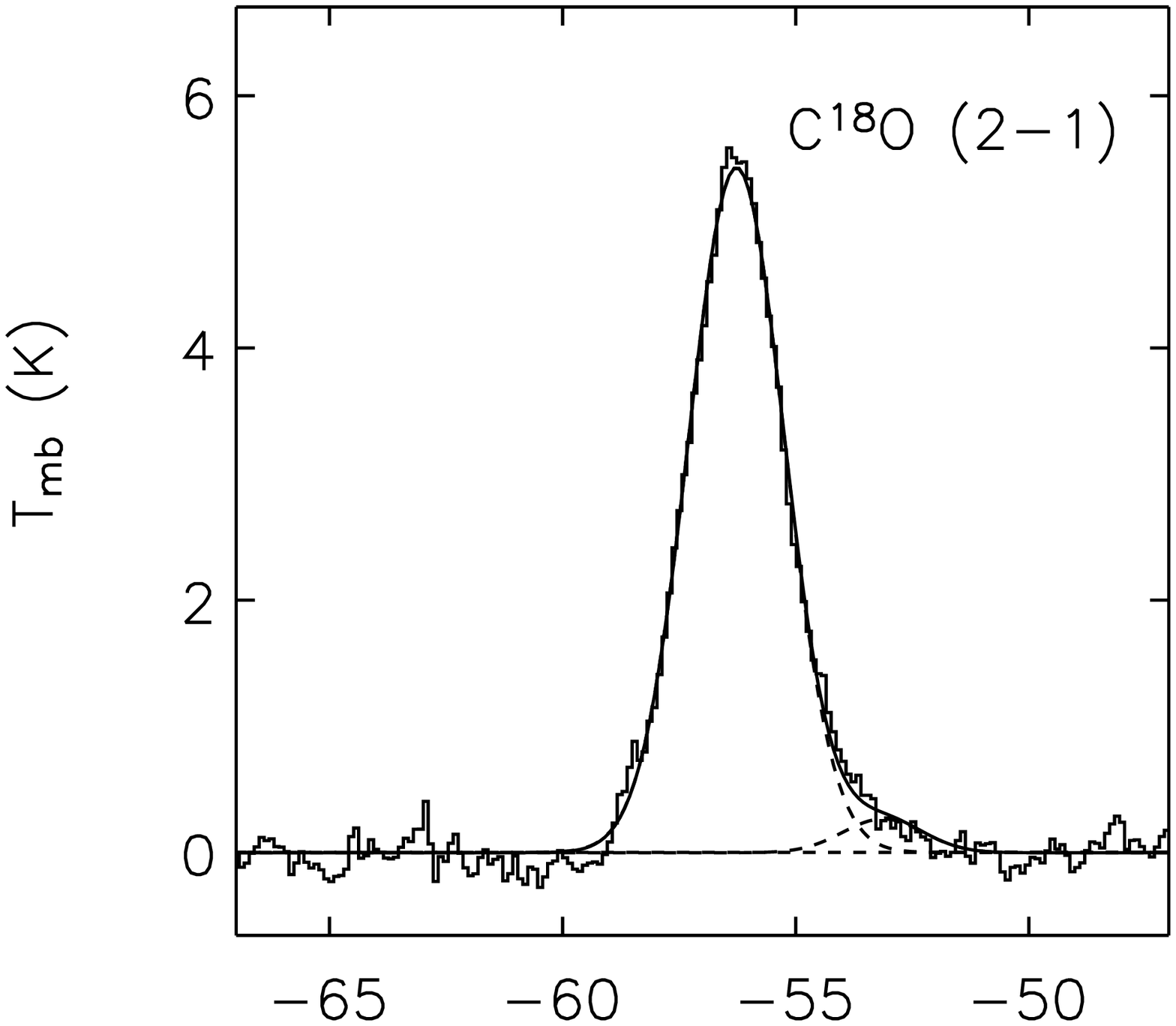}}
\hspace{-0.04\linewidth}
\subfloat{
\includegraphics[bb=75 375 660 900,width=0.25\linewidth,clip]{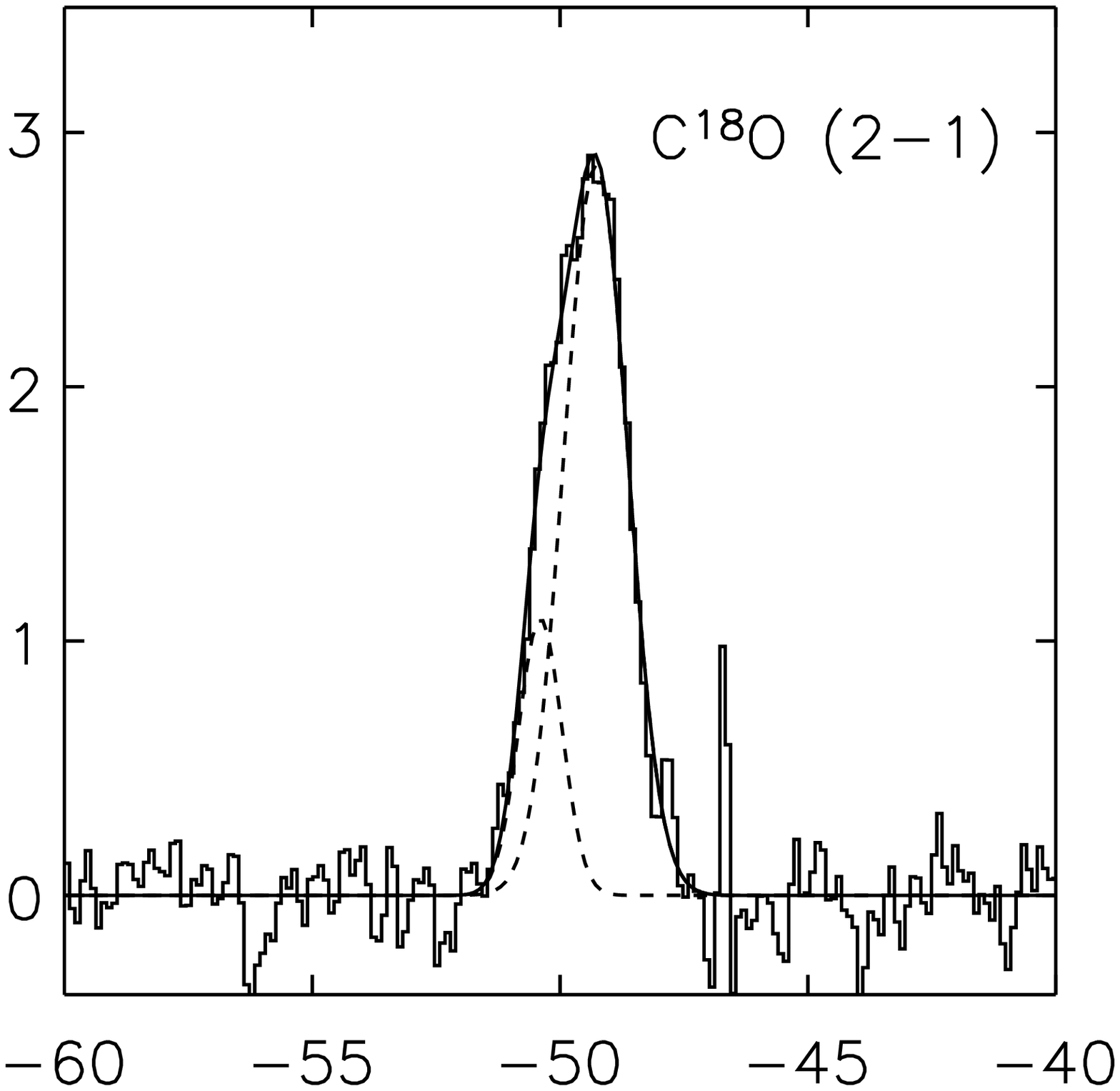}}
\hspace{-0.04\linewidth}
\subfloat{
\includegraphics[bb=75 375 660 900,width=0.25\linewidth,clip]{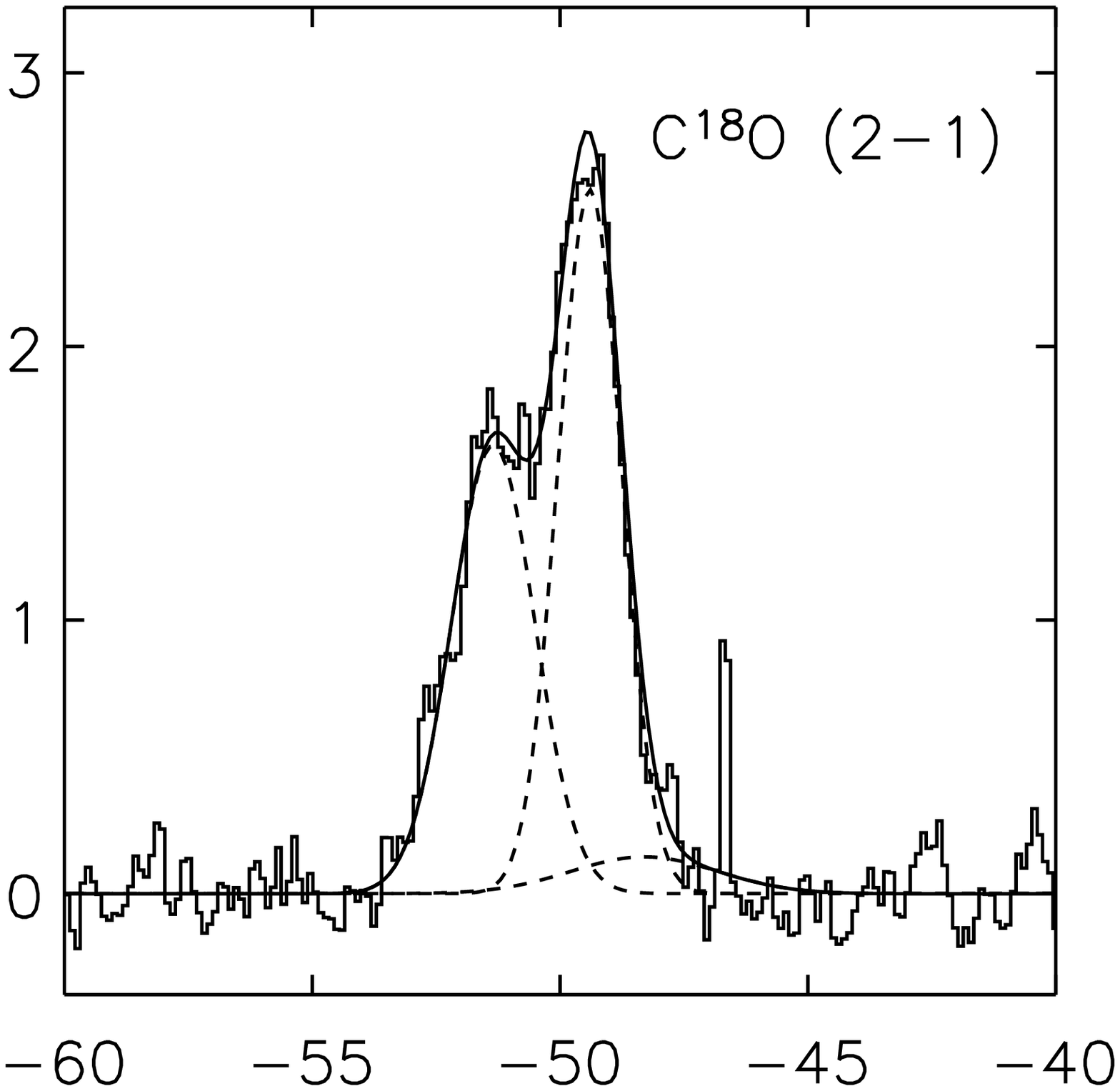}}
\hspace{-0.04\linewidth}
\subfloat{
\includegraphics[bb=75 375 660 900,width=0.25\linewidth,clip]{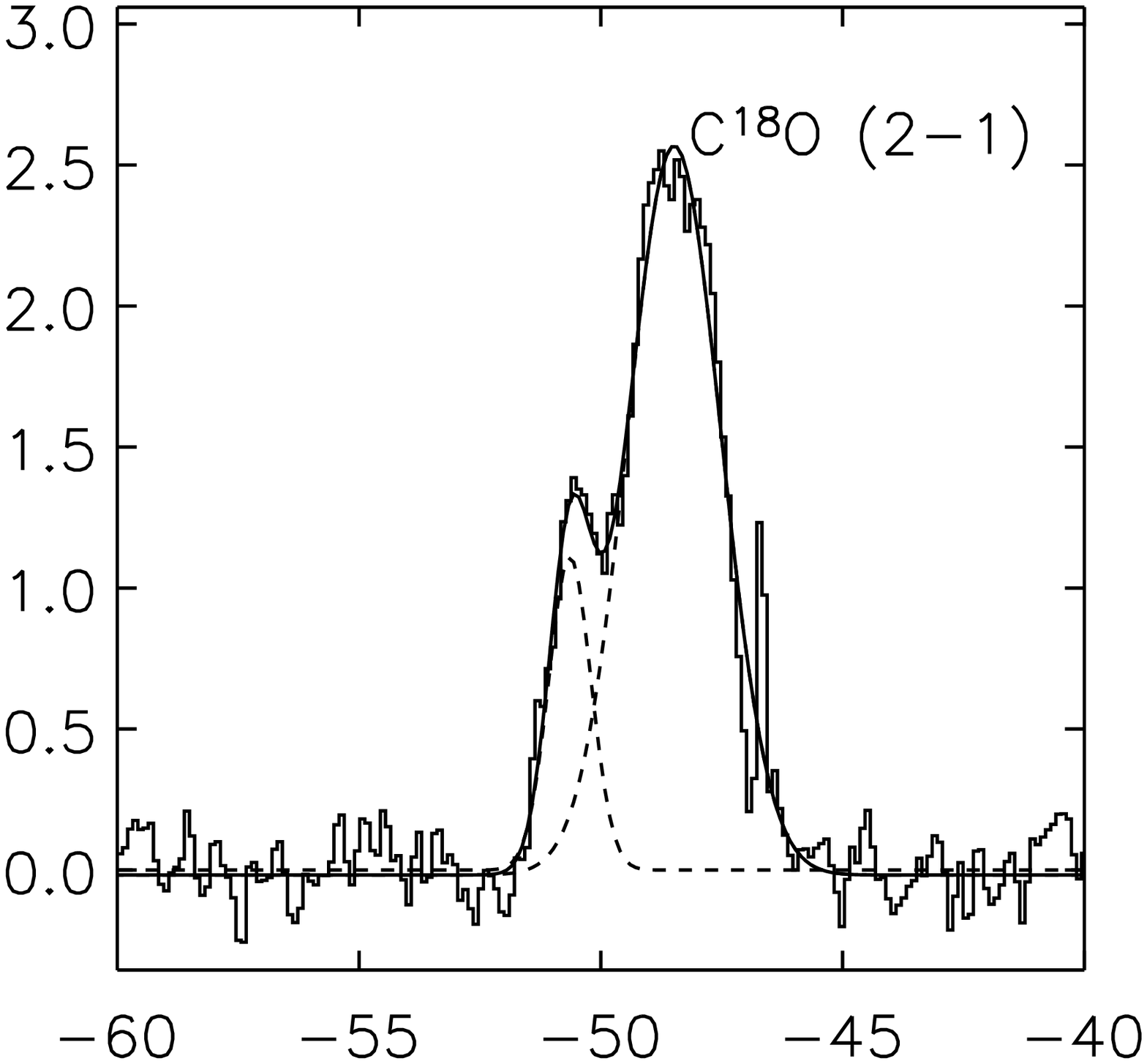}}
\\[-10pt]
\subfloat{
\includegraphics[bb=75 375 660 900,width=0.25\linewidth,clip]{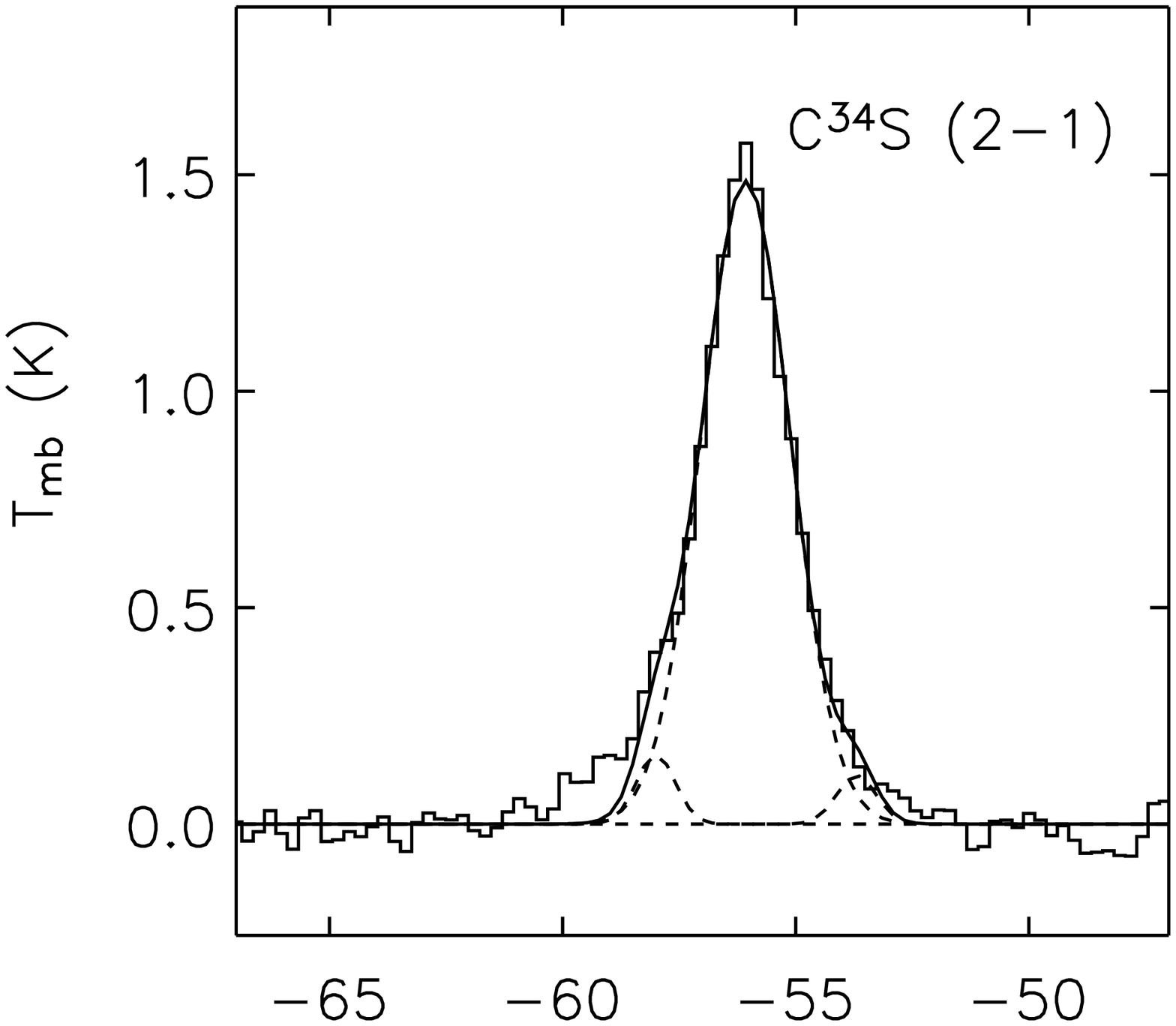}}
\hspace{-0.04\linewidth}
\subfloat{
\includegraphics[bb=75 375 660 900,width=0.25\linewidth,clip]{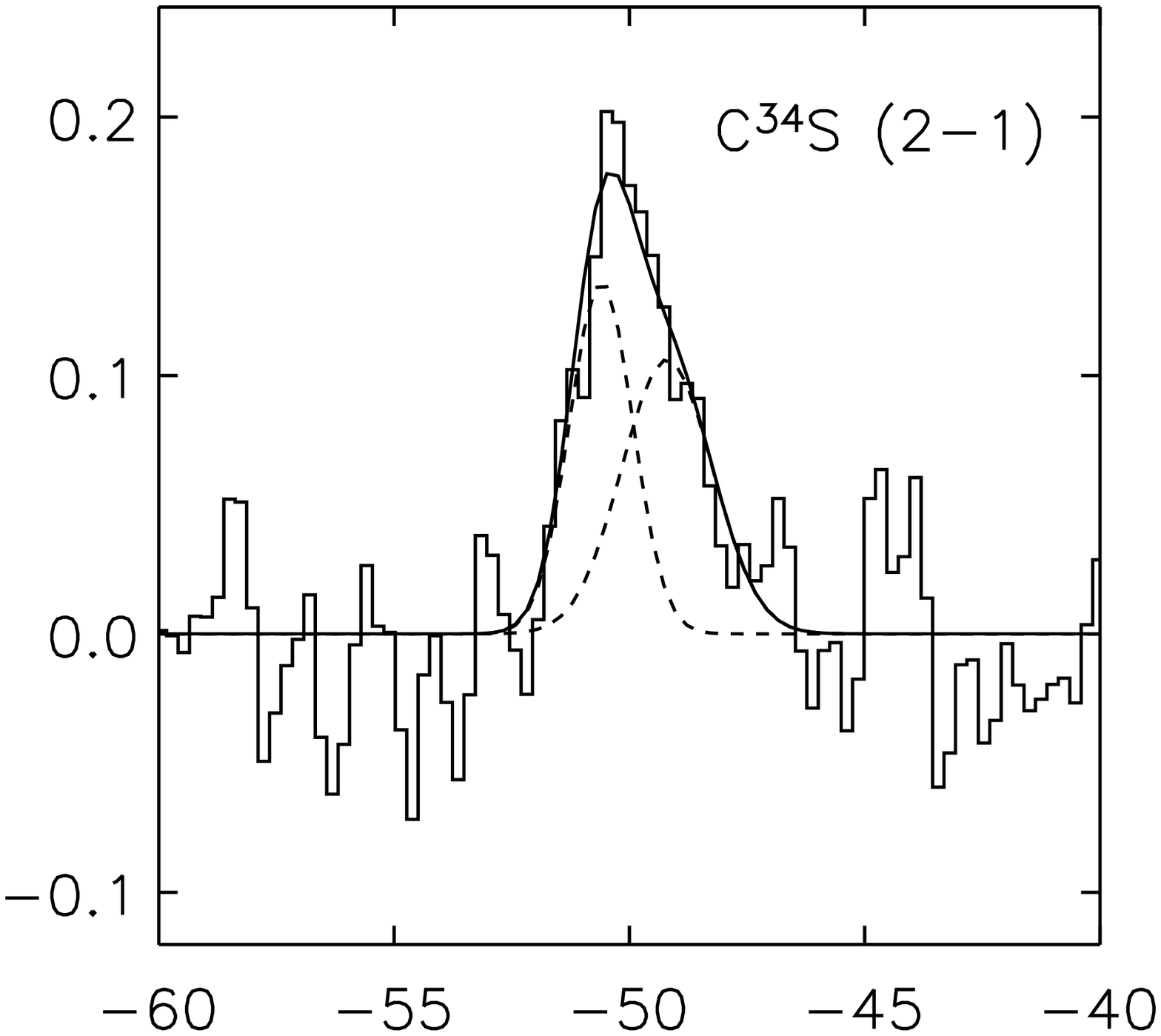}}
\hspace{-0.04\linewidth}
\subfloat{
\includegraphics[bb=75 375 660 900,width=0.25\linewidth,clip]{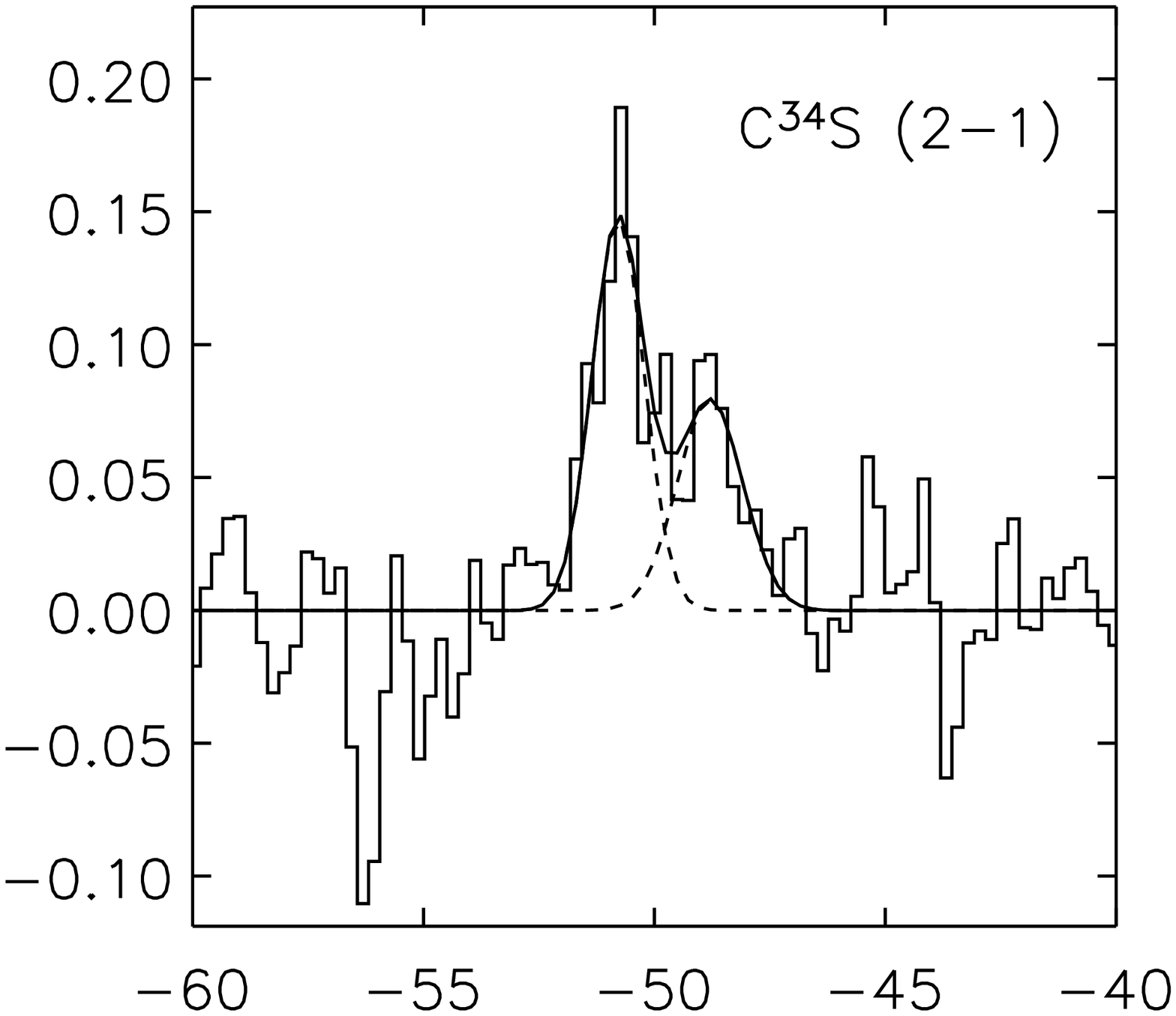}}
\hspace{-0.04\linewidth}
\subfloat{
\includegraphics[bb=75 375 660 900,width=0.25\linewidth,clip]{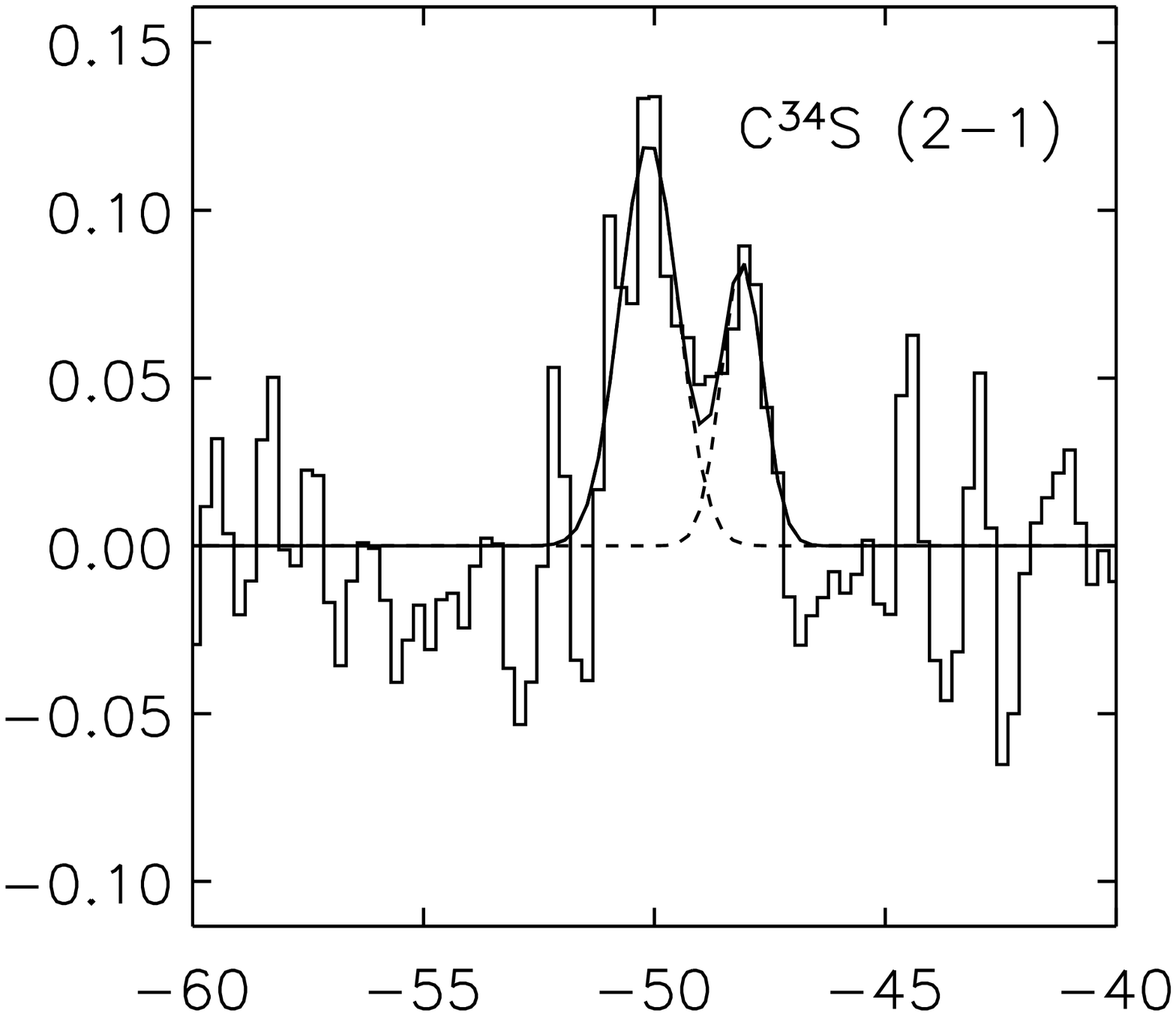}}
\\[-10pt]
\subfloat{
\includegraphics[bb=75 375 660 900,width=0.25\linewidth,clip]{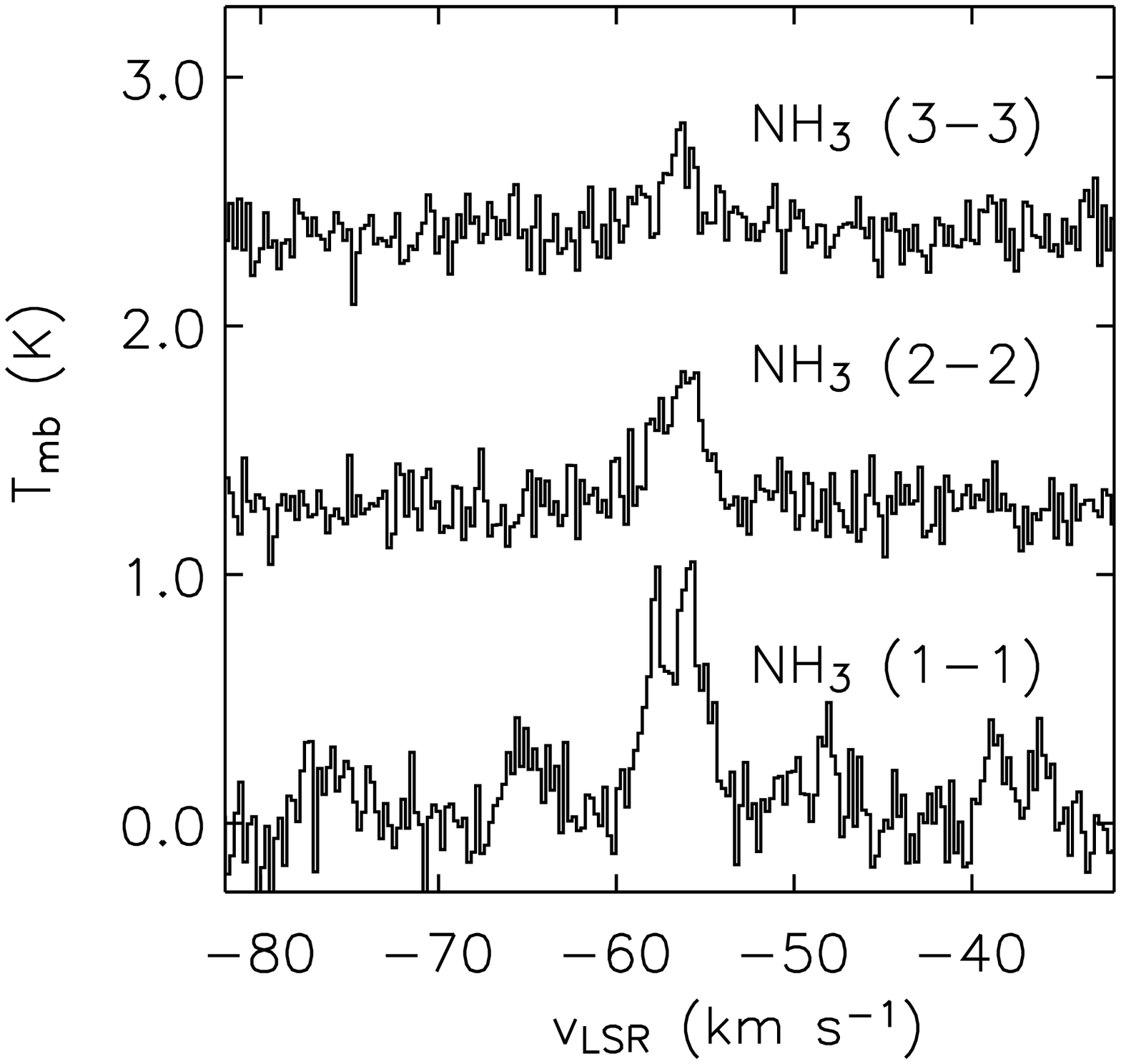}}
\hspace{-0.04\linewidth}
\subfloat{
\includegraphics[bb=75 375 660 900,width=0.25\linewidth,clip]{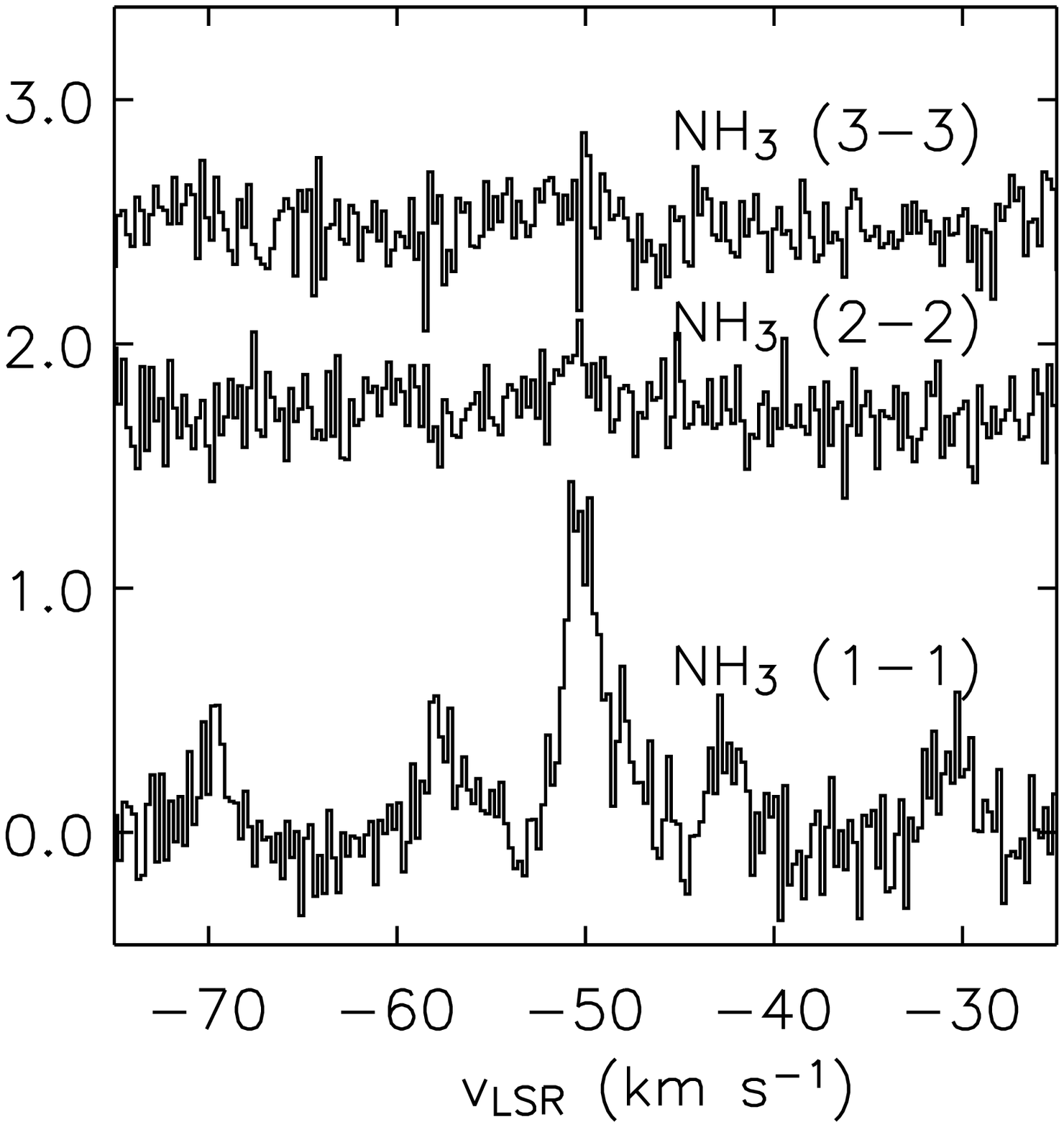}}
\hspace{-0.04\linewidth}
\subfloat{
\includegraphics[bb=75 375 660 900,width=0.25\linewidth,clip]{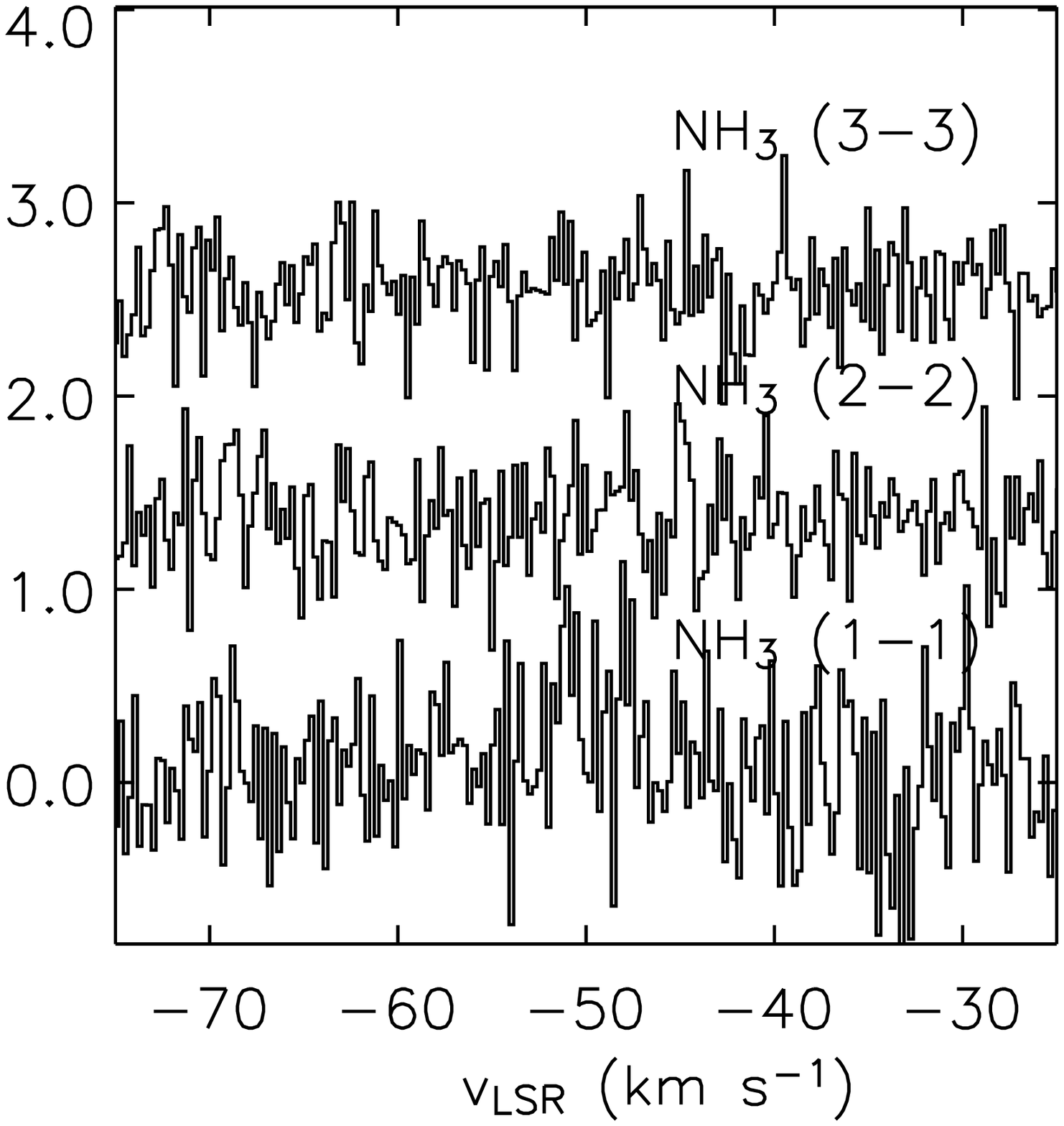}}
\hspace{-0.04\linewidth}
\subfloat{
\includegraphics[bb=75 375 660 900,width=0.25\linewidth,clip]{dummy.eps}}
\caption{Observed spectra (solid lines) and for the IRAM data the Gaussian fits (dashed lines) for the positions 
listed in Table \ref{tab1}. Each column contains the spectra of the position labeled at the top in the following 
order:  \tco\ 1-0, \tco\ 2-1, \ceo\ 1-0, \ceo\ 2-1, \cts\ 2-1 and \nht. Missing panels were 
not observed.} 
\label{figA1} %% label for entire figure 
\end{figure*} 
\begin{figure*}[!H]
\centering
\subfloat{
\includegraphics[bb=75 375 660 900,width=0.25\linewidth,clip]{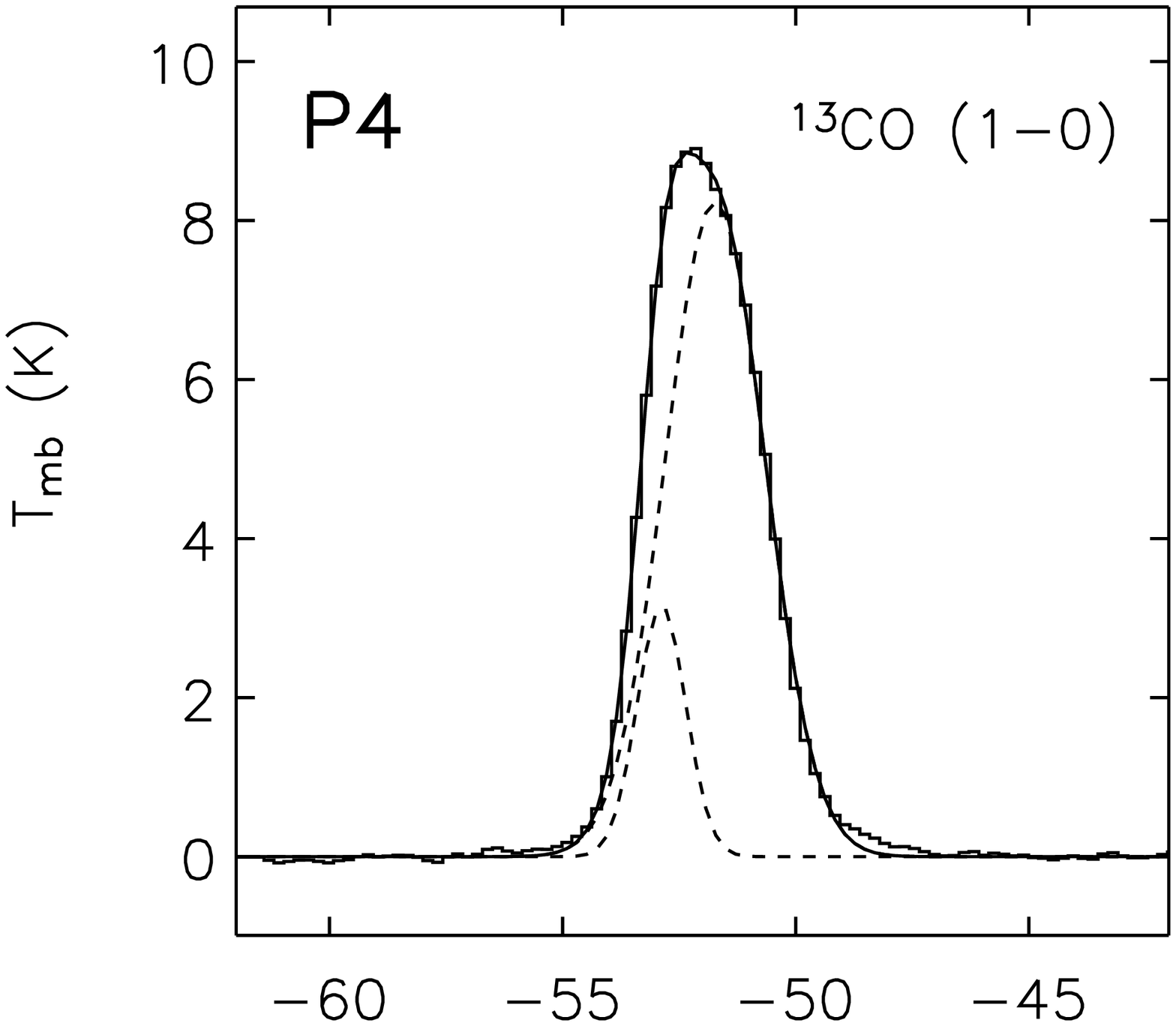}}
\hspace{-0.04\linewidth}
\subfloat{
\includegraphics[bb=75 375 660 900,width=0.25\linewidth,clip]{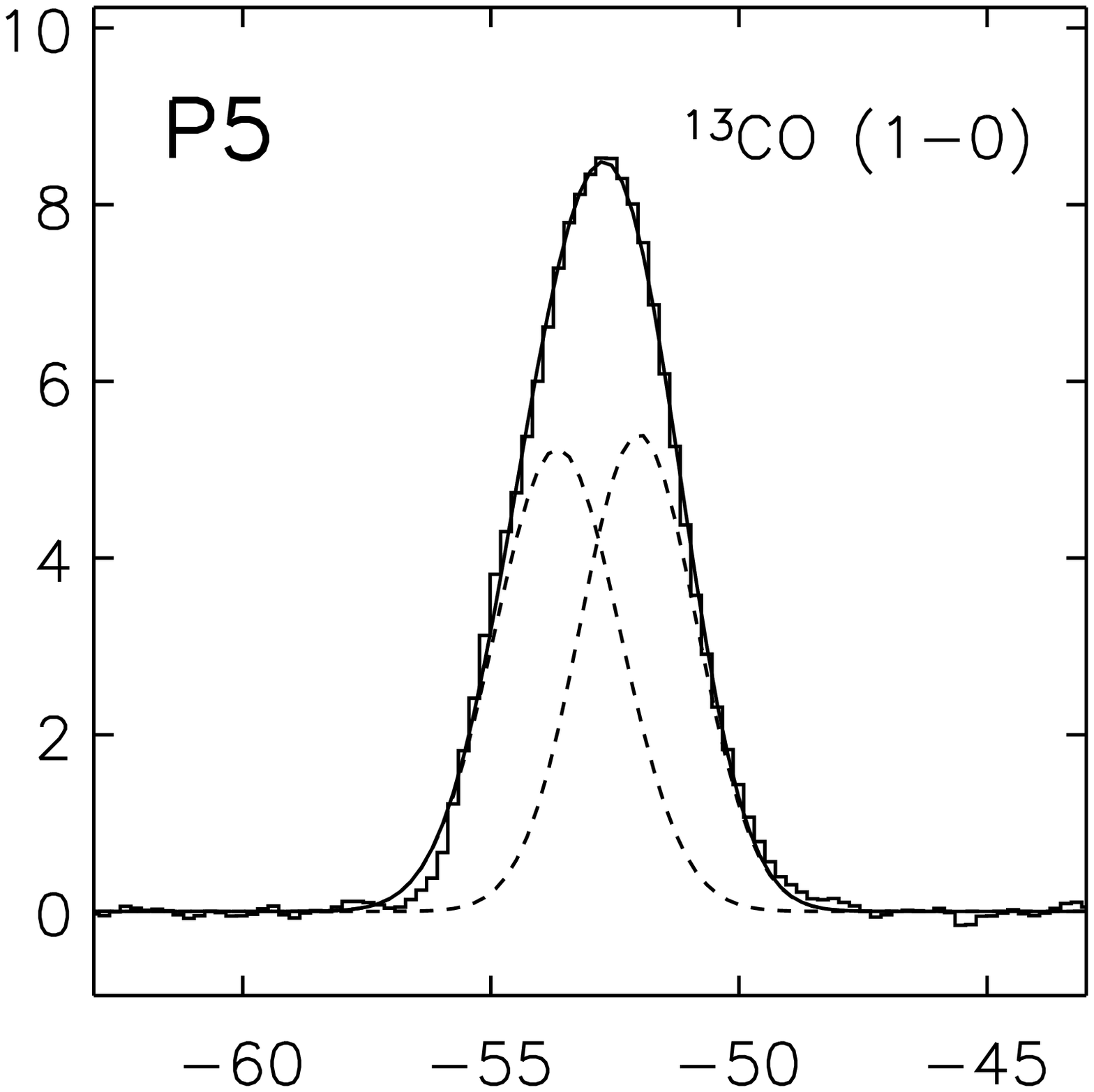}}
\hspace{-0.04\linewidth}
\subfloat{
\includegraphics[bb=75 375 660 900,width=0.25\linewidth,clip]{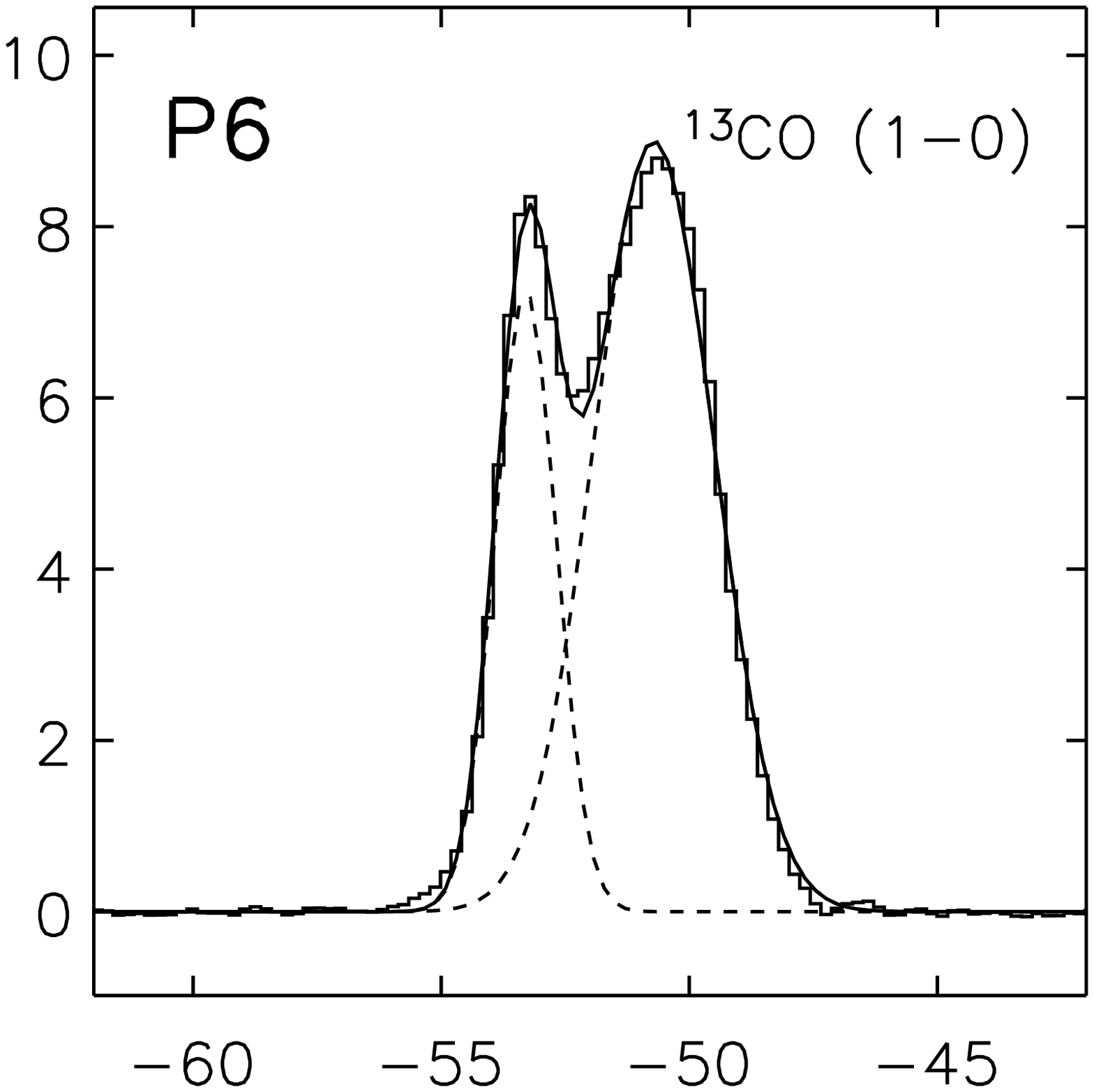}}
\hspace{-0.04\linewidth}
\subfloat{
\includegraphics[bb=75 375 660 900,width=0.25\linewidth,clip]{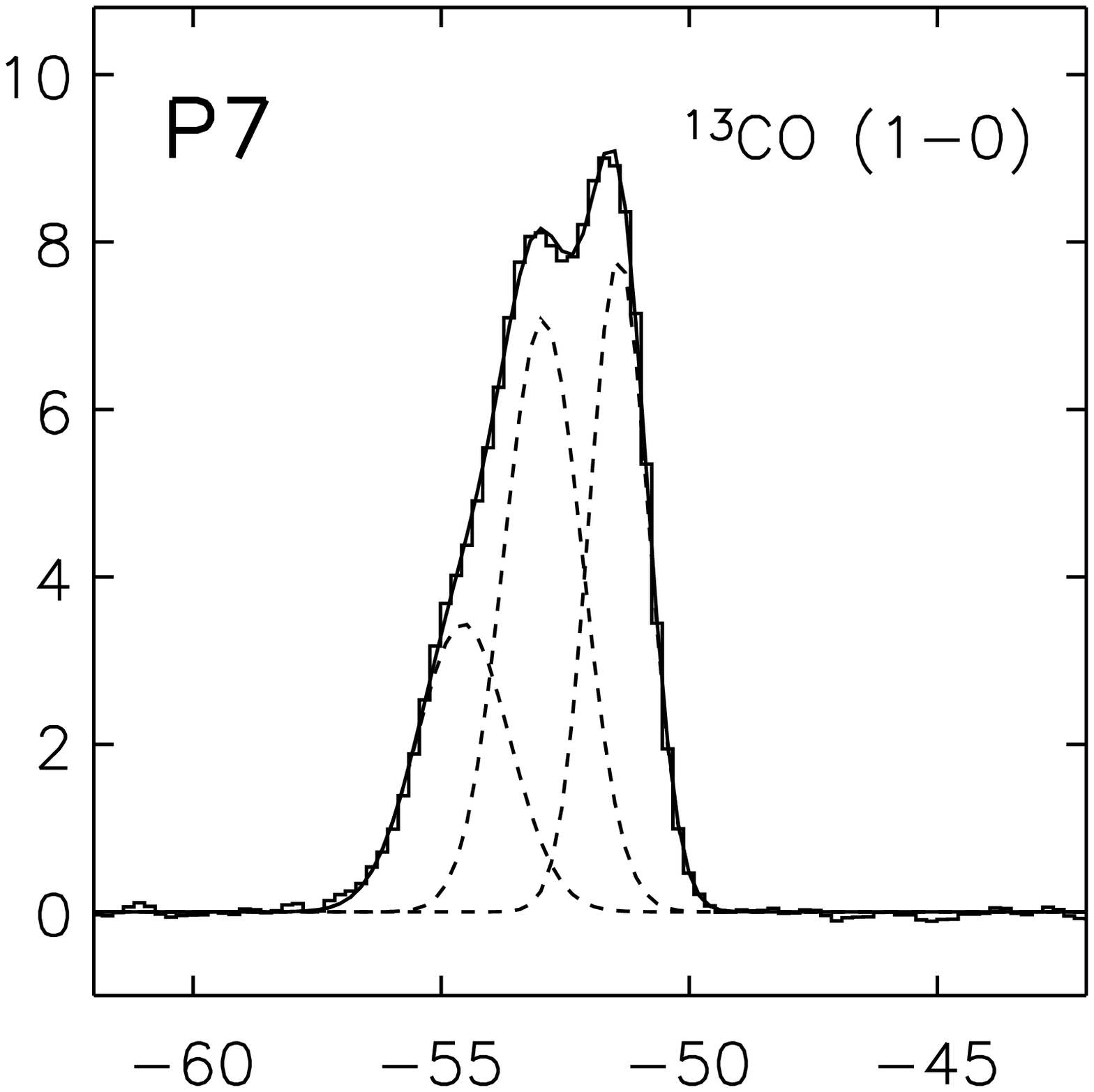}}
\\[-10pt]
\subfloat{
\includegraphics[bb=75 375 660 900,width=0.25\linewidth,clip]{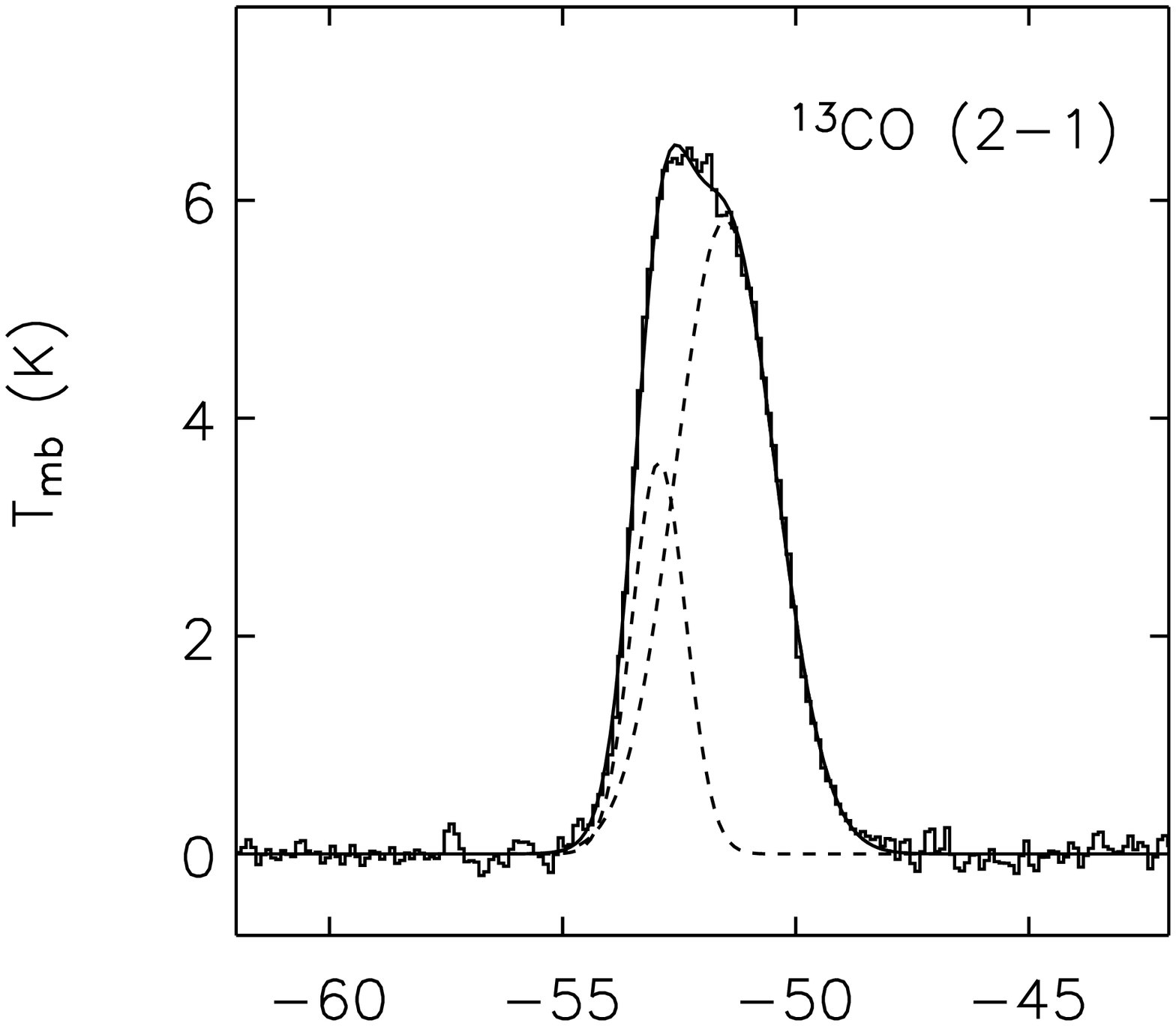}}
\hspace{-0.04\linewidth}
\subfloat{
\includegraphics[bb=75 375 660 900,width=0.25\linewidth,clip]{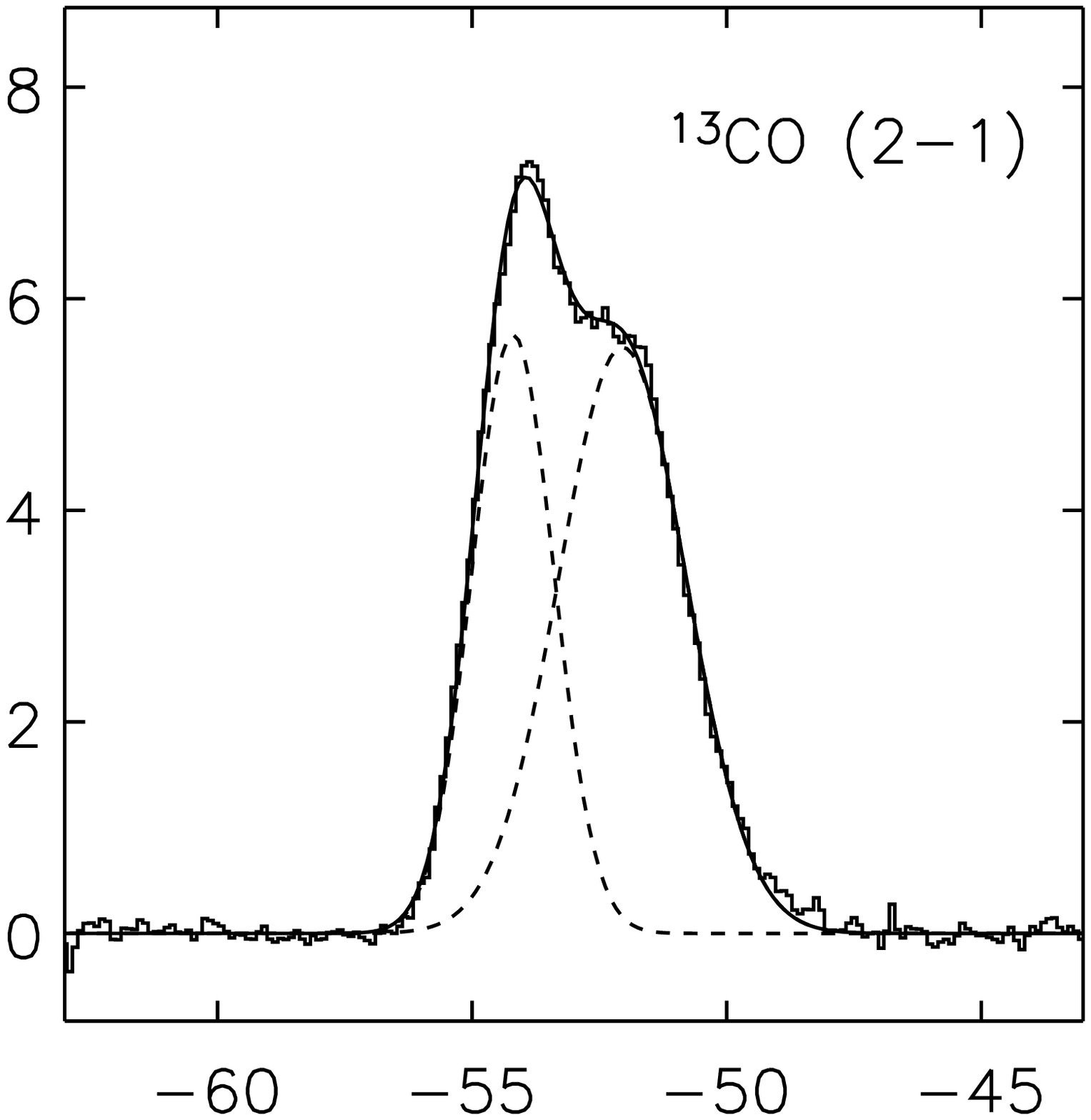}}
\hspace{-0.04\linewidth}
\subfloat{
\includegraphics[bb=75 375 660 900,width=0.25\linewidth,clip]{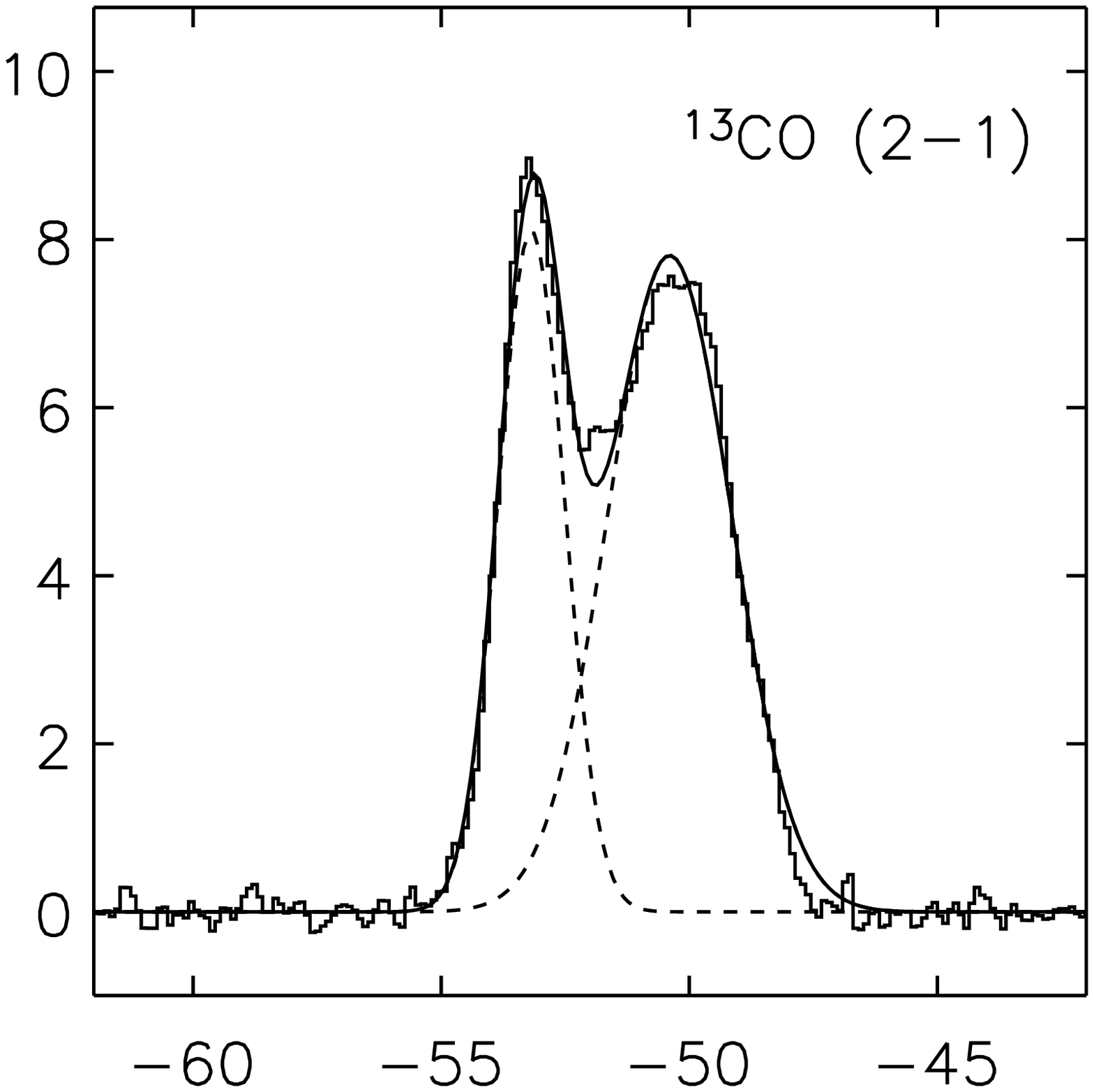}}
\hspace{-0.04\linewidth}
\subfloat{
\includegraphics[bb=75 375 660 900,width=0.25\linewidth,clip]{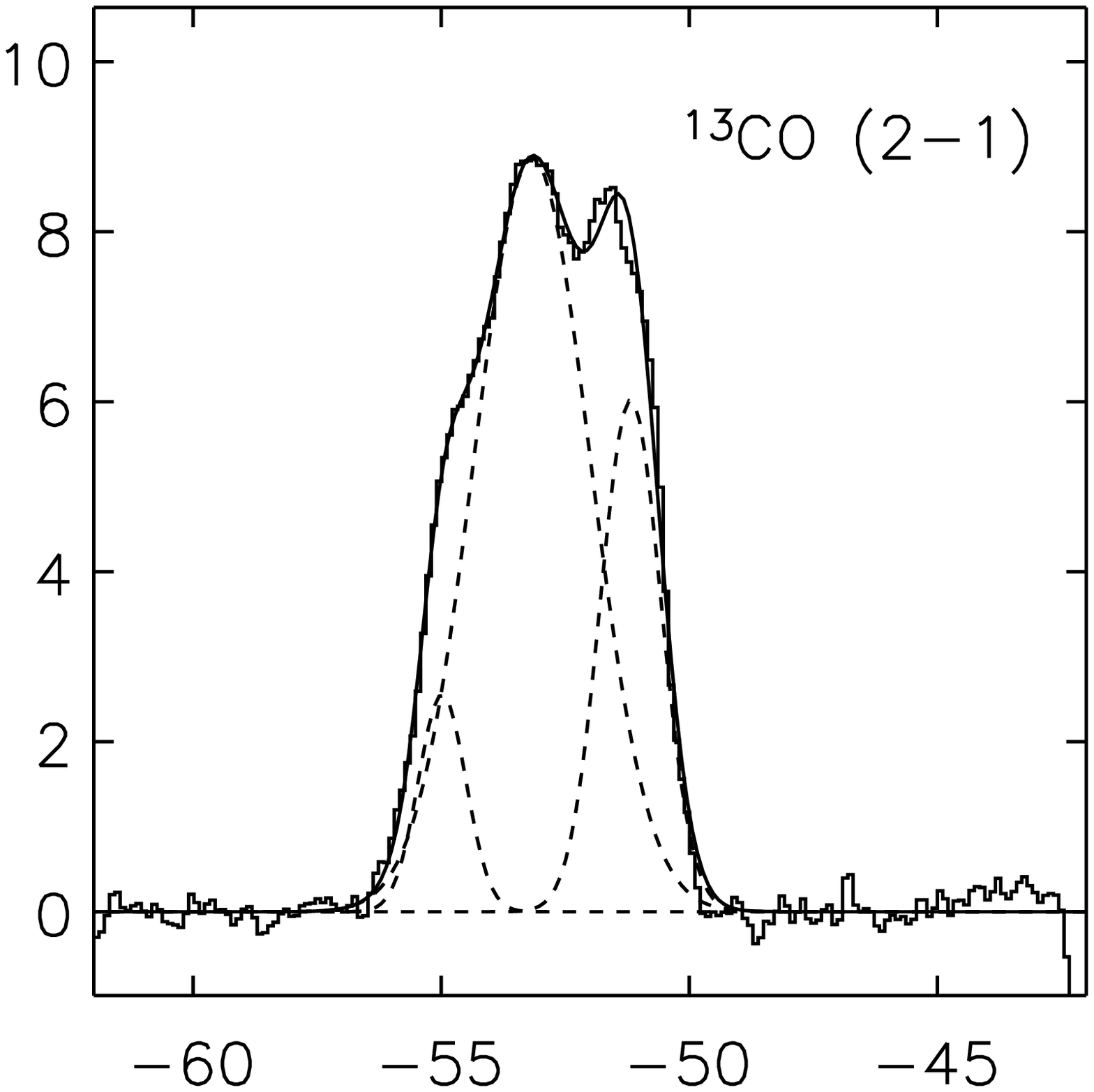}}
\\[-10pt]
\subfloat{
\includegraphics[bb=75 375 660 900,width=0.25\linewidth,clip]{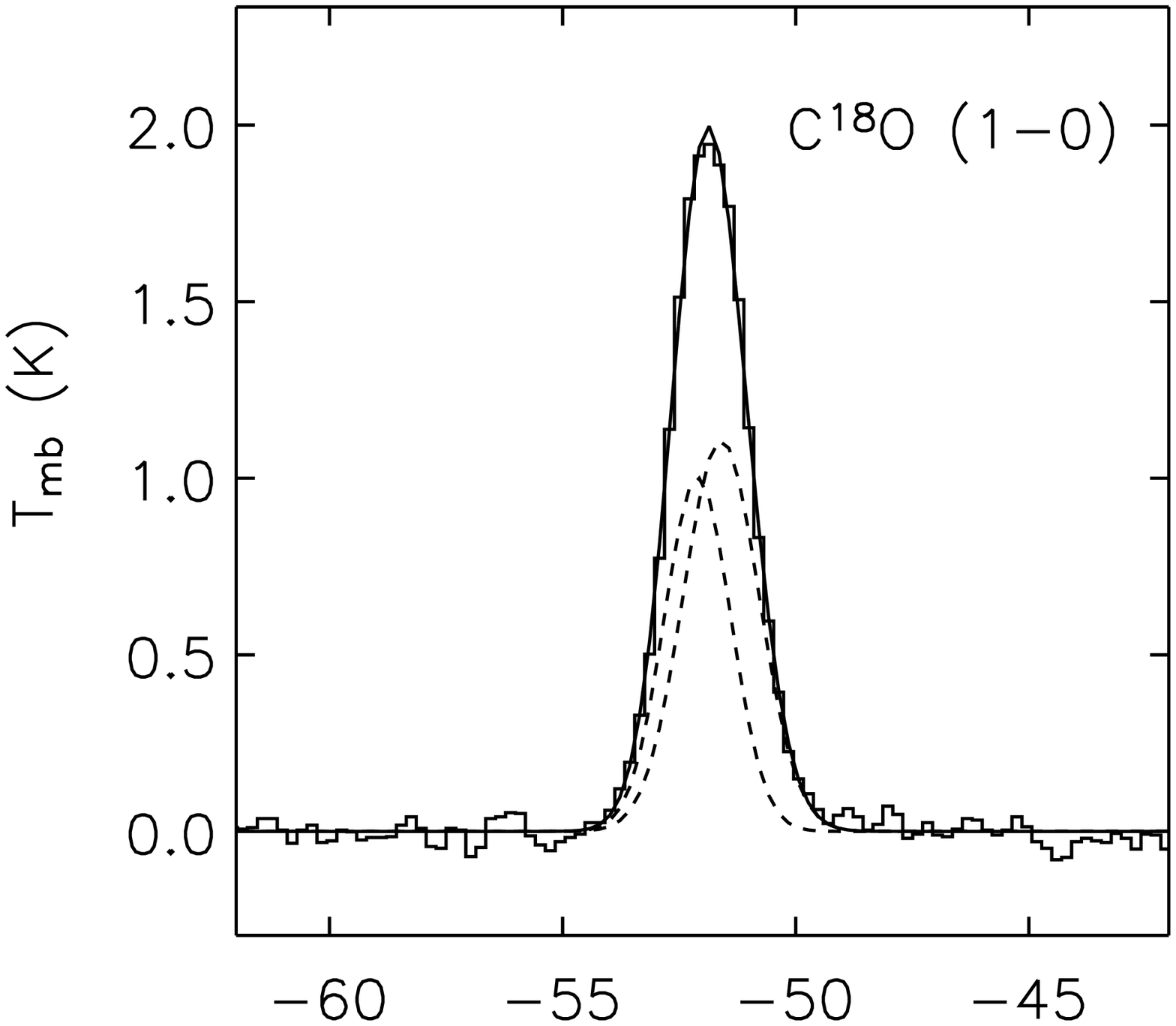}}
\hspace{-0.04\linewidth}
\subfloat{
\includegraphics[bb=75 375 660 900,width=0.25\linewidth,clip]{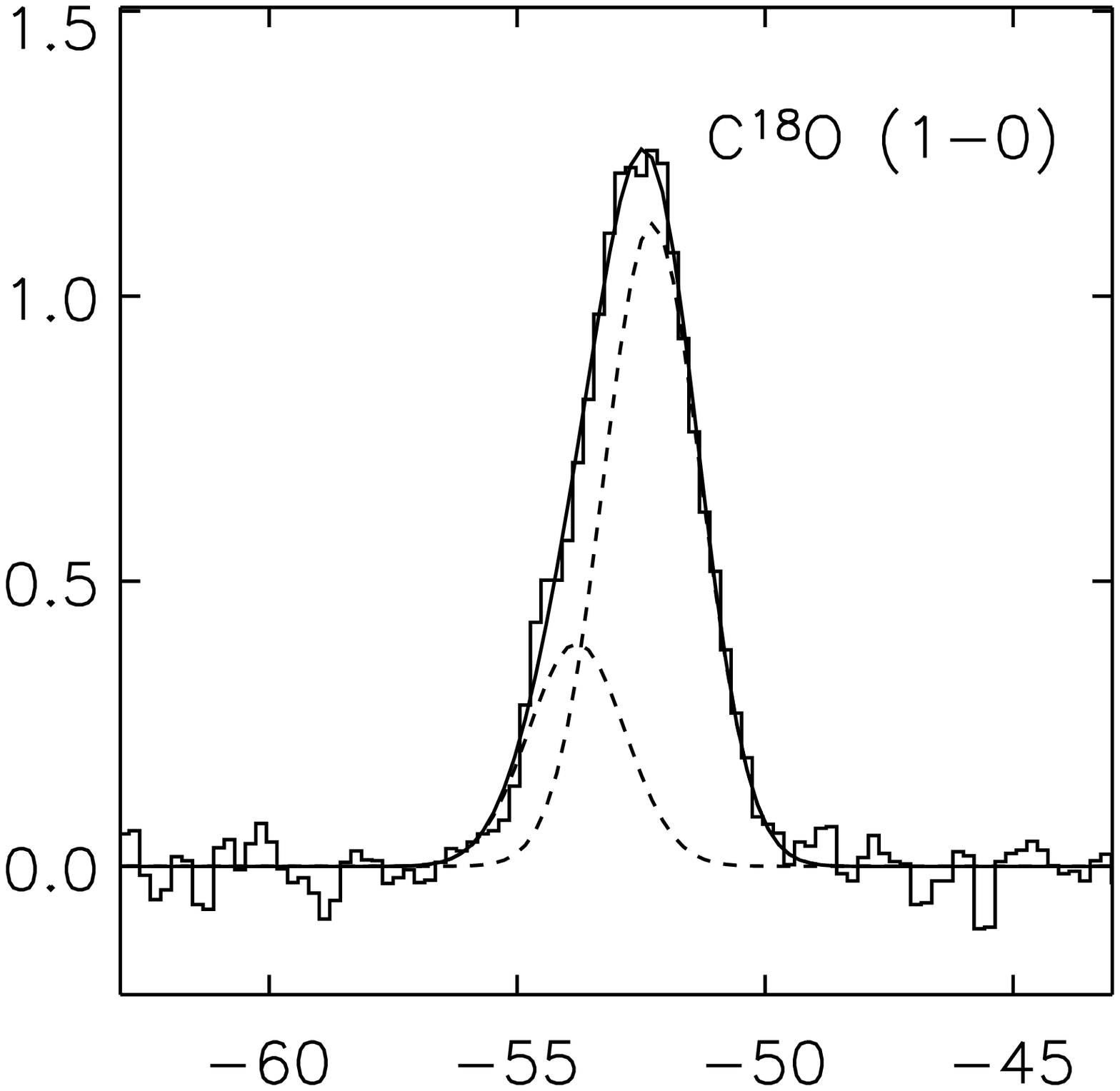}}
\hspace{-0.04\linewidth}
\subfloat{
\includegraphics[bb=75 375 660 900,width=0.25\linewidth,clip]{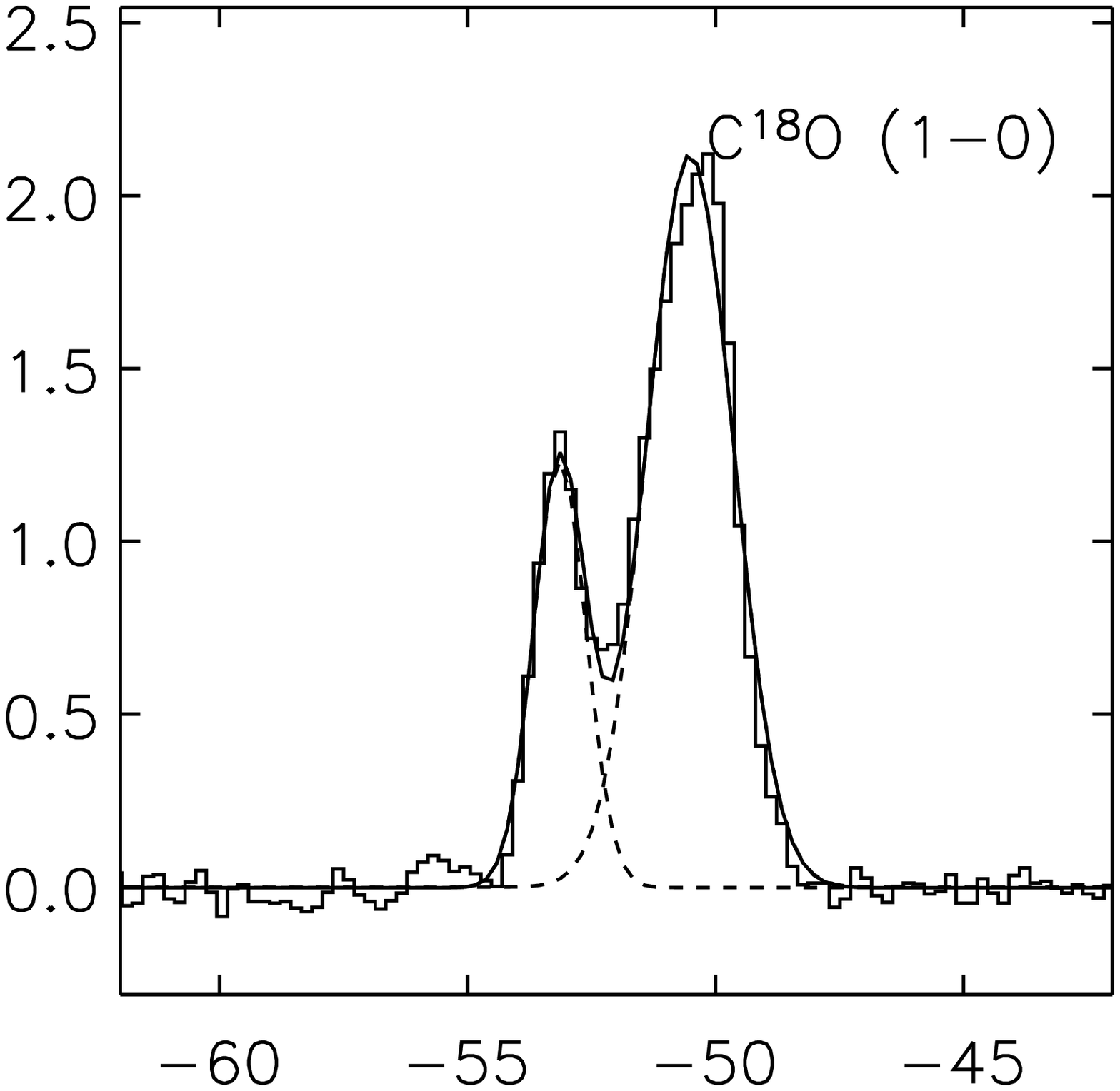}}
\hspace{-0.04\linewidth}
\subfloat{
\includegraphics[bb=75 375 660 900,width=0.25\linewidth,clip]{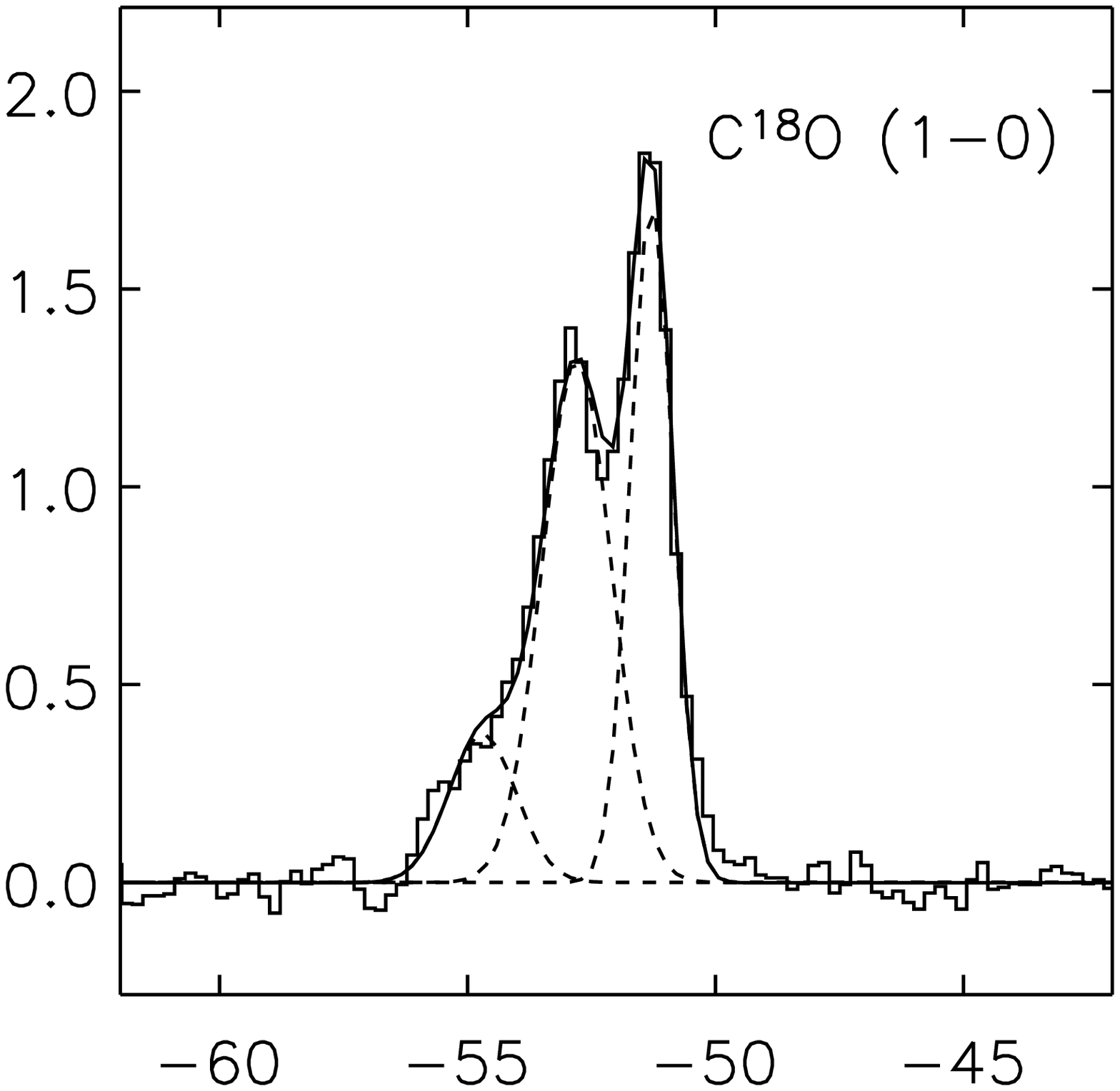}}
\\[-10pt]
\subfloat{
\includegraphics[bb=75 375 660 900,width=0.25\linewidth,clip]{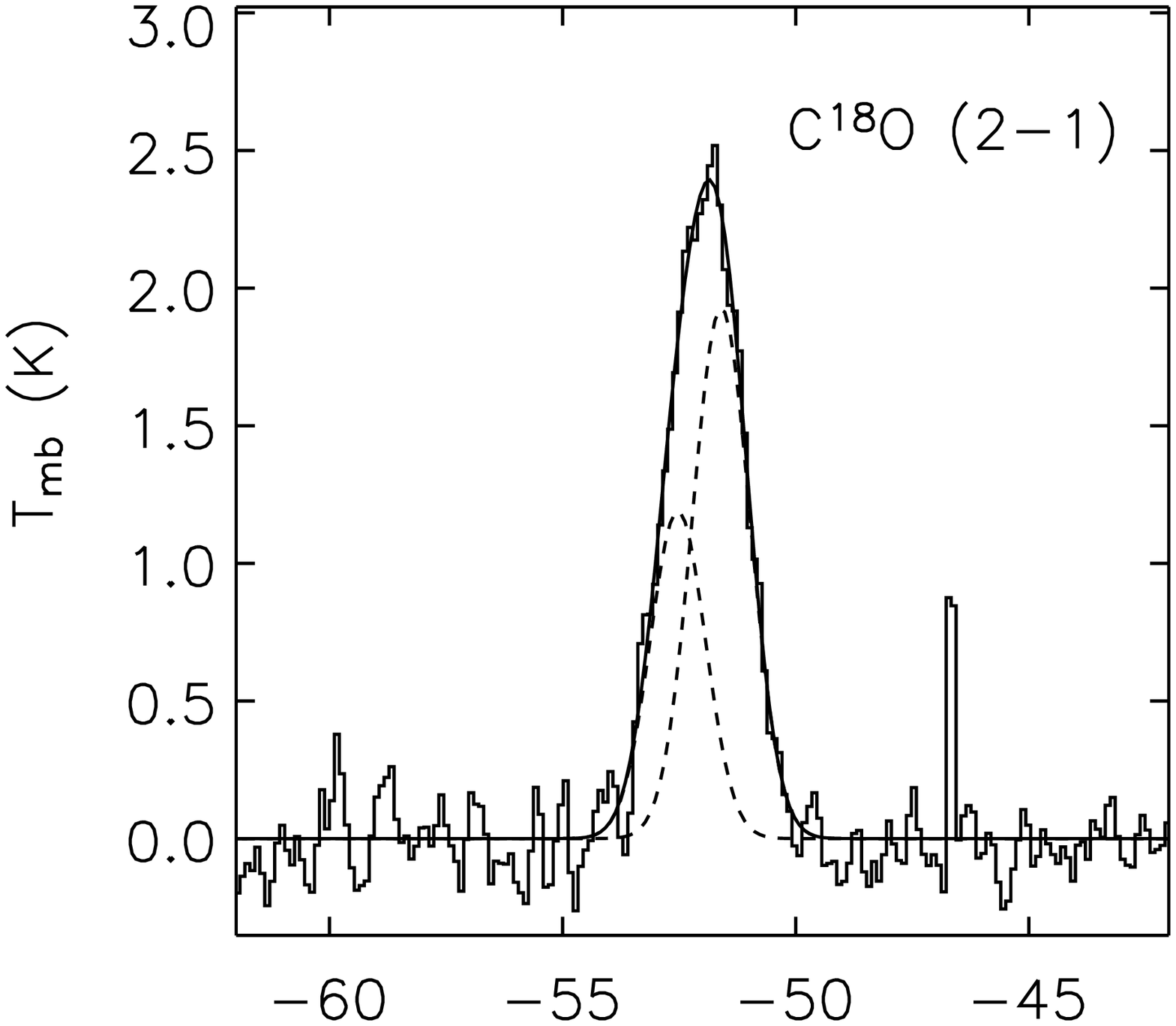}}
\hspace{-0.04\linewidth}
\subfloat{
\includegraphics[bb=75 375 660 900,width=0.25\linewidth,clip]{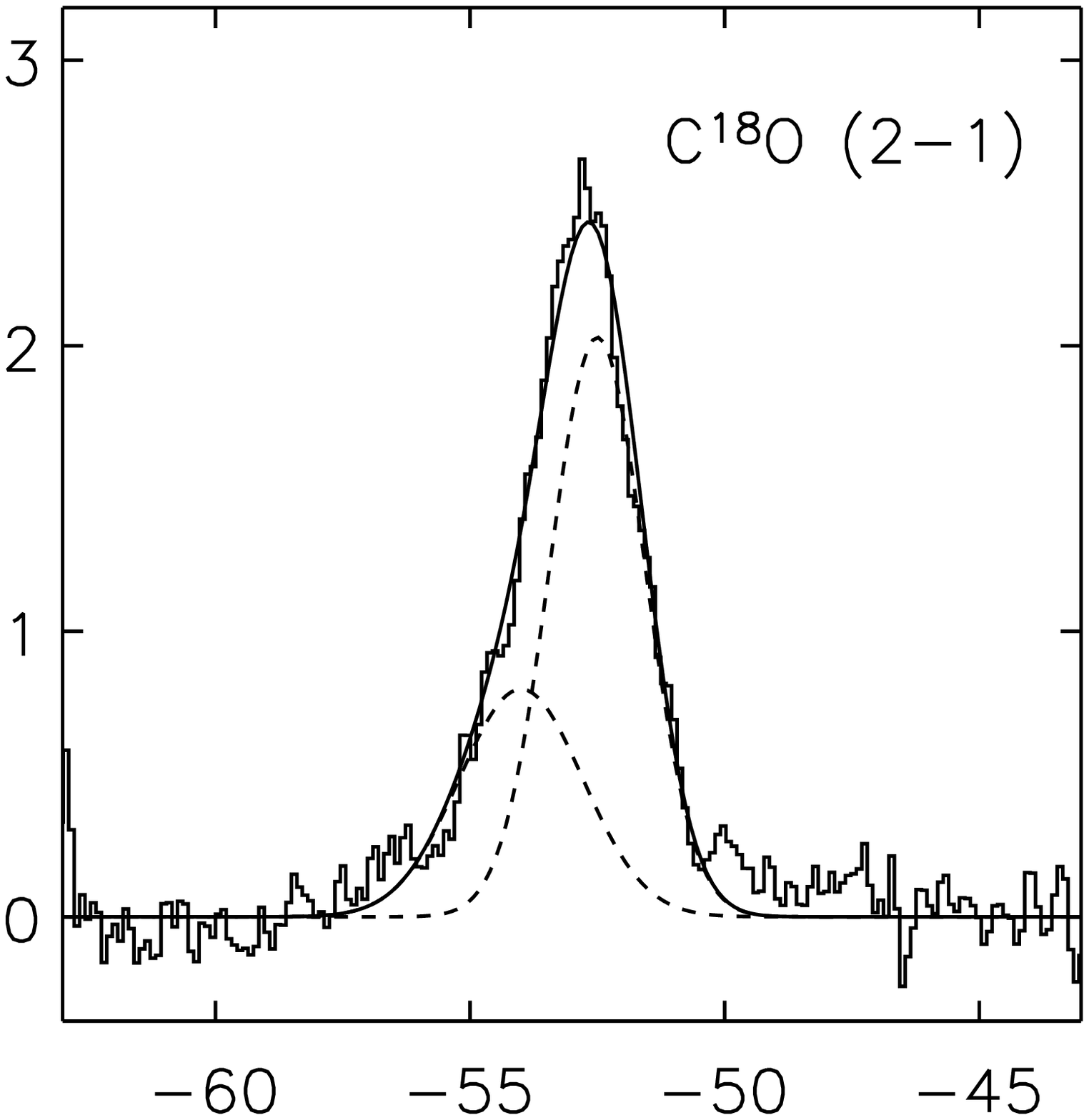}}
\hspace{-0.04\linewidth}
\subfloat{
\includegraphics[bb=75 375 660 900,width=0.25\linewidth,clip]{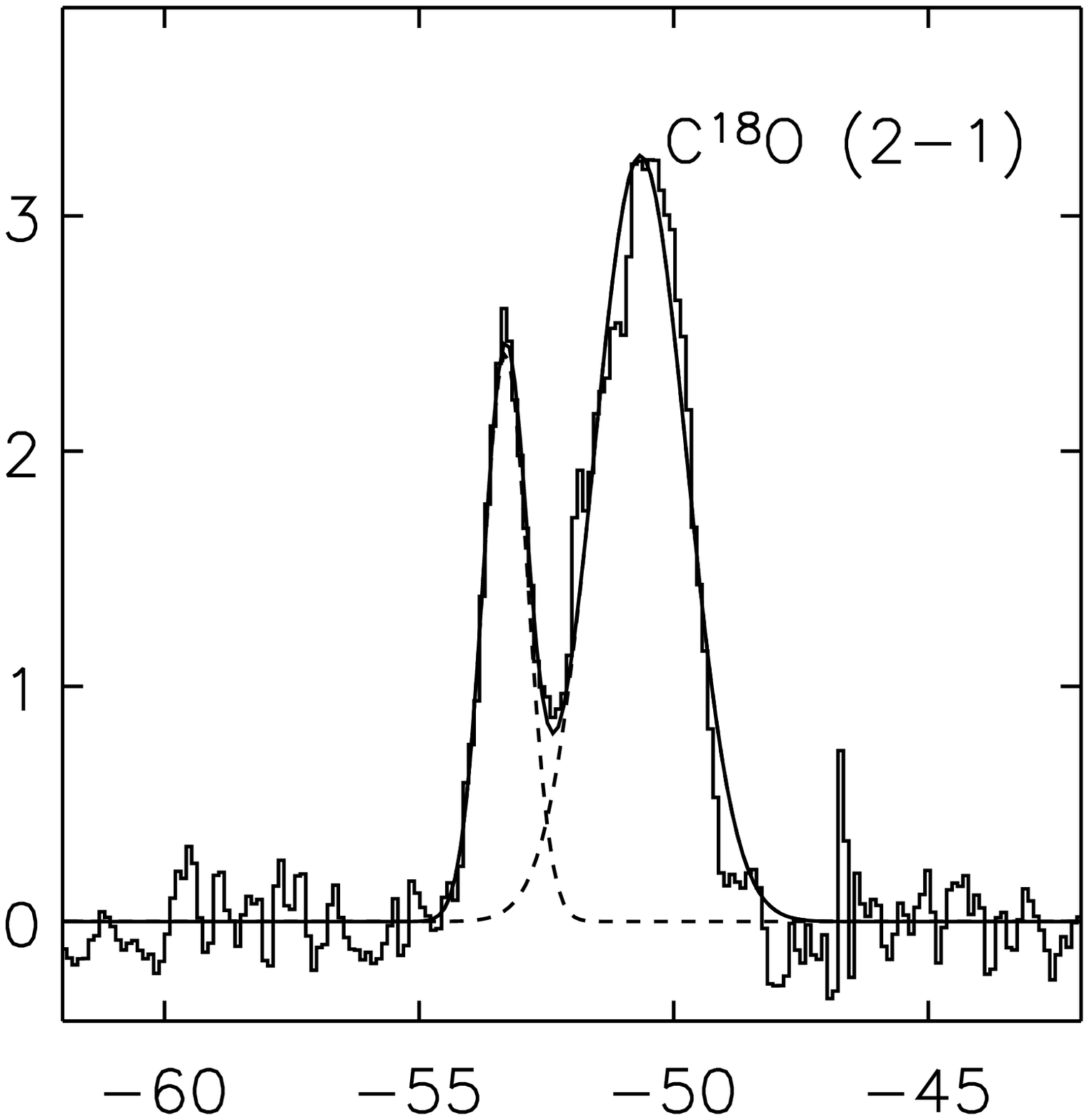}}
\hspace{-0.04\linewidth}
\subfloat{
\includegraphics[bb=75 375 660 900,width=0.25\linewidth,clip]{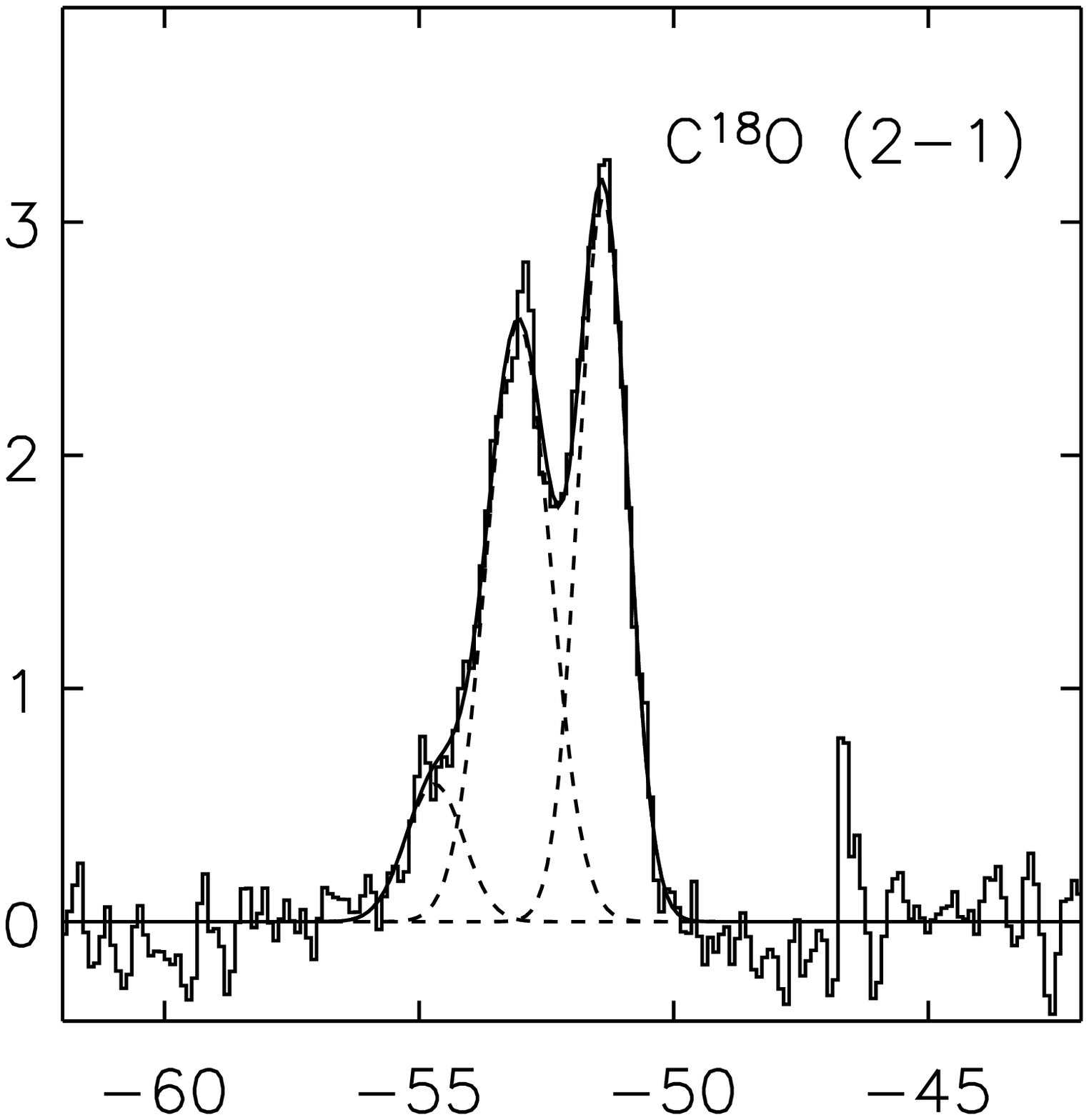}}
\\[-10pt]
\subfloat{
\includegraphics[bb=75 375 660 900,width=0.25\linewidth,clip]{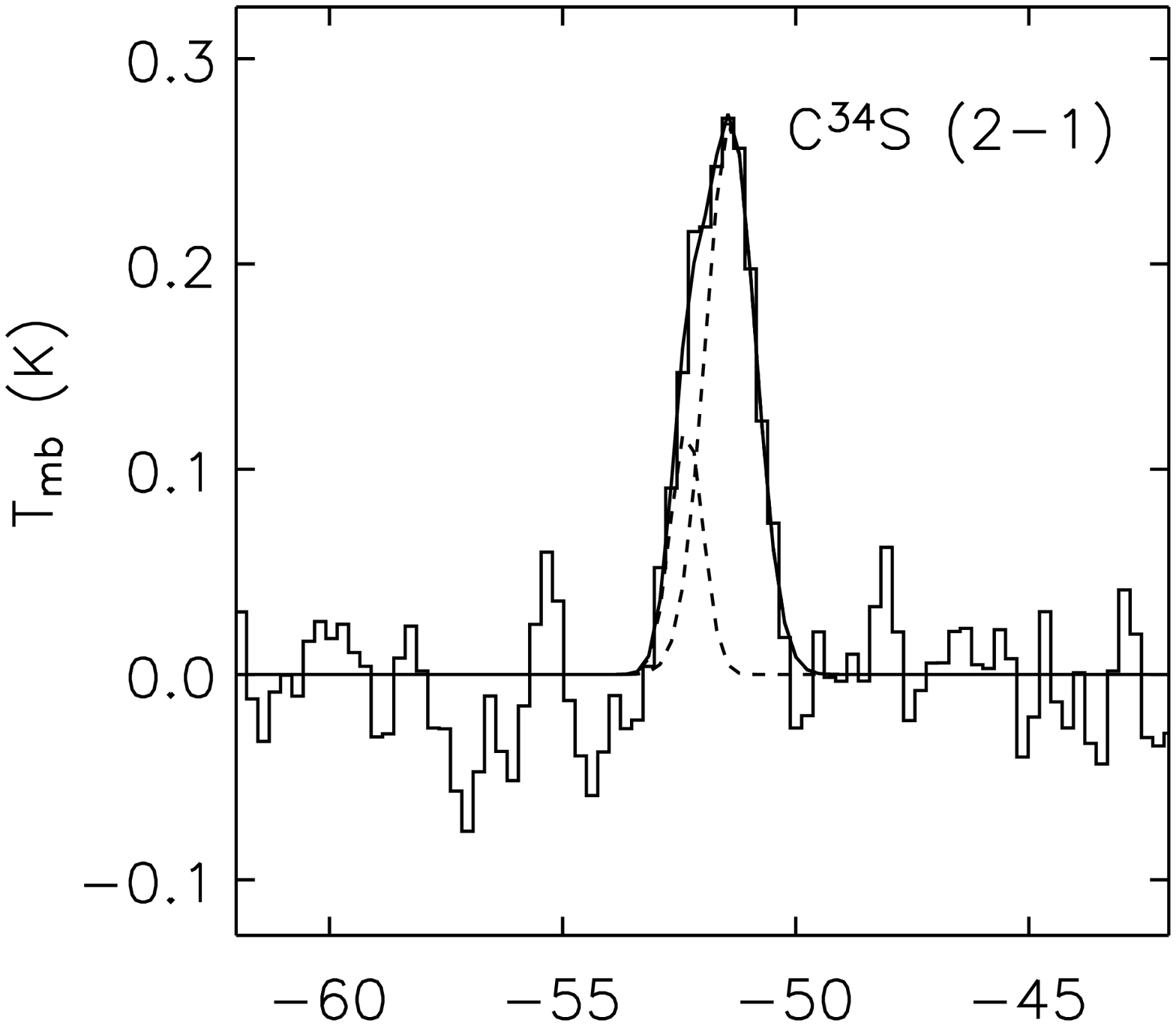}}
\hspace{-0.04\linewidth}
\subfloat{
\includegraphics[bb=75 375 660 900,width=0.25\linewidth,clip]{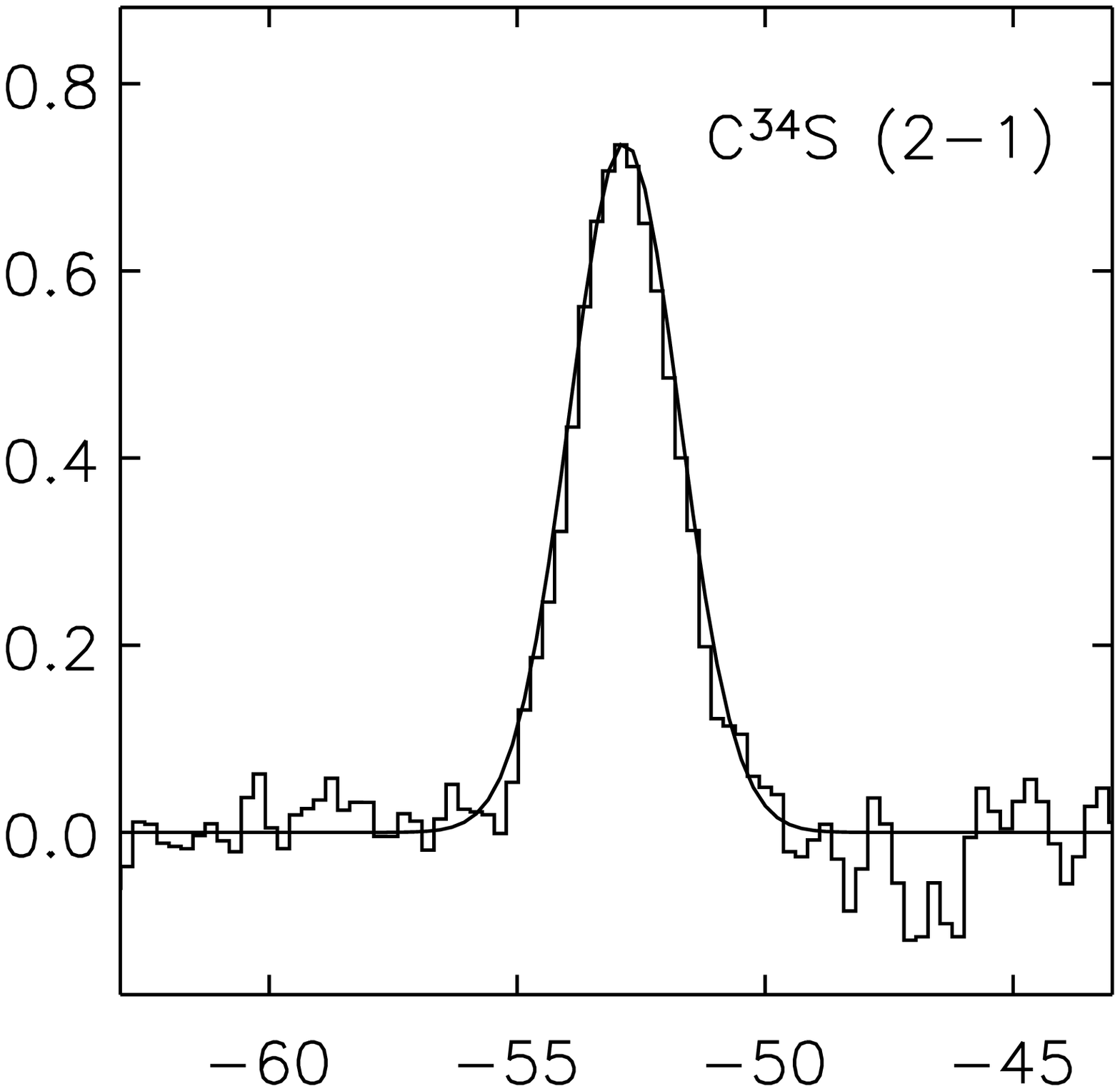}}
\hspace{-0.04\linewidth}
\subfloat{
\includegraphics[bb=75 375 660 900,width=0.25\linewidth,clip]{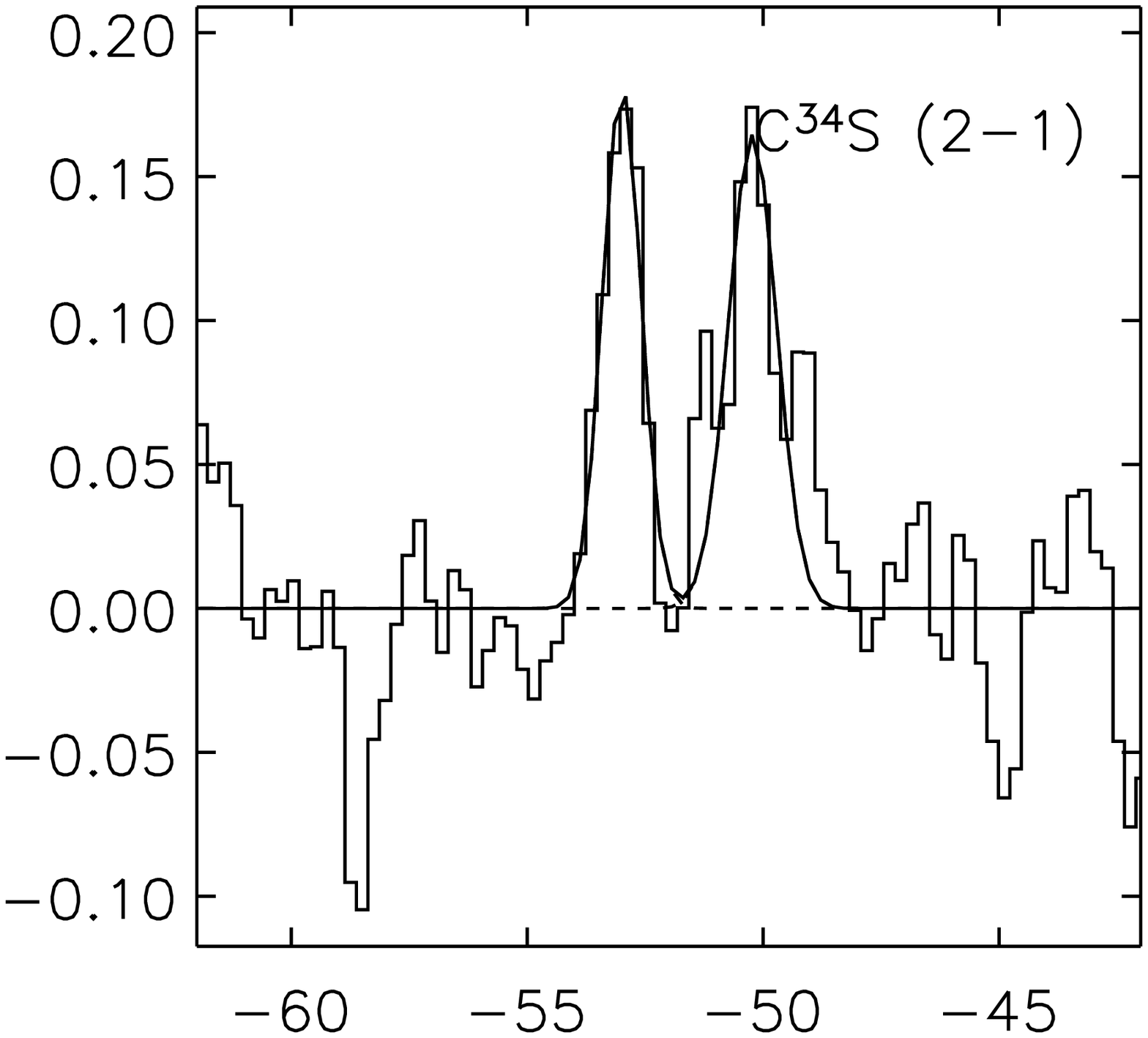}}
\hspace{-0.04\linewidth}
\subfloat{
\includegraphics[bb=75 375 660 900,width=0.25\linewidth,clip]{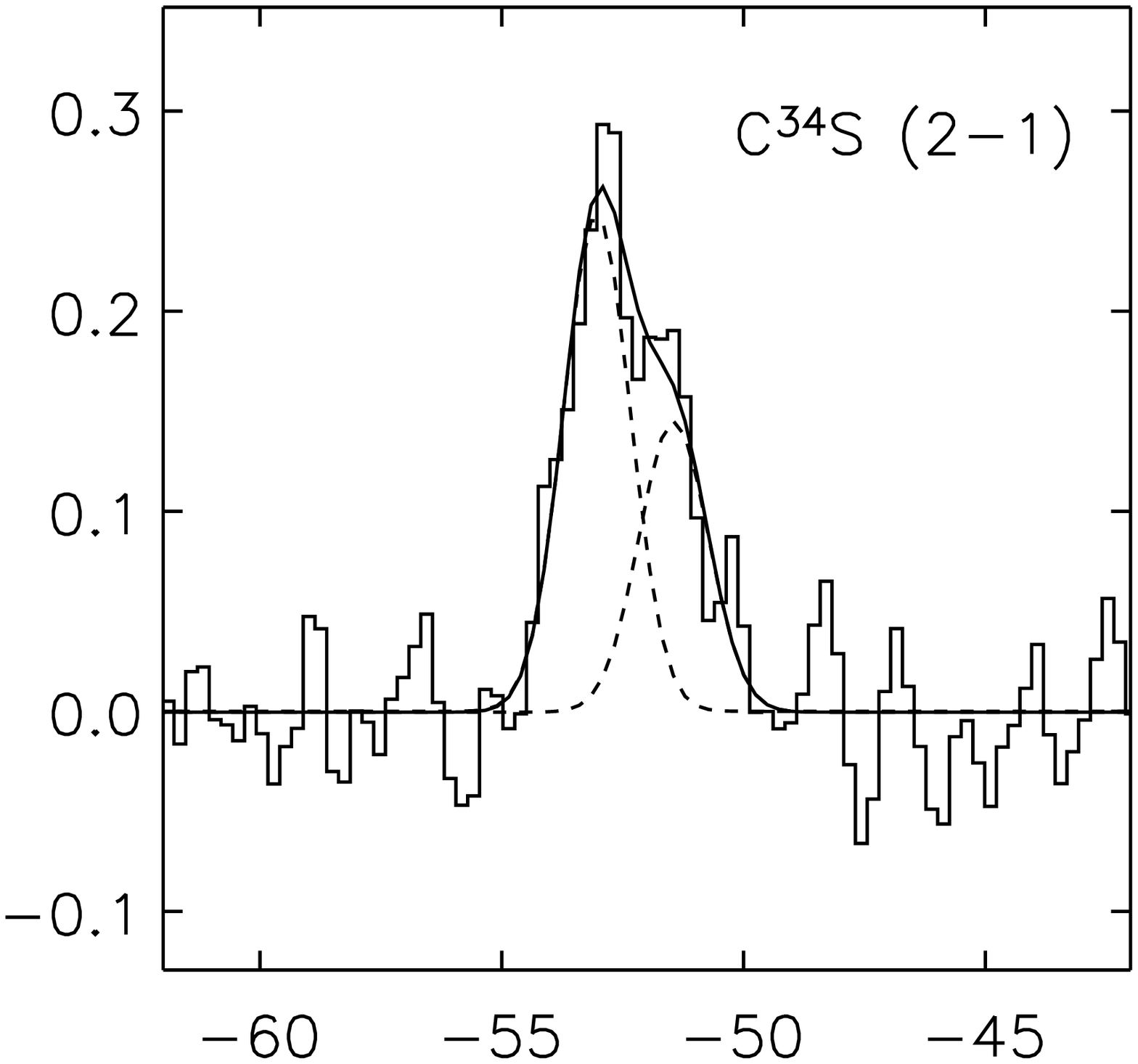}}
\\[-10pt]
\subfloat{
\includegraphics[bb=75 375 660 900,width=0.25\linewidth,clip]{dummy.eps}}
\hspace{-0.04\linewidth}
\subfloat{
\includegraphics[bb=75 375 660 900,width=0.25\linewidth,clip]{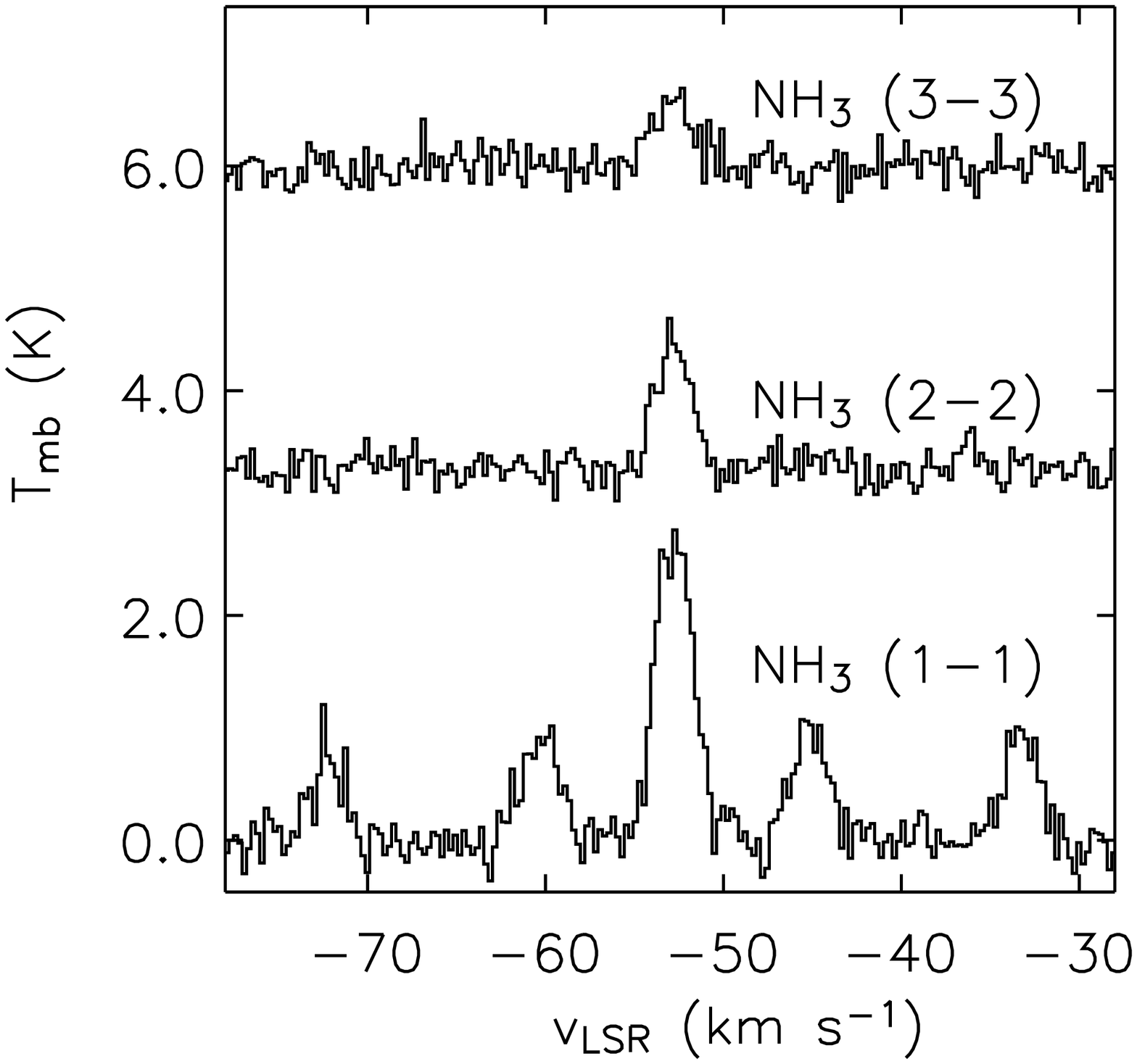}}
\hspace{-0.04\linewidth}
\subfloat{
\includegraphics[bb=75 375 660 900,width=0.25\linewidth,clip]{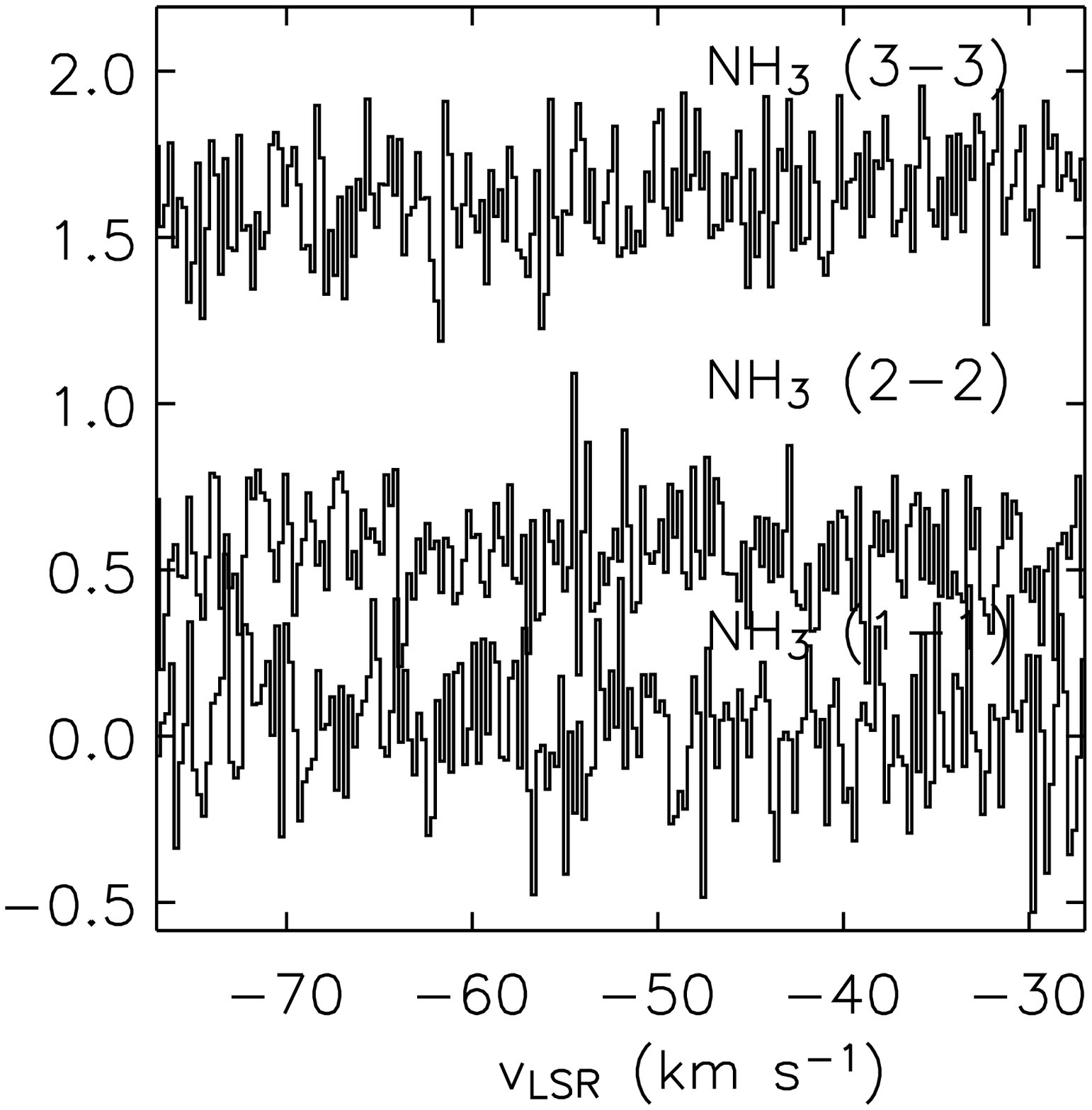}}
\hspace{-0.04\linewidth}
\subfloat{
\includegraphics[bb=75 375 660 900,width=0.25\linewidth,clip]{dummy.eps}}
\caption{Same as Figure \ref{figA1}.} 
\label{figA2} %% label for entire figure 
\end{figure*} 
\begin{figure*}[!H]
\centering
\subfloat{
\includegraphics[bb=75 375 660 900,width=0.25\linewidth,clip]{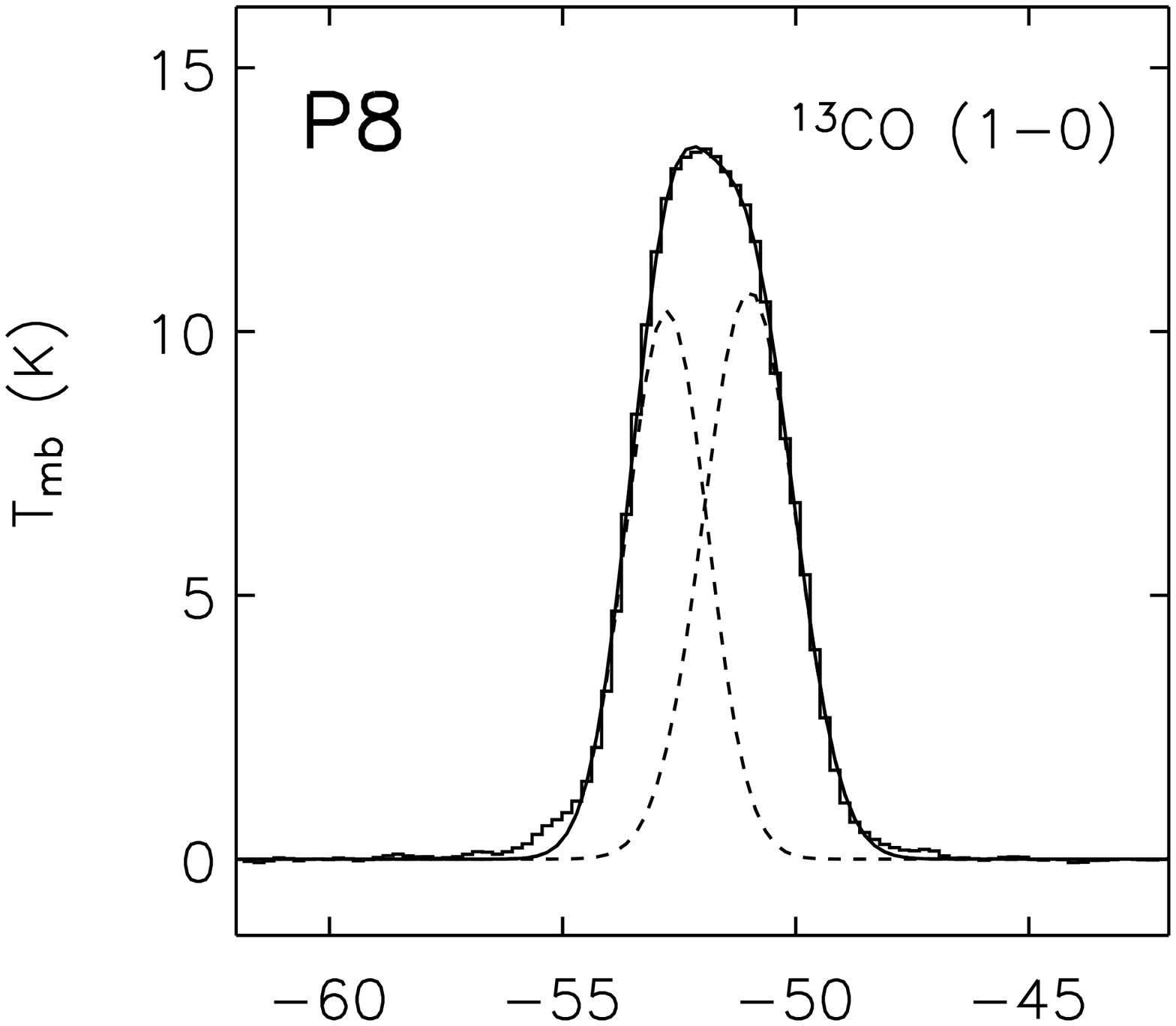}}
\hspace{-0.04\linewidth}
\subfloat{
\includegraphics[bb=75 375 660 900,width=0.25\linewidth,clip]{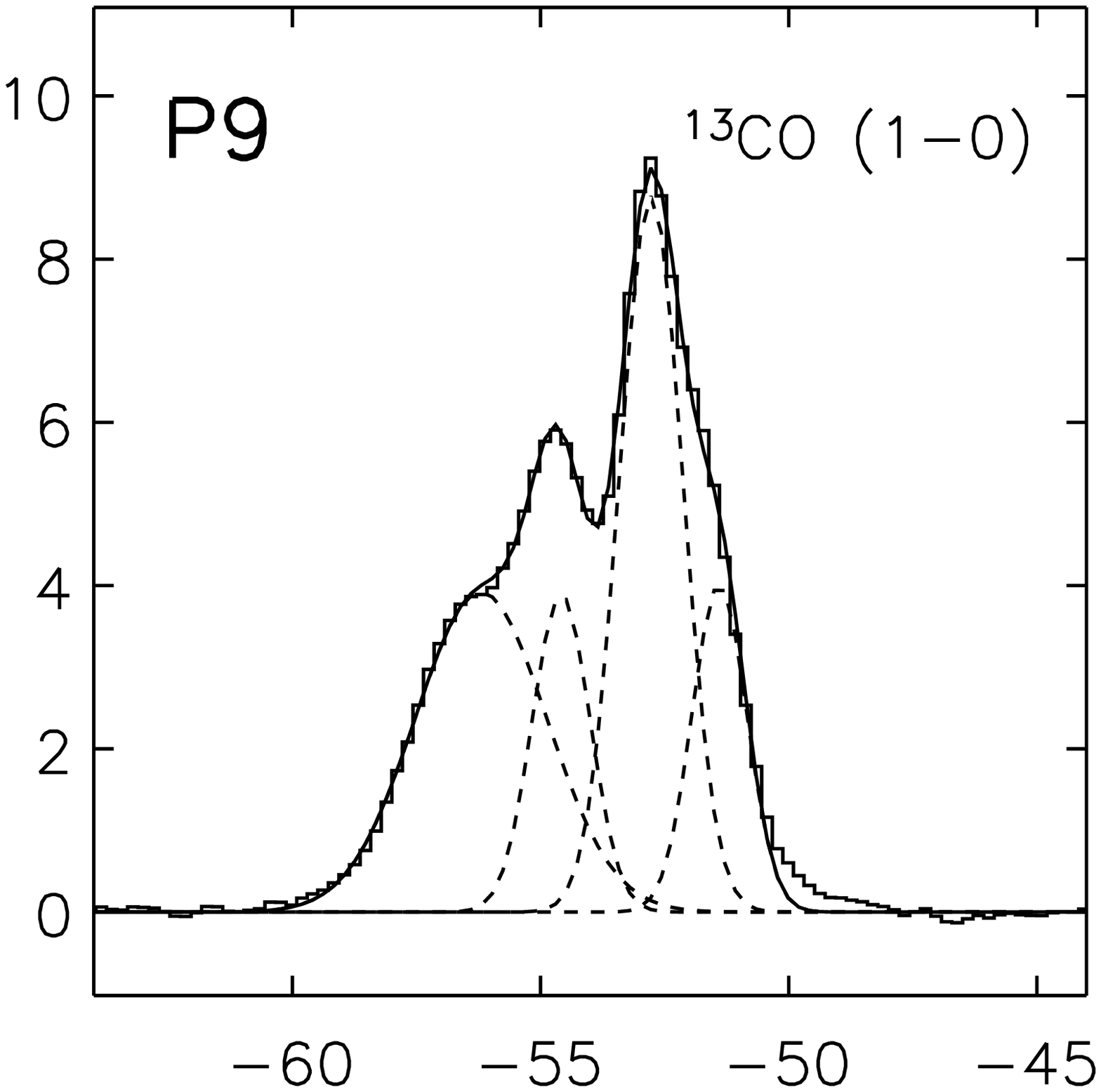}}
\hspace{-0.04\linewidth}
\subfloat{
\includegraphics[bb=75 375 660 900,width=0.25\linewidth,clip]{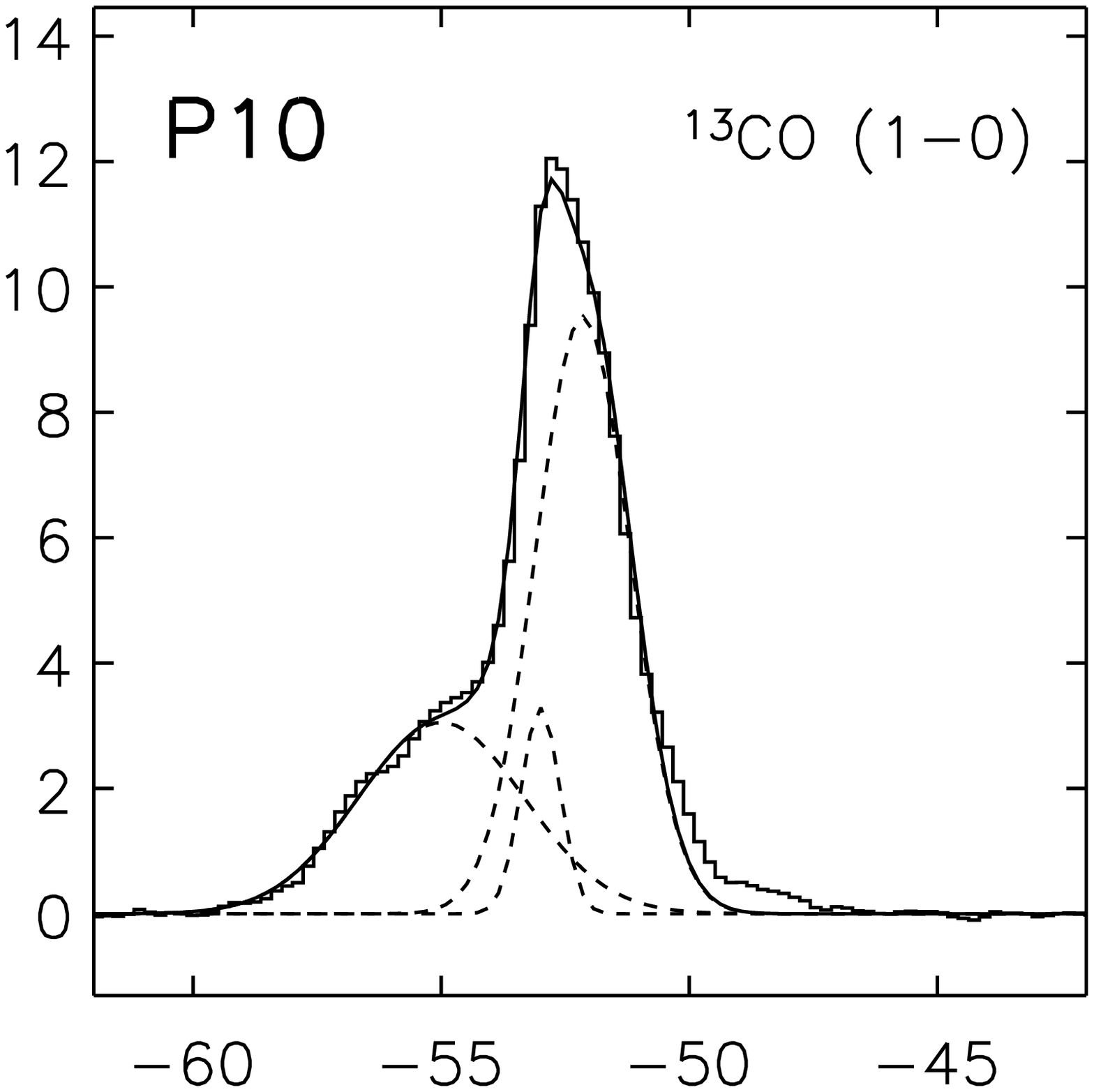}}
\hspace{-0.04\linewidth}
\subfloat{
\includegraphics[bb=75 375 660 900,width=0.25\linewidth,clip]{dummy.eps}}
\\[-10pt]
\subfloat{
\includegraphics[bb=75 375 660 900,width=0.25\linewidth,clip]{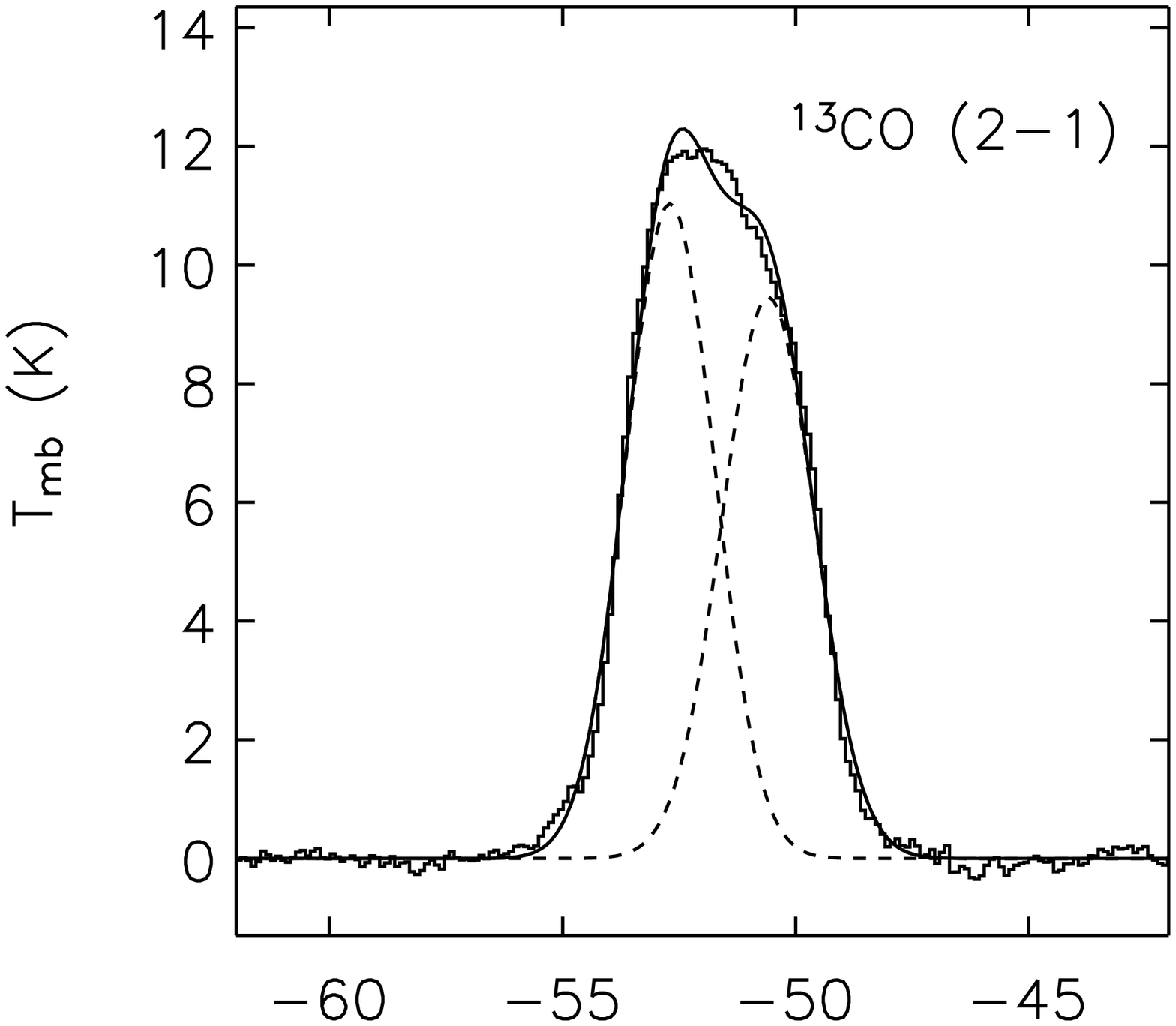}}
\hspace{-0.04\linewidth}
\subfloat{
\includegraphics[bb=75 375 660 900,width=0.25\linewidth,clip]{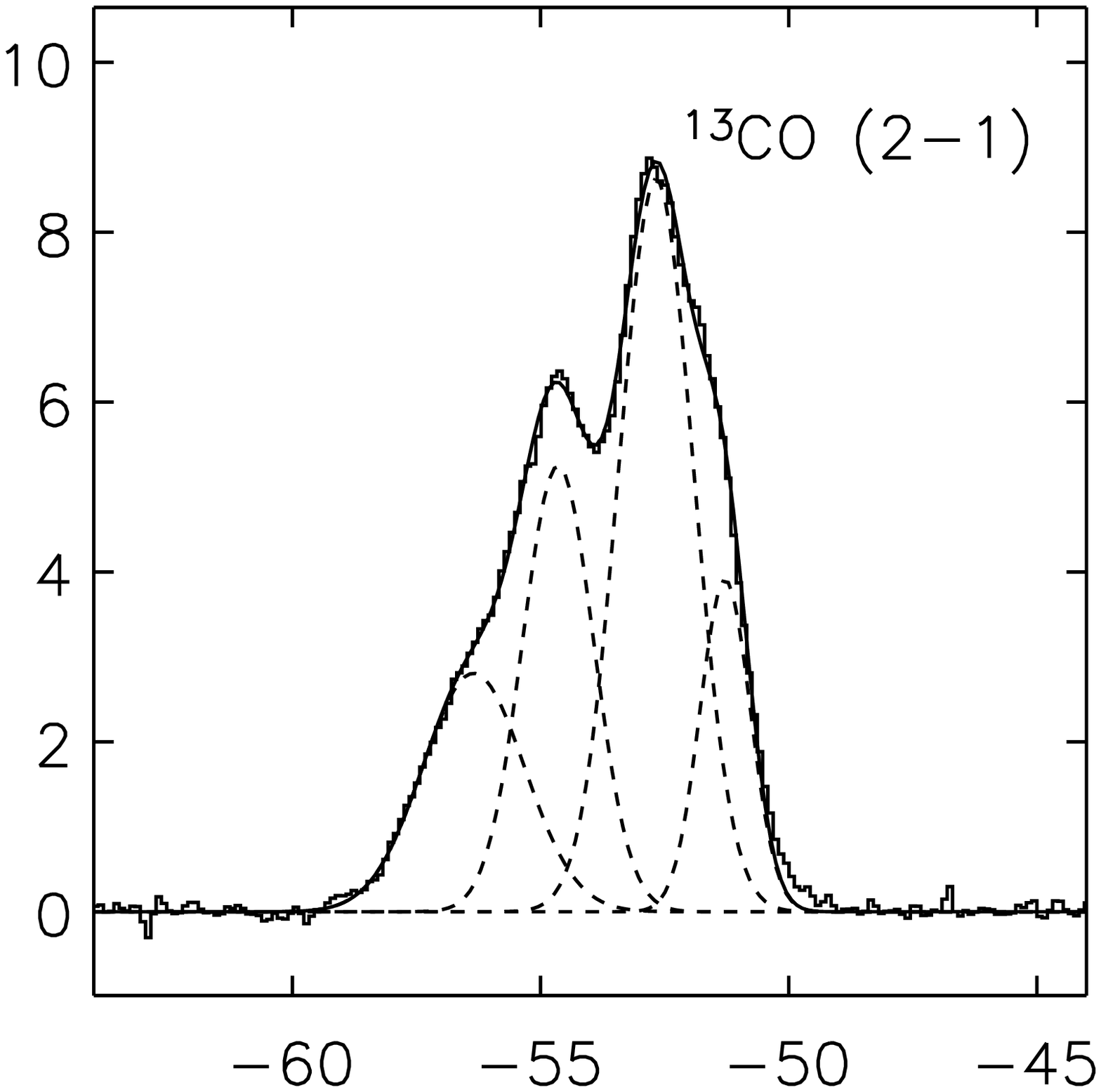}}
\hspace{-0.04\linewidth}
\subfloat{
\includegraphics[bb=75 375 660 900,width=0.25\linewidth,clip]{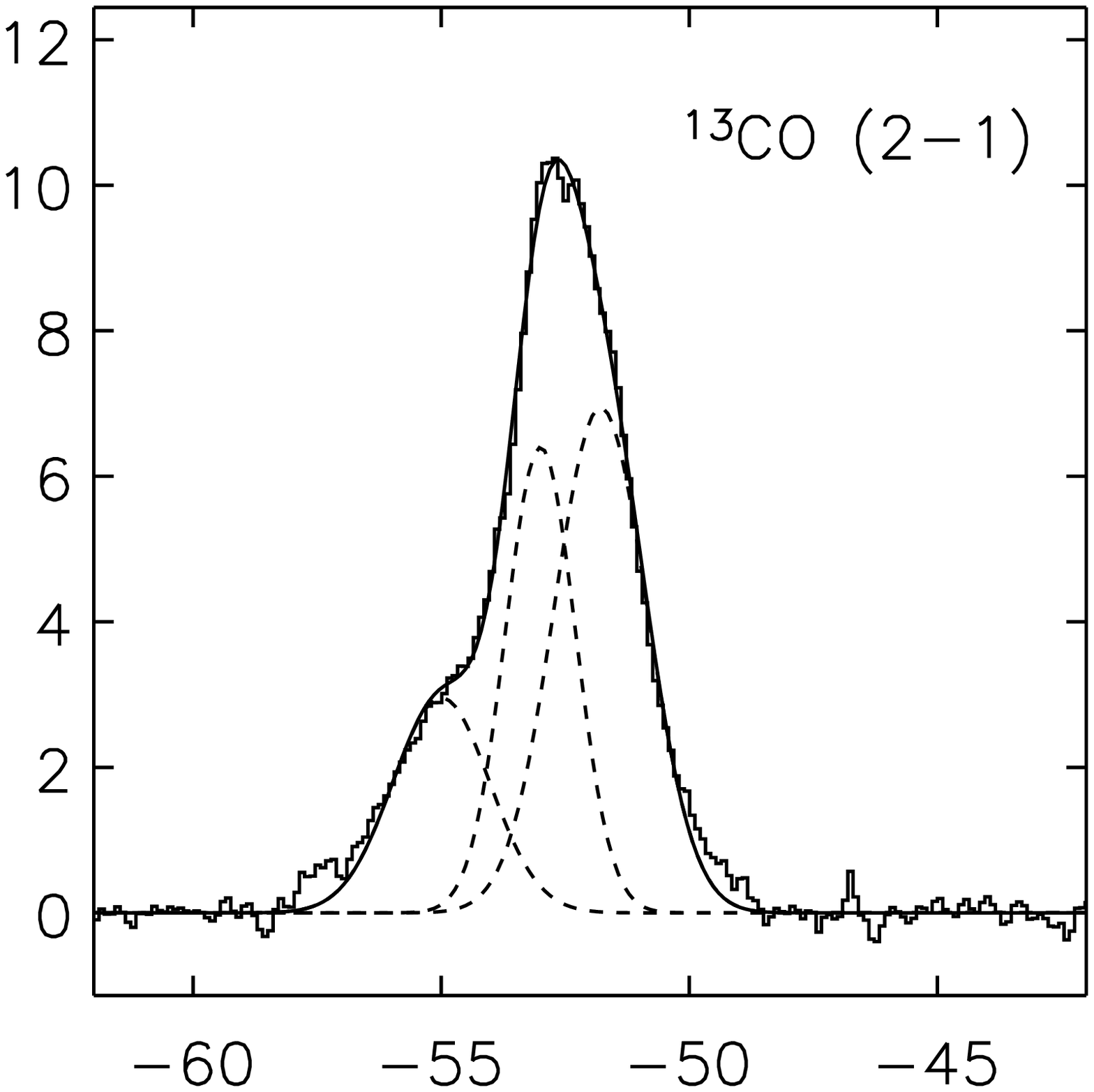}}
\hspace{-0.04\linewidth}
\subfloat{
\includegraphics[bb=75 375 660 900,width=0.25\linewidth,clip]{dummy.eps}}
\\[-10pt]
\subfloat{
\includegraphics[bb=75 375 660 900,width=0.25\linewidth,clip]{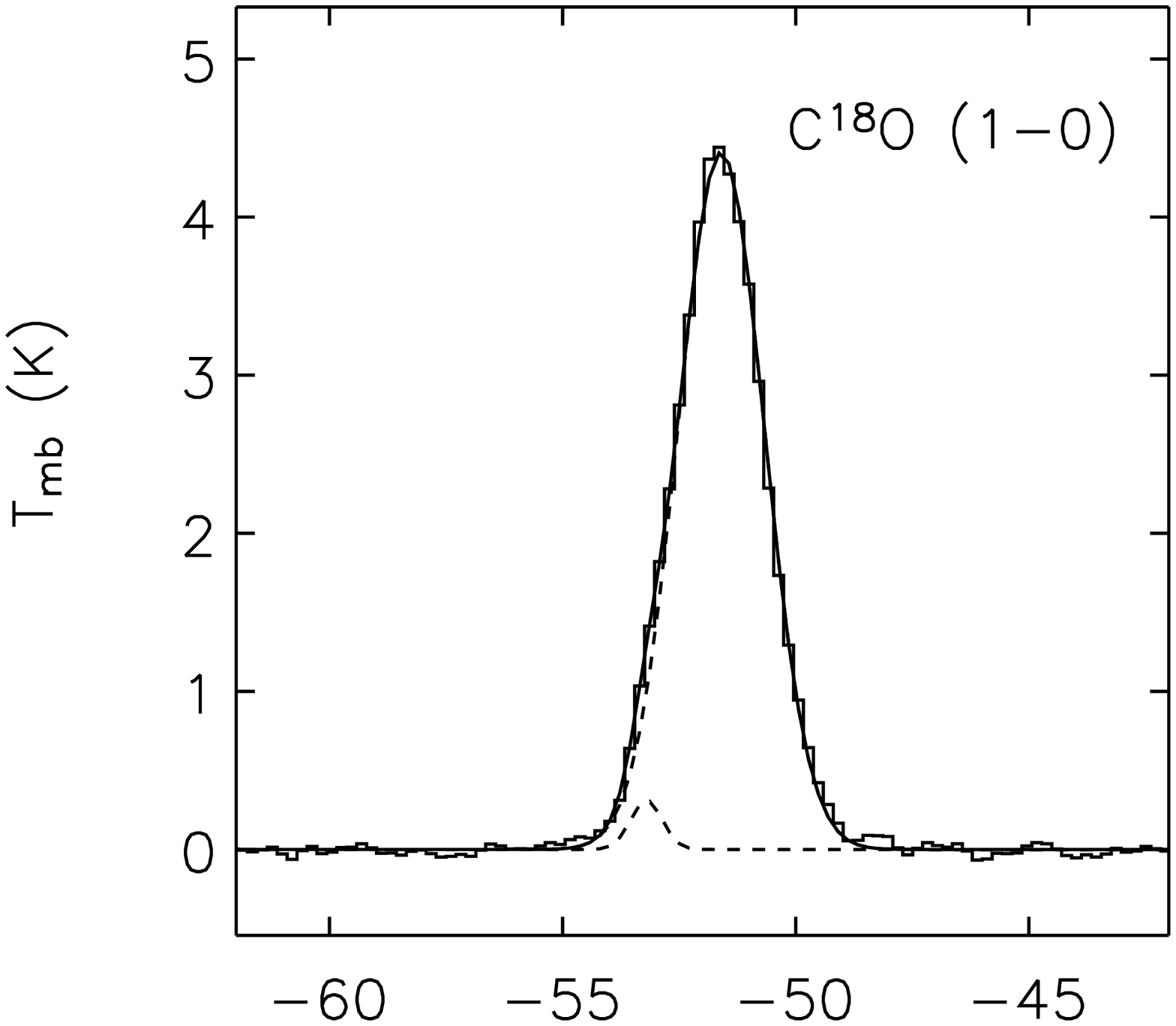}}
\hspace{-0.04\linewidth}
\subfloat{
\includegraphics[bb=75 375 660 900,width=0.25\linewidth,clip]{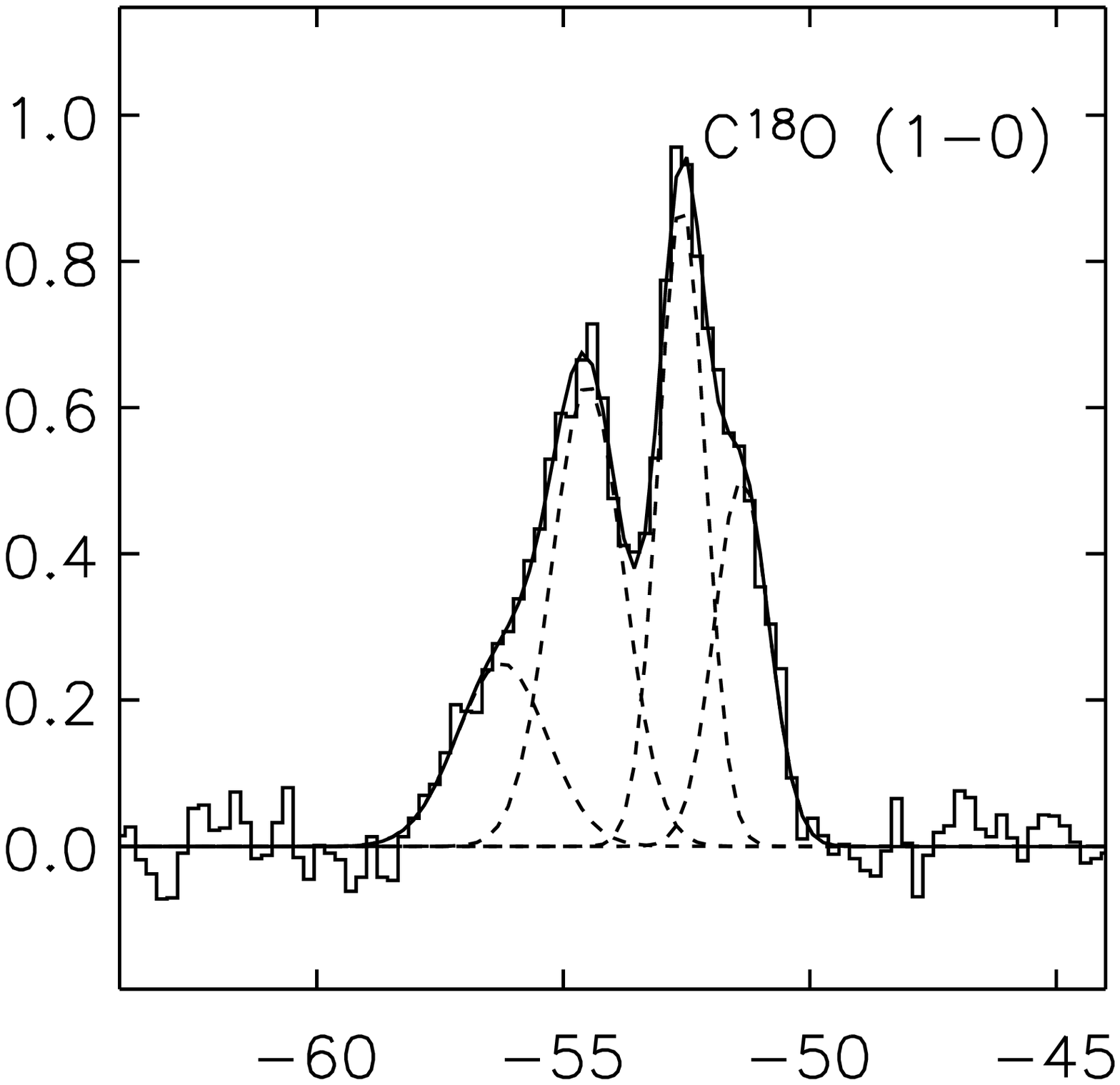}}
\hspace{-0.04\linewidth}
\subfloat{
\includegraphics[bb=75 375 660 900,width=0.25\linewidth,clip]{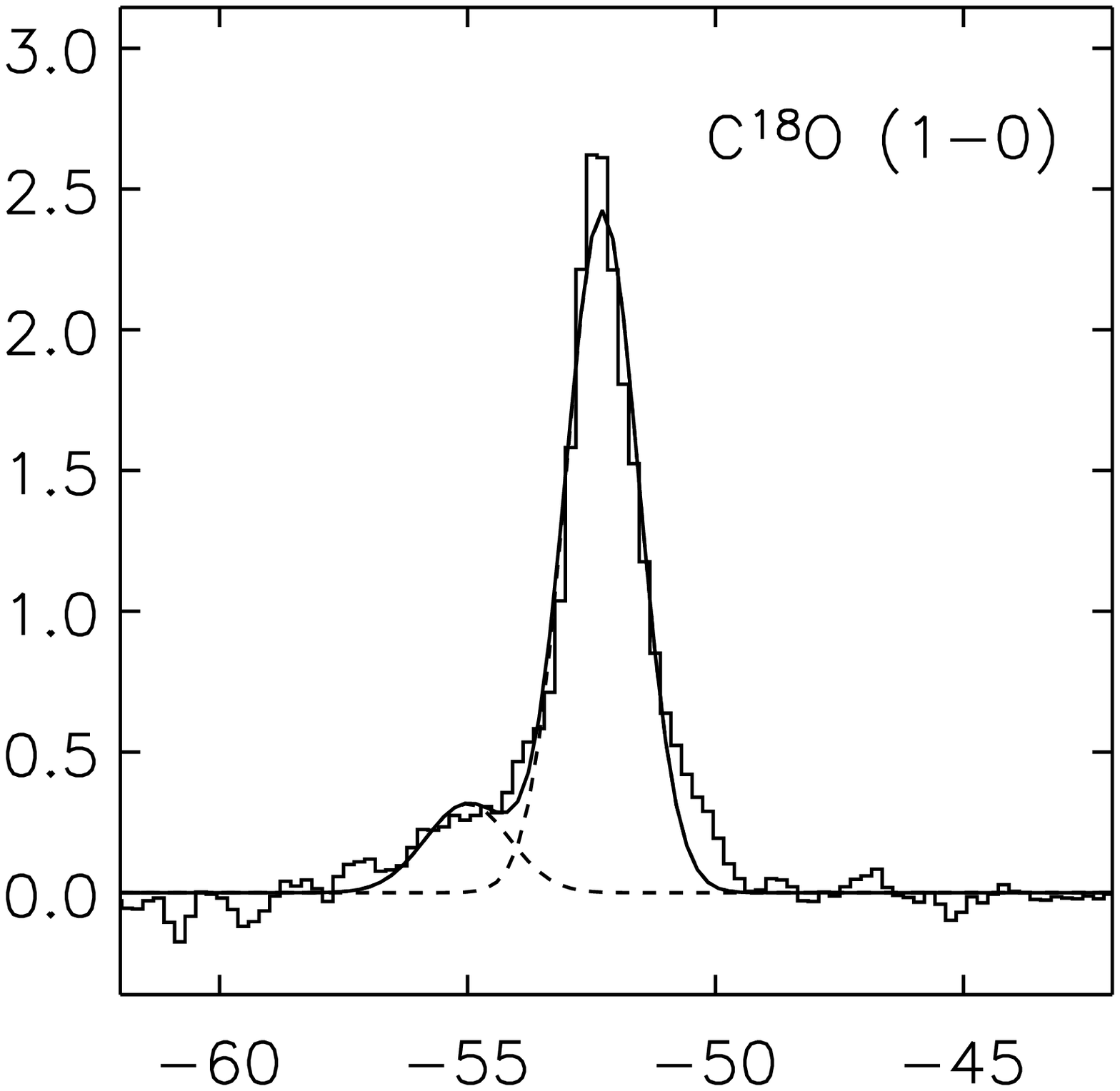}}
\hspace{-0.04\linewidth}
\subfloat{
\includegraphics[bb=75 375 660 900,width=0.25\linewidth,clip]{dummy.eps}}
\\[-10pt]
\subfloat{
\includegraphics[bb=75 375 660 900,width=0.25\linewidth,clip]{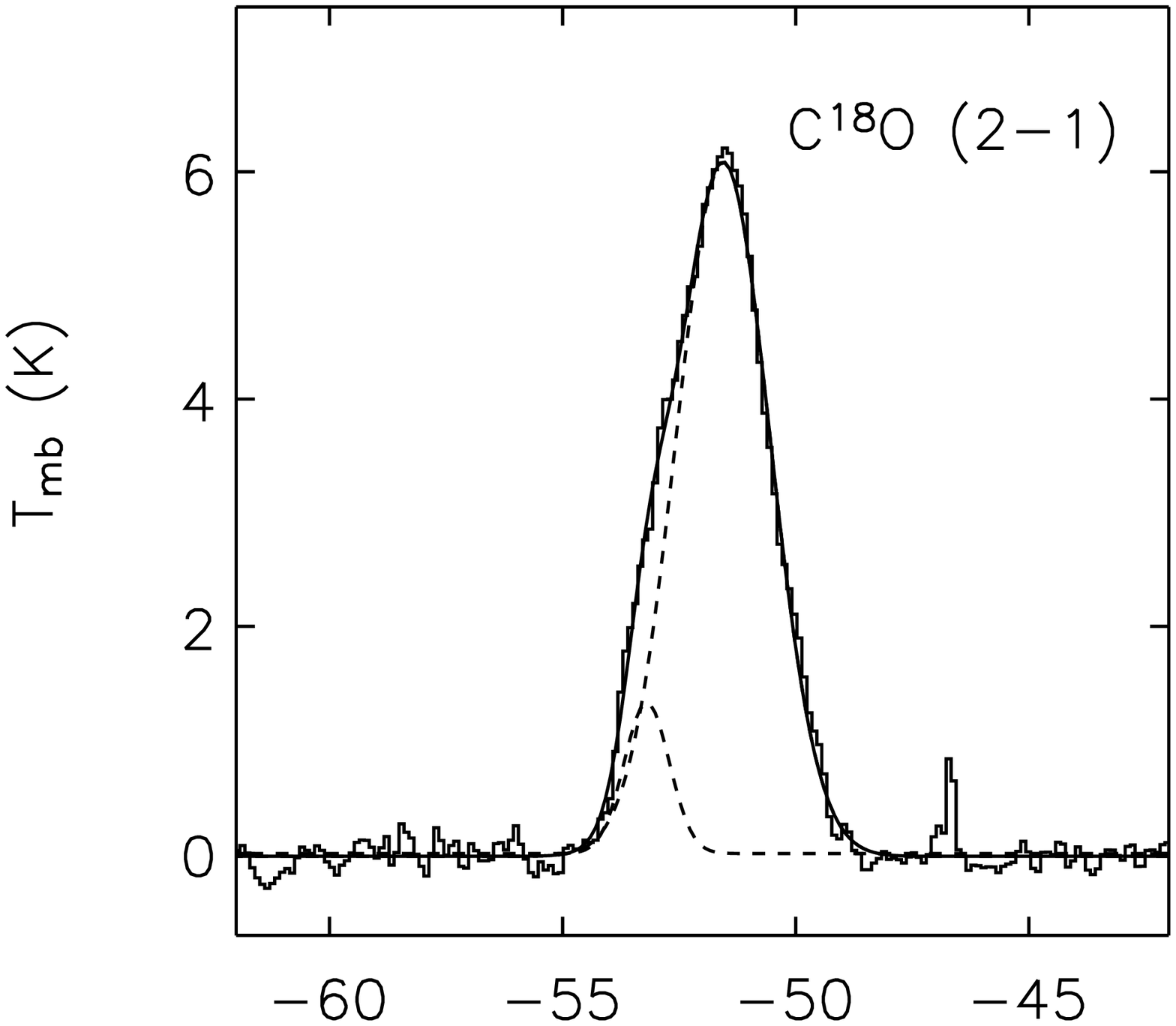}}
\hspace{-0.04\linewidth}
\subfloat{
\includegraphics[bb=75 375 660 900,width=0.25\linewidth,clip]{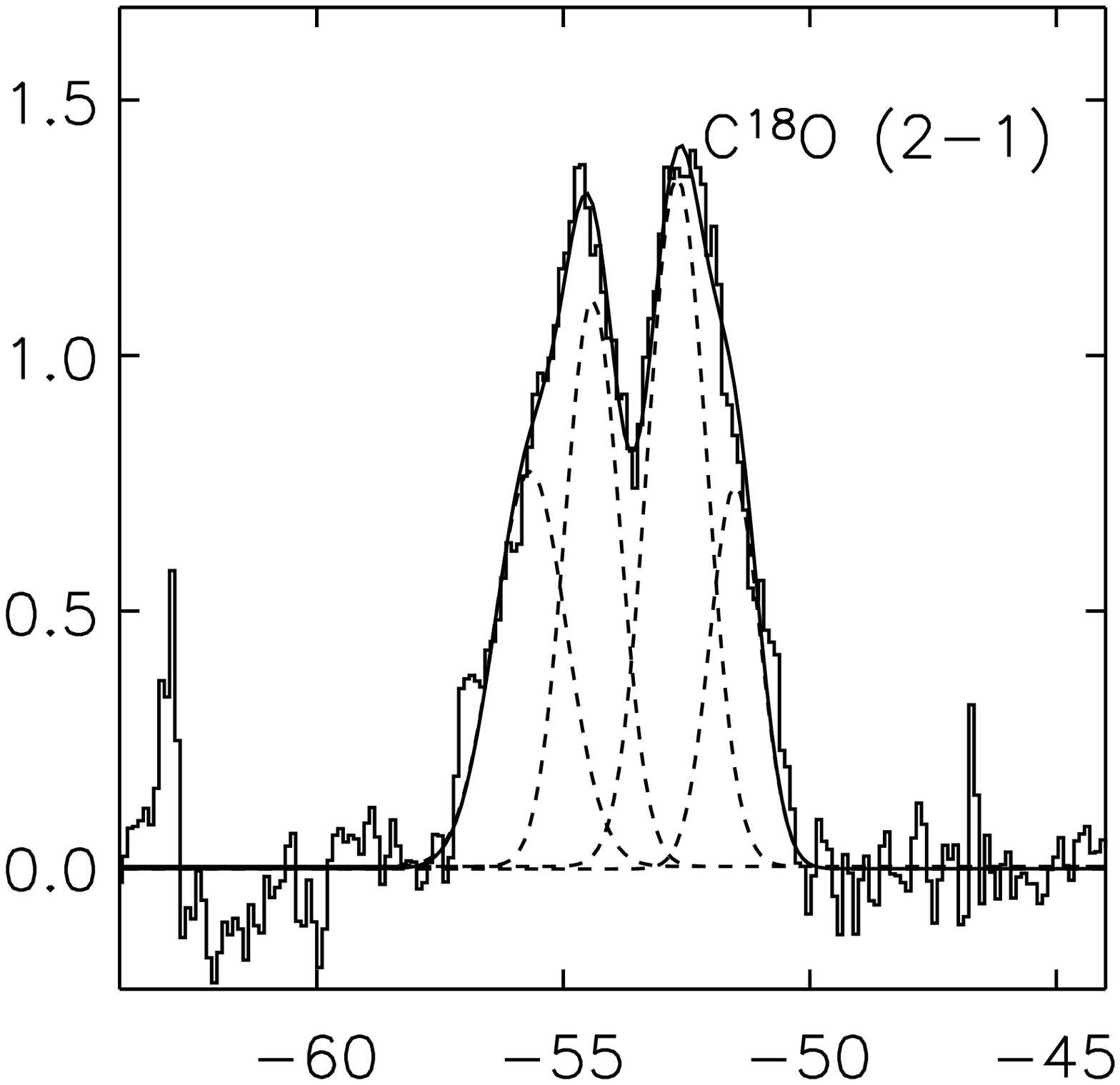}}
\hspace{-0.04\linewidth}
\subfloat{
\includegraphics[bb=75 375 660 900,width=0.25\linewidth,clip]{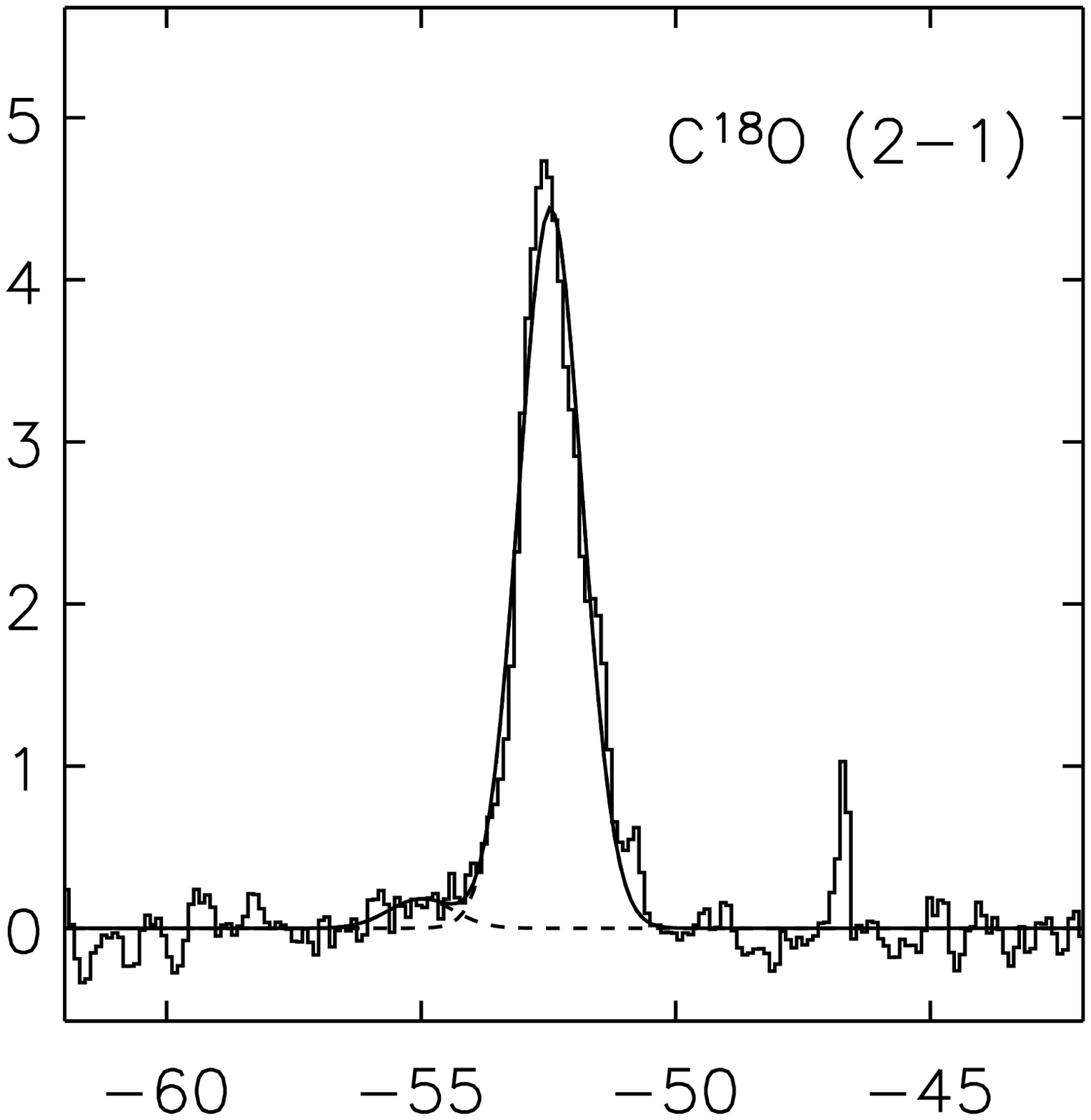}}
\hspace{-0.04\linewidth}
\subfloat{
\includegraphics[bb=75 375 660 900,width=0.25\linewidth,clip]{dummy.eps}}
\\[-10pt]
\subfloat{
\includegraphics[bb=75 375 660 900,width=0.25\linewidth,clip]{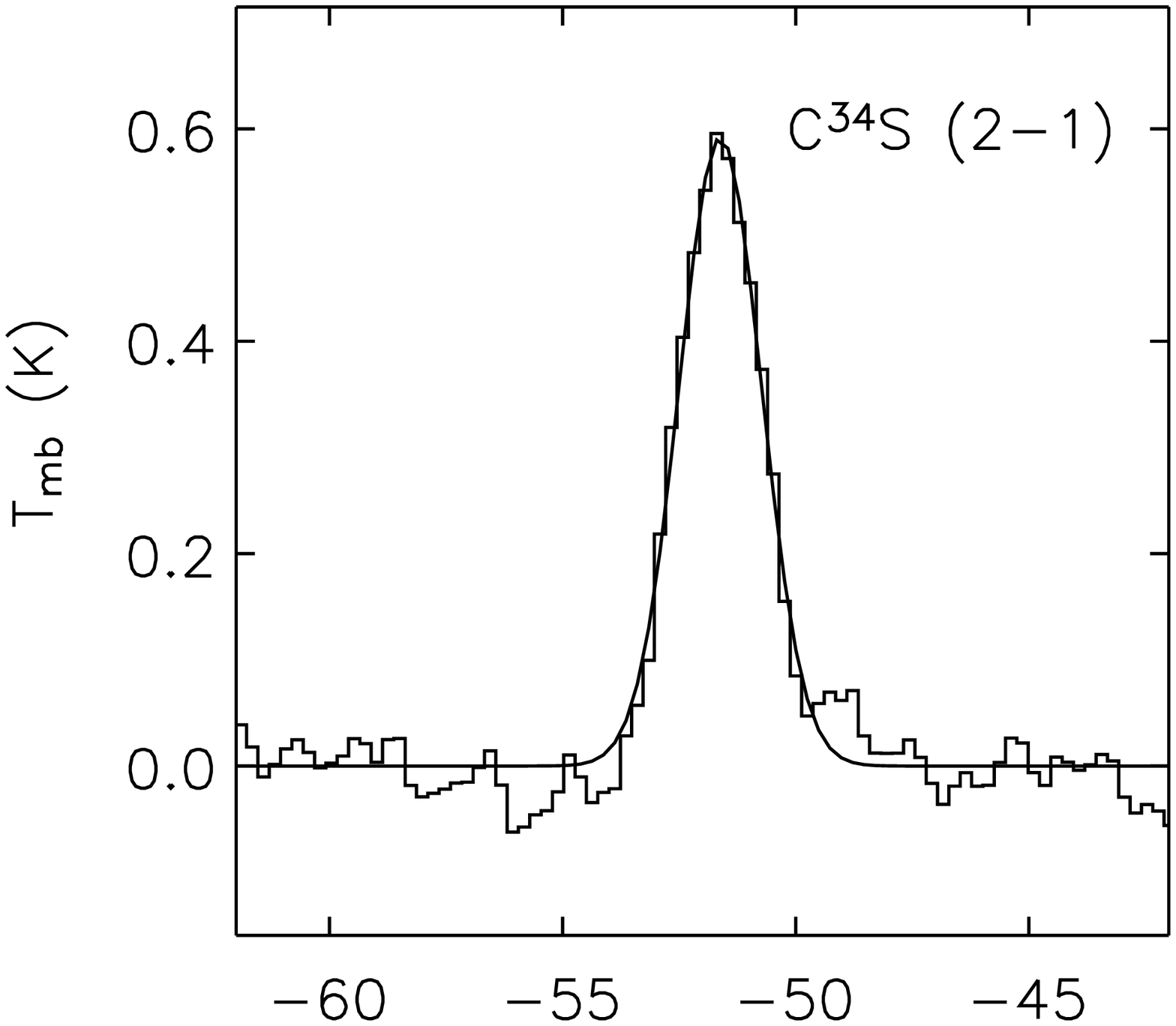}}
\hspace{-0.04\linewidth}
\subfloat{
\includegraphics[bb=75 375 660 900,width=0.25\linewidth,clip]{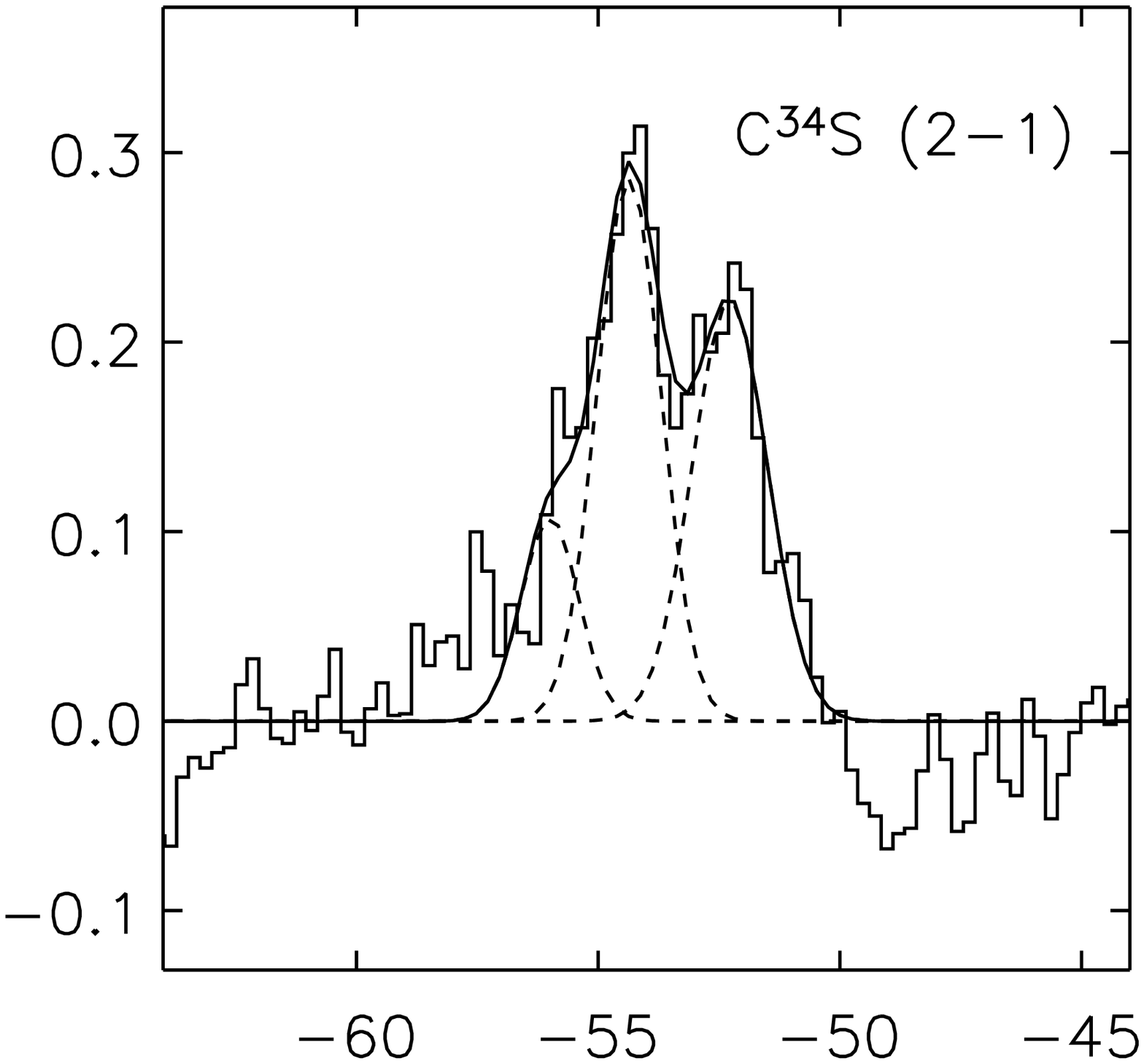}}
\hspace{-0.04\linewidth}
\subfloat{
\includegraphics[bb=75 375 660 900,width=0.25\linewidth,clip]{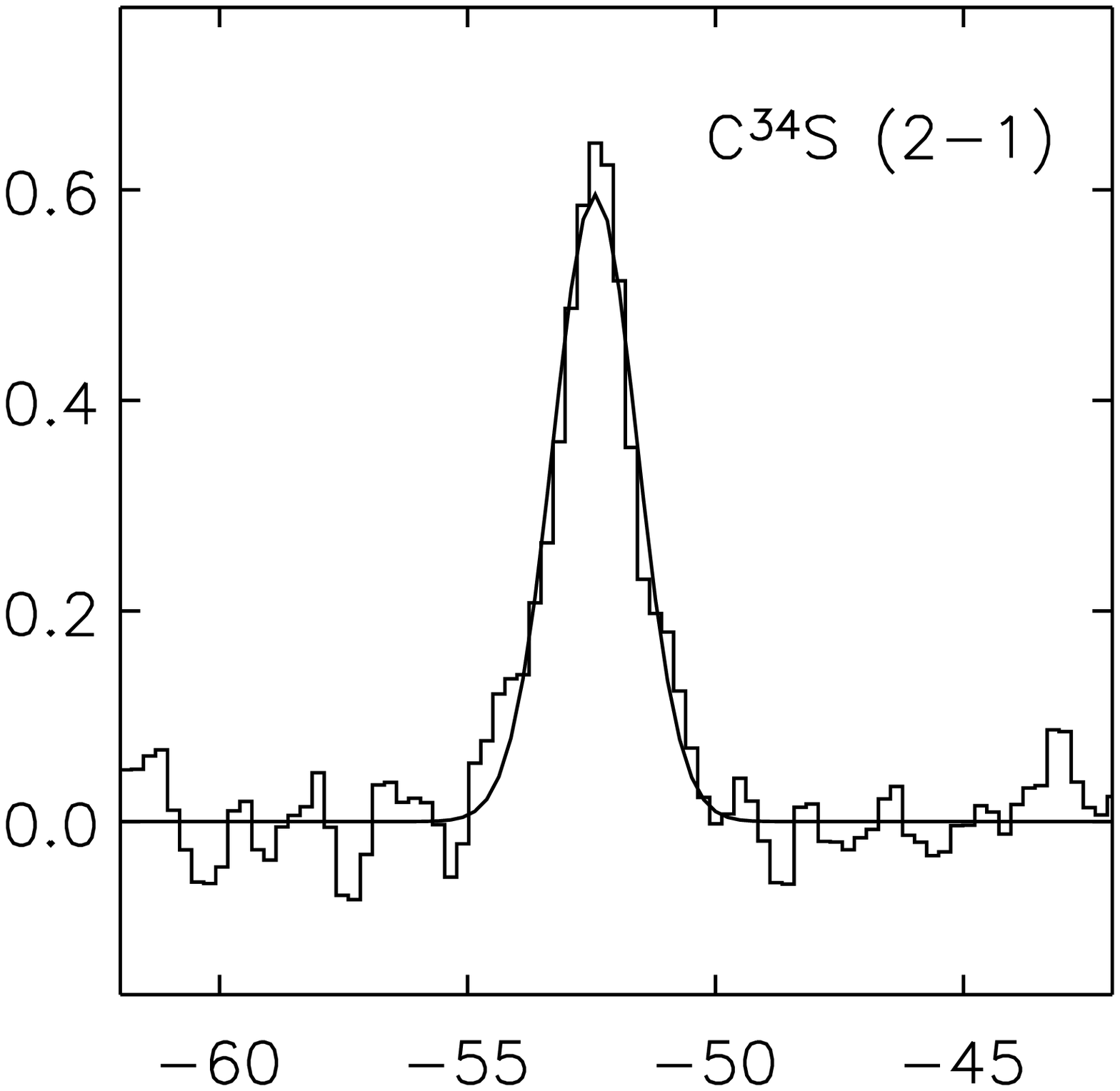}}
\hspace{-0.04\linewidth}
\subfloat{
\includegraphics[bb=75 375 660 900,width=0.25\linewidth,clip]{dummy.eps}}
\\[-10pt]
\subfloat{
\includegraphics[bb=75 375 660 900,width=0.25\linewidth,clip]{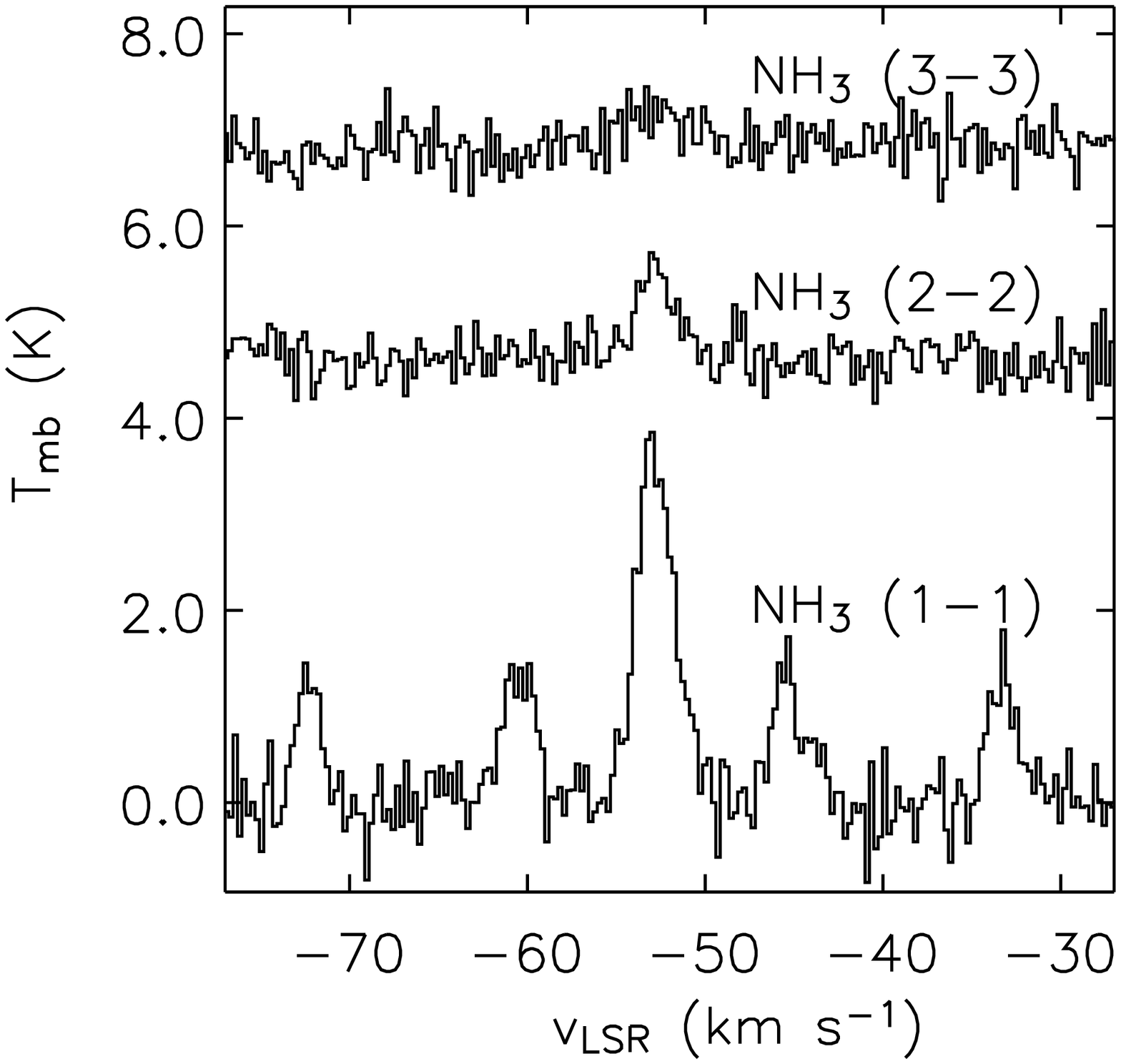}}
\hspace{-0.04\linewidth}
\subfloat{
\includegraphics[bb=75 375 660 900,width=0.25\linewidth,clip]{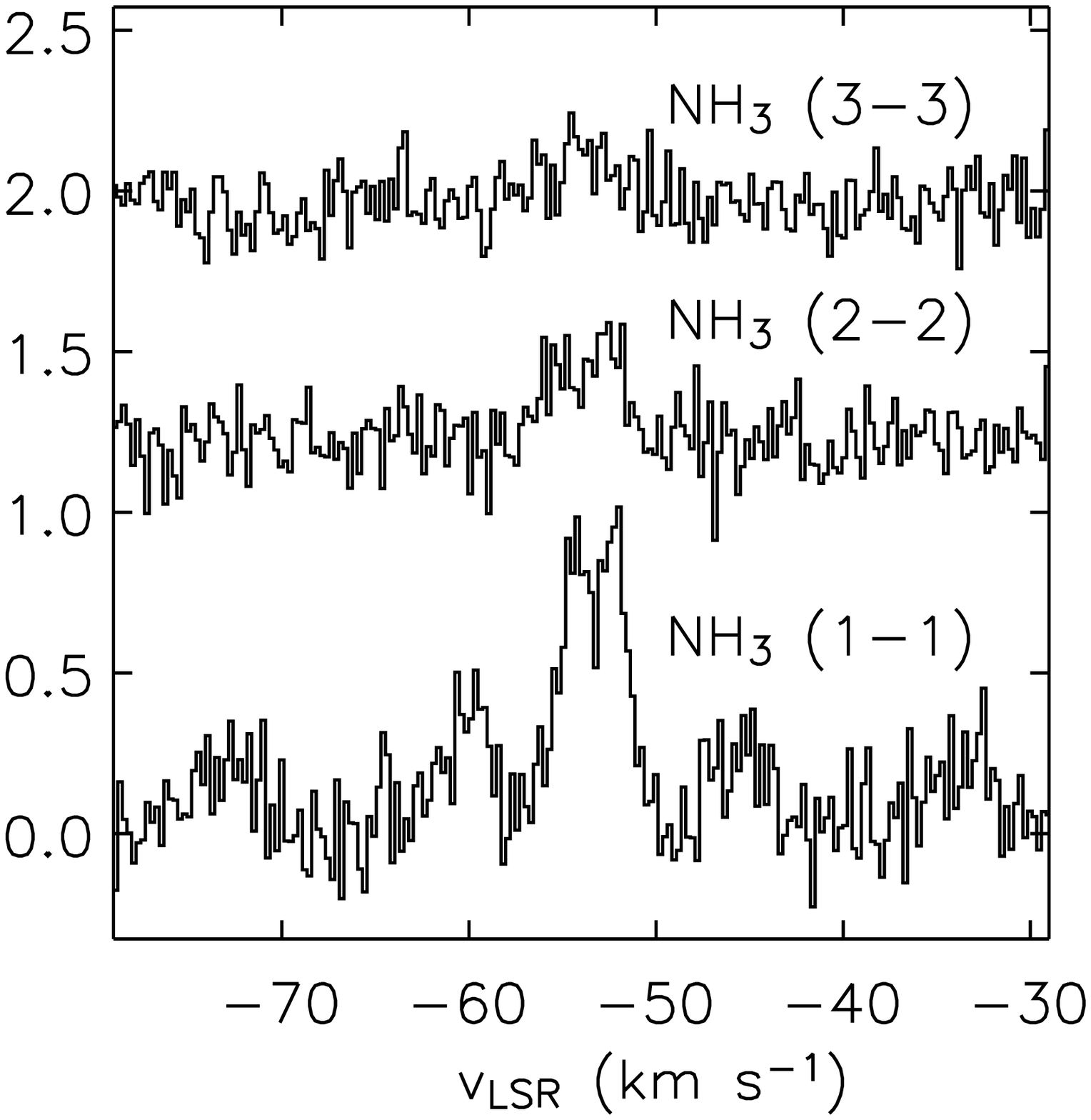}}
\hspace{-0.04\linewidth}
\subfloat{
\includegraphics[bb=75 375 660 900,width=0.25\linewidth,clip]{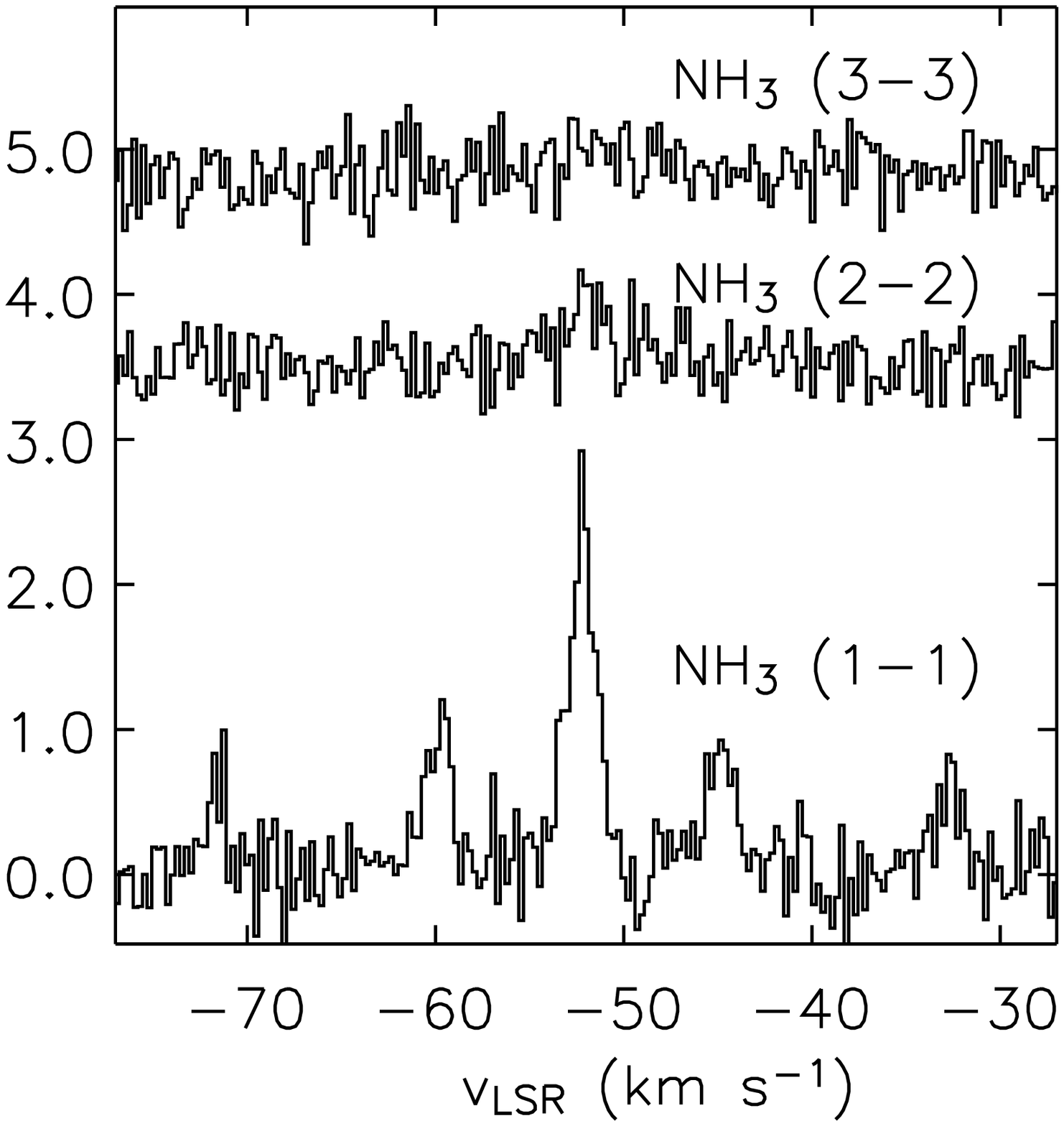}}
\hspace{-0.04\linewidth}
\subfloat{
\includegraphics[bb=75 375 660 900,width=0.25\linewidth,clip]{dummy.eps}}
\caption{Same as Figure \ref{figA1}.} 
\label{figA3} %% label for entire figure 
\end{figure*}

\end{document}